\documentclass[12pt]{article}
\usepackage{epsfig,latexsym,amsmath,amssymb,bm,cite}
\usepackage{graphicx}
\usepackage{color}
\newcommand{\be}{\begin{equation}}
\newcommand{\ee}{\end{equation}}
\newcommand{\bea}{\begin{eqnarray}}
\newcommand{\eea}{\end{eqnarray}}
\newcommand{\bem}{\begin{multline}}
\newcommand{\eem}{\end{multline}}
\newcommand{\beg}{\begin{gather}}
\newcommand{\eeg}{\end{gather}}

\newcommand{\as}{\alpha_s}

\def\eq#1{{Eq.~(\ref{#1})}}
\def\fig#1{{Fig.~\ref{#1}}}
\newcommand{\ben}{\begin{eqnarray*}}
\newcommand{\een}{\end{eqnarray*}}

\newcommand{\am}{\alpha_\mu}
\newcommand{\tr}{\text{tr}}
\newcommand{\oone}{
\begin{picture}(10,8)
\put(5,5){\circle{8}}
\put(2.9,2.5){{\scriptsize 1}}
\end{picture}
}
\newcommand{\otwo}{
\begin{picture}(10,8)
\put(5,5){\circle{8}}
\put(2.9,2.5){{\scriptsize 2}}
\end{picture}
}

\begin{document}
\title{{\bf Triumvirate of Running Couplings \\[.5cm] in Small-$x$ Evolution
\\[1.5cm] }}
\author{
{\bf Yuri V.\ Kovchegov and Heribert Weigert}
\\[1cm] {\it\small Department of Physics, The Ohio State University}\\ 
{\it\small Columbus, OH 43210,USA}\\[5mm]}

\date{September 2006}

\maketitle

\thispagestyle{empty}

\begin{abstract}
  We study the inclusion of running coupling corrections into the
  non-linear small-$x$ JIMWLK and BK evolution equations by resumming
  all powers of $\as N_f$ in the evolution kernels. We demonstrate
  that the running coupling corrections are included in the JIMWLK/BK
  evolution kernel by replacing the fixed coupling constant $\as$ in
  it with $\frac{\as (1/r_1^2) \, \as (1/r_2^2)}{\as (1/R^2)}$, where
  $r_1$ and $r_2$ are transverse distances between the emitted gluon
  and the harder gluon (or quark) off of which it was emitted to the
  left and to the right of the interaction with the target. In the
  formalism of Mueller's dipole model $r_1$ and $r_2$ are the
  transverse sizes of ``daughter'' dipoles produced in one step of the
  dipole evolution. The scale $R$ is a function of two-dimensional
  vectors ${\bf r}_1$ and ${\bf r}_2$, the exact form of which is
  scheme-dependent. We propose using a particular scheme which gives
  us $R$ as an explicit function of $r_1$ and $r_2$.
\end{abstract}

\thispagestyle{empty}

\newpage

\setcounter{page}{1}

%%%%%%%%%%%%%%%%%%%%%%%%%%%%%%%%%%%%%%%%%%%%%%%%%%%%%%%%%%%%%%%%%%%%%%%%%%%%%%%%%

\section{Introduction}

In the recent years there has been a lot of progress in small-$x$ physics due
to developments in the area of parton saturation and Color Glass Condensate
(CGC) \cite{Gribov:1981ac,Mueller:1986wy,McLerran:1994vd,McLerran:1994ka,
  McLerran:1994ni,Kovchegov:1996ty,Kovchegov:1997pc,Jalilian-Marian:1997xn,
  Jalilian-Marian:1997jx, Jalilian-Marian:1997gr, Jalilian-Marian:1997dw,
  Jalilian-Marian:1998cb, Kovner:2000pt, Weigert:2000gi,
  Iancu:2000hn,Ferreiro:2001qy,Kovchegov:1999yj, Kovchegov:1999ua,
  Balitsky:1996ub, Balitsky:1997mk, Balitsky:1998ya}. Among other things the
CGC led to a new way of calculating the hadronic and nuclear structure
functions and total cross sections in deep inelastic scattering (DIS) at small
values of Bjorken $x$ variable. According to the CGC approach to high energy
processes, one first has to calculate an observable in question in the
quasi-classical limit of the McLerran-Venugopalan model
\cite{McLerran:1994vd,McLerran:1994ka, McLerran:1994ni} which resums all
multiple rescatterings in the target hadron or nucleus. After that one has to
include the quantum evolution corrections resumming all powers of $\as \, \ln
1/x_{Bj}$ along with all the multiple rescatterings.  Such corrections are
included in the general case of a large target by the
Jalilian-Marian--Iancu--McLerran--Weigert--Leonidov--Kovner (JIMWLK)
functional integro-differential equation \cite{Jalilian-Marian:1997jx,
  Jalilian-Marian:1997gr, Jalilian-Marian:1997dw, Jalilian-Marian:1998cb,
  Kovner:2000pt, Weigert:2000gi, Iancu:2000hn,Ferreiro:2001qy}, or, if the
large-$N_c$ limit is imposed, by the Balitsky-Kovchegov (BK)
integro-differential evolution equation \cite{Balitsky:1996ub,
  Balitsky:1997mk, Balitsky:1998ya,Kovchegov:1999yj, Kovchegov:1999ua} based
on Mueller's dipole model
\cite{Mueller:1994rr,Mueller:1994jq,Mueller:1995gb,Chen:1995pa}. The JIMWLK
and BK evolution equations unitarize the Balitsky-Fadin-Kuraev-Lipatov (BFKL)
linear evolution equation \cite{Kuraev:1977fs,Bal-Lip}. For detailed reviews
of the physics of the Color Glass Condensate we refer the reader to
\cite{Iancu:2003xm,Weigert:2005us,Jalilian-Marian:2005jf}.

Both the JIMWLK and BK evolution equations resum leading logarithmic
$\as \, \ln 1/x_{Bj}$ corrections with $\as$ the coupling constant. At
this leading order the running coupling corrections to the JIMWLK and
BK evolution kernels are negligible next-to-leading order (NLO)
corrections. A running coupling correction would bring in powers of,
for instance, $\as^2 \, \ln 1/x_{Bj}$, which are not leading
logarithms anymore. Hence both JIMWLK and BK evolution equations do
not include any running coupling corrections in their kernels. The
drawback of this lack of running coupling corrections is that the
scale of the coupling constant to be used in solving these evolution
equations is not known.  Indeed, as was argued originally by McLerran
and Venugopalan \cite{McLerran:1994vd,McLerran:1994ka,
  McLerran:1994ni} and confirmed by the numerical solutions of JIMWLK
and BK equations
\cite{Braun:2000wr,Golec-Biernat:2003ym,Rummukainen:2003ns,Lublinsky:2001bc},
the high parton density in the small-$x$ hadronic and nuclear wave
functions gives rise to a hard momentum scale --- the saturation scale
$Q_s$. For small enough $x$ and for large enough nuclei this scale
becomes much larger than the QCD confinement scale, $Q_s \gg
\Lambda_{\text{QCD}}$. The existence of a large intrinsic momentum
scale leads to the expectation that this scale would enter in the
argument of the running coupling constant making it small and allowing
for a perturbative description of the relevant physical processes.
However, until now this expectation has never been confirmed by
explicit calculations.

In the past there have been several good guesses of the scale of the
running coupling in the JIMWLK and BK kernels in the literature
\cite{Albacete:2004gw,Rummukainen:2003ns}. A resummation of all-order
running coupling corrections for the linear BFKL equation in momentum
space was first performed by Levin in~\cite{Levin:1994di} by imposing
the conformal bootstrap condition. There it was first observed that to
set the scale of the running coupling constant in the BFKL kernel one
has to replace a single factor of $\as$ by the ``triumvirate'' of
couplings $\as \, \as / \as$ with each coupling having a different
argument~\cite{Levin:1994di}.

In this paper we calculate the scale of the running coupling in the
JIMWLK and BK evolution kernels. Our strategy is similar to
\cite{Gardi:2006}: we note however, that \cite{Gardi:2006} relies on
the dispersive method to determine the running coupling corrections,
while below we use a purely diagrammatic approach.  We concentrate on
corrections due to fermion (quark) bubble diagrams, which bring in
factors of $\as \, N_f$. Indeed some factors of $N_f$ may come from
the QCD beta-function (see \eq{beta} below), while other factors of
$N_f$ may come in from conformal (non-running coupling) NLO (and
higher order) corrections
\cite{Fadin:1998py,Ciafaloni:1998gs,Brodsky:1998kn}. While we do not
know how to separate the two contributions uniquely, we propose a way
of distinguishing them guided by UV divergences. This leaves us with
an uncertainty with respect to finite contributions in separating the
conformal and the running coupling factors of $\as \, N_f$ that
influence the scale of the obtained running coupling constant in a way
reminiscent of the scheme dependence. Once we pick a certain way of
singling out the factors of $\as \, N_f$ coming from the QCD
beta-function, we replace $N_f \rightarrow - \, 6 \, \pi\, \beta_2$
(``completing'' $N_f$ to the full beta-function) and obtain all the
running coupling corrections to the JIMWLK and BK kernels at the
one-loop beta-function level.

The paper is structured as follows. We begin in Section \ref{LObb} by
calculating the lowest order fermion bubble correction to the JIMWLK
and BK kernels, as shown in Figs. \ref{fig:NLO1} and
\ref{fig:NLO1_inst}, in the framework of the light cone perturbation
theory (LCPT) \cite{Lepage:1980fj,Brodsky:1997de}. We note that the
diagrams in Figs.  \ref{fig:NLO1}A and \ref{fig:NLO1_inst}A$^\prime$
give a new kind of evolution kernel, which does not look like a higher
order correction to the LO JIMWLK or BK kernels. We analyze the
problem in Section \ref{Subtr}, where we propose a subtraction
procedure to single out the part of these diagrams' contribution
giving the running coupling correction. There we show that this
subtraction procedure is not unique and introduces a scheme dependence
into the scale of the running coupling.

In Section \ref{all_orders} we resum fermion bubble corrections to all orders,
and, after the $N_f \rightarrow - 6 \, \, \pi\, \beta_2$ replacement obtain
the JIMWLK evolution kernel with the running coupling correction given by
\eq{Kall} in transverse momentum space as a double Fourier transform.  The
corresponding BK kernel is obtained from \eq{Kall} using \eq{KrcBK}. Notice
that the running coupling comes in as a ``triumvirate'' originally derived by
Levin for the BFKL evolution equation~\cite{Levin:1994di}.
Fourier-transforming the running couplings into transverse coordinate space is
more involved since one encounters integration over Landau pole leading to
power corrections. A careful treatment of the uncertainties associated with
power corrections in small-$x$ evolution was performed in~\cite{Gardi:2006}.
Here we calculate the Fourier transforms by simply ignoring those corrections
and by using the Brodsky-Lepage-Mackenzie (BLM) method \cite{BLM} to set the
scale of the running coupling. The JIMWLK kernel with the running coupling
corrections in the transverse coordinate space is given in \eq{Krc}.

We conclude in Section \ref{conc} by explicitly writing down the full
JIMWLK Hamiltonian with the running coupling corrections in
\eq{JIMWLKrc} and the full BK evolution equation with the running
coupling corrections in \eq{eqNrc} and by discussing various limits of
the obtained result.

We note that our analysis is complimentary to \cite{Gardi:2006}, where
the running coupling correction to the JIMWLK and BK kernels was
determined using the dispersive method. Our result for the all-order
series of $\as \, N_f$-terms is the same as in \cite{Gardi:2006}.
However, using the diagrammatic approach, we have been able to
identify the structure of that series as coming from a ``triumvirate''
of the coupling constants in \eq{Krc}, which is an exact result in the
transverse momentum space and a better approximation of the full
answer in the transverse coordinate space.

%%%%%%%%%%%%%%%%%%%%%%%%%%%%%%%%%%%%%%%%%%%%%%%%%%%%%%%%%%%%%%%%%%%%%%%%%%%%%%%%%

\section{Leading Order Fermion Bubbles}
\label{LObb}

Our goal in this work is to resum all $\as N_f$ corrections to the
leading logarithmic non-linear JIMWLK and BK small-$x$ evolution
equations \cite{Kovchegov:1999yj, Kovchegov:1999ua, Balitsky:1996ub,
  Balitsky:1997mk, Balitsky:1998ya,Jalilian-Marian:1997xn,
  Jalilian-Marian:1997jx, Jalilian-Marian:1997gr,
  Jalilian-Marian:1997dw, Jalilian-Marian:1998cb, Kovner:2000pt,
  Weigert:2000gi, Iancu:2000hn,Ferreiro:2001qy} (for review
see~\cite{Iancu:2002xk, Iancu:2003xm, Weigert:2005us,
  Jalilian-Marian:2005jf}). After extracting the running coupling $\as
N_f$-corrections out of all possible $\as N_f$ terms, the complete
running coupling correction to the JIMWLK and BK evolution kernels
would then be easy to obtain by replacing
\begin{align}\label{repl}
N_f \rightarrow - 6 \, \pi \, \beta_2
\end{align} 
in the former, where
\begin{align}\label{beta}
\beta_2 = \frac{11 N_c - 2 N_f}{12 \, \pi}.  
\end{align}

\begin{figure}[htbp]
  \centering
  \begin{minipage}{5.2cm}
\centering
\includegraphics[width=5.2cm]{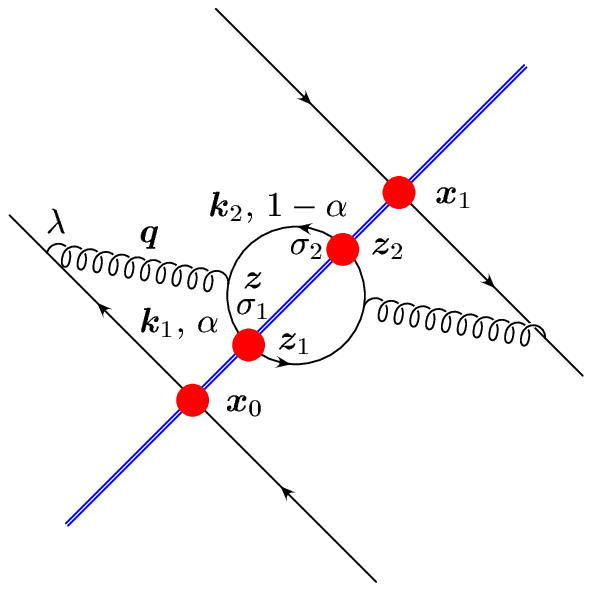}\\
A
  \end{minipage}
  \begin{minipage}{5.2cm}  
\centering
  \includegraphics[width=5.2cm]{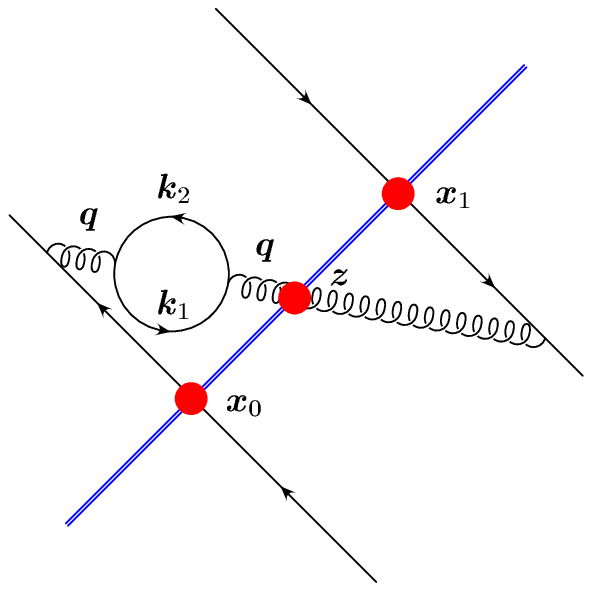}  
\\ B
\end{minipage}  
\begin{minipage}{5.2cm} 
\centering 
  \includegraphics[width=5.2cm]{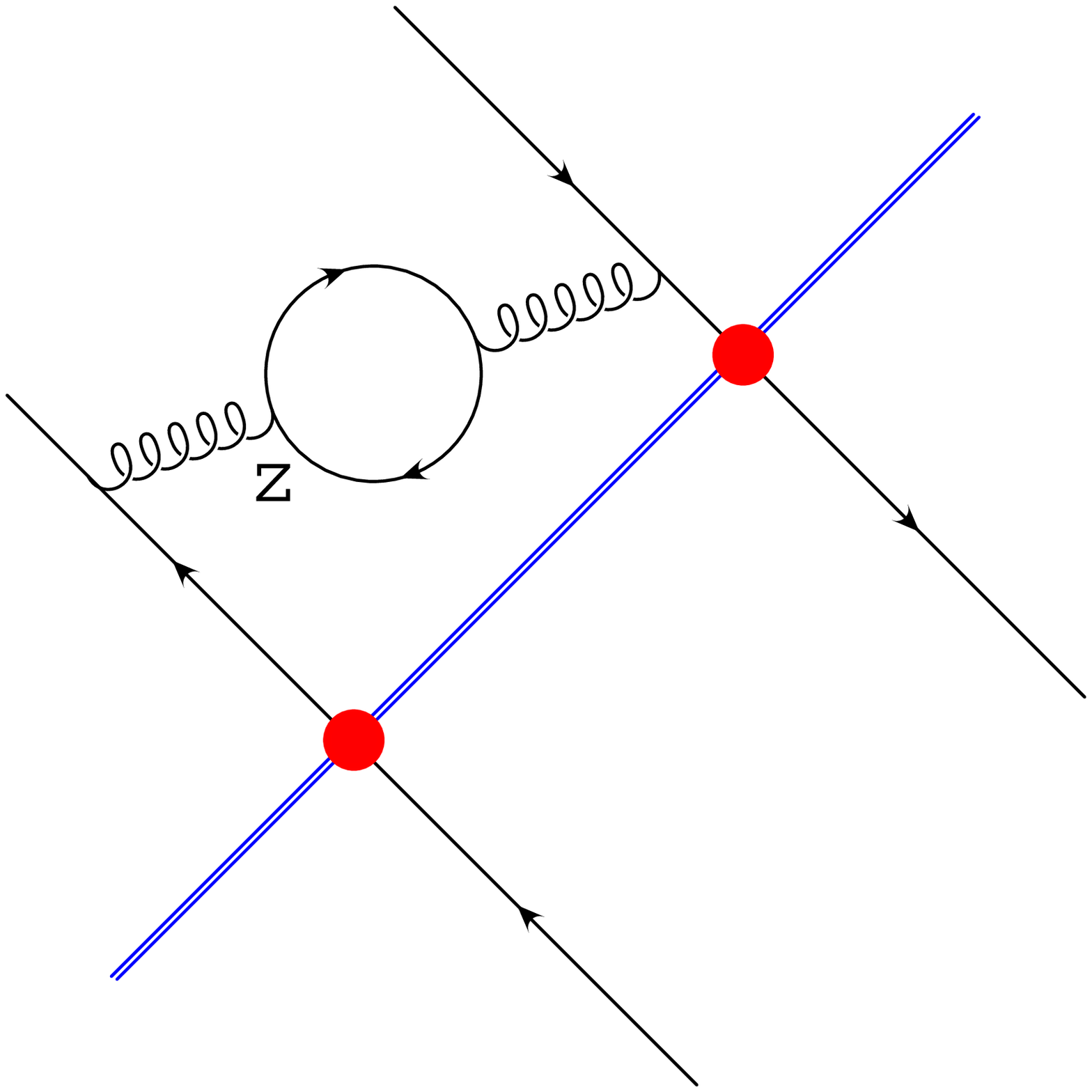}
\\ C
\end{minipage} 
  \caption{\em Diagrams giving the leading $\as N_f$ correction 
    to the kernels of JIMWLK and BK small-$x$ evolution equations. The
    thick dots on gluon and quark lines denote interactions with the
    target.}
  \label{fig:NLO1}
\end{figure}

\begin{figure}[htbp]
  \centering
  \begin{minipage}{5.2cm}
\centering
\includegraphics[width=5.2cm]{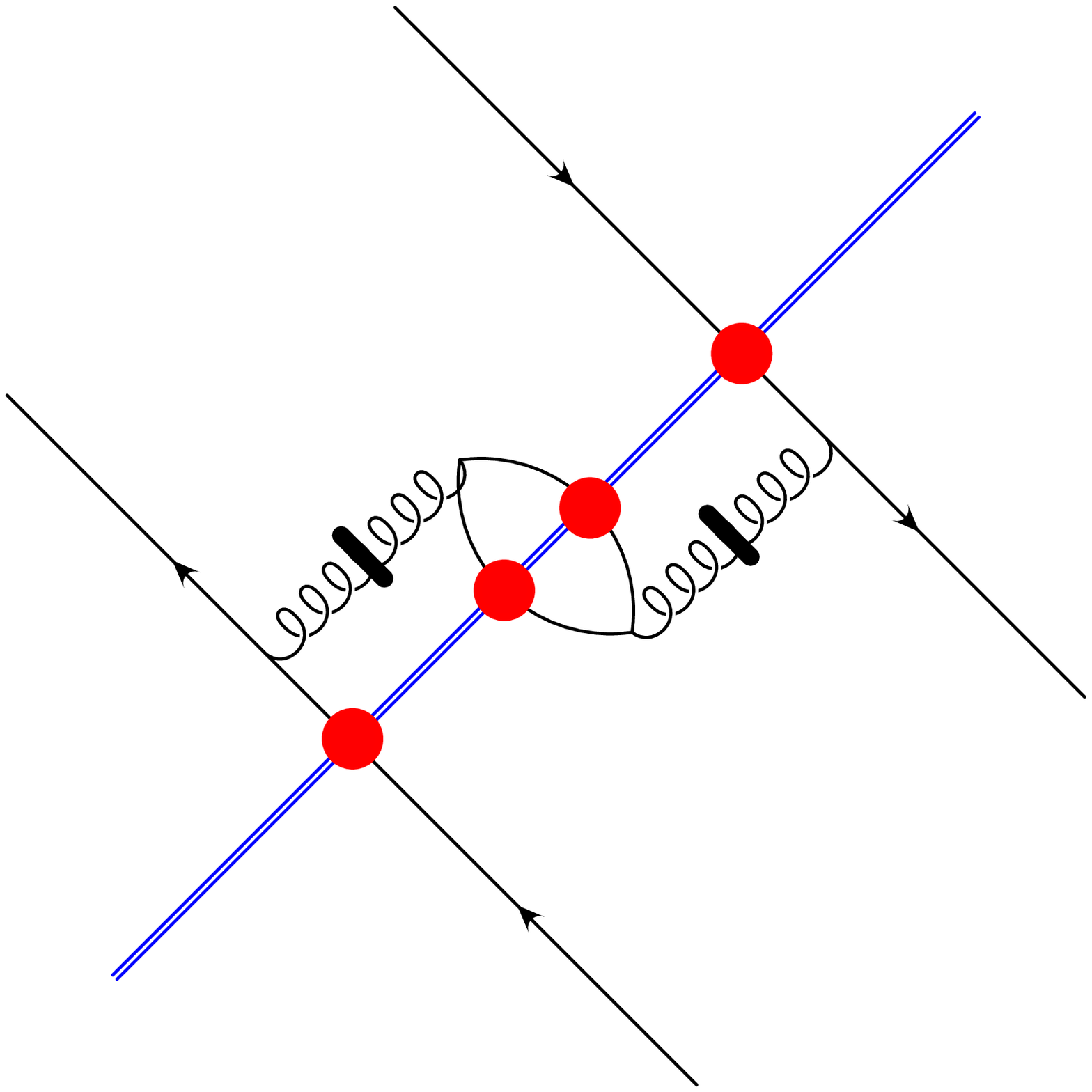}\\
A$^\prime$
  \end{minipage}
\hspace{5.2cm}
\begin{minipage}{5.2cm} 
\centering 
  \includegraphics[width=5.2cm]{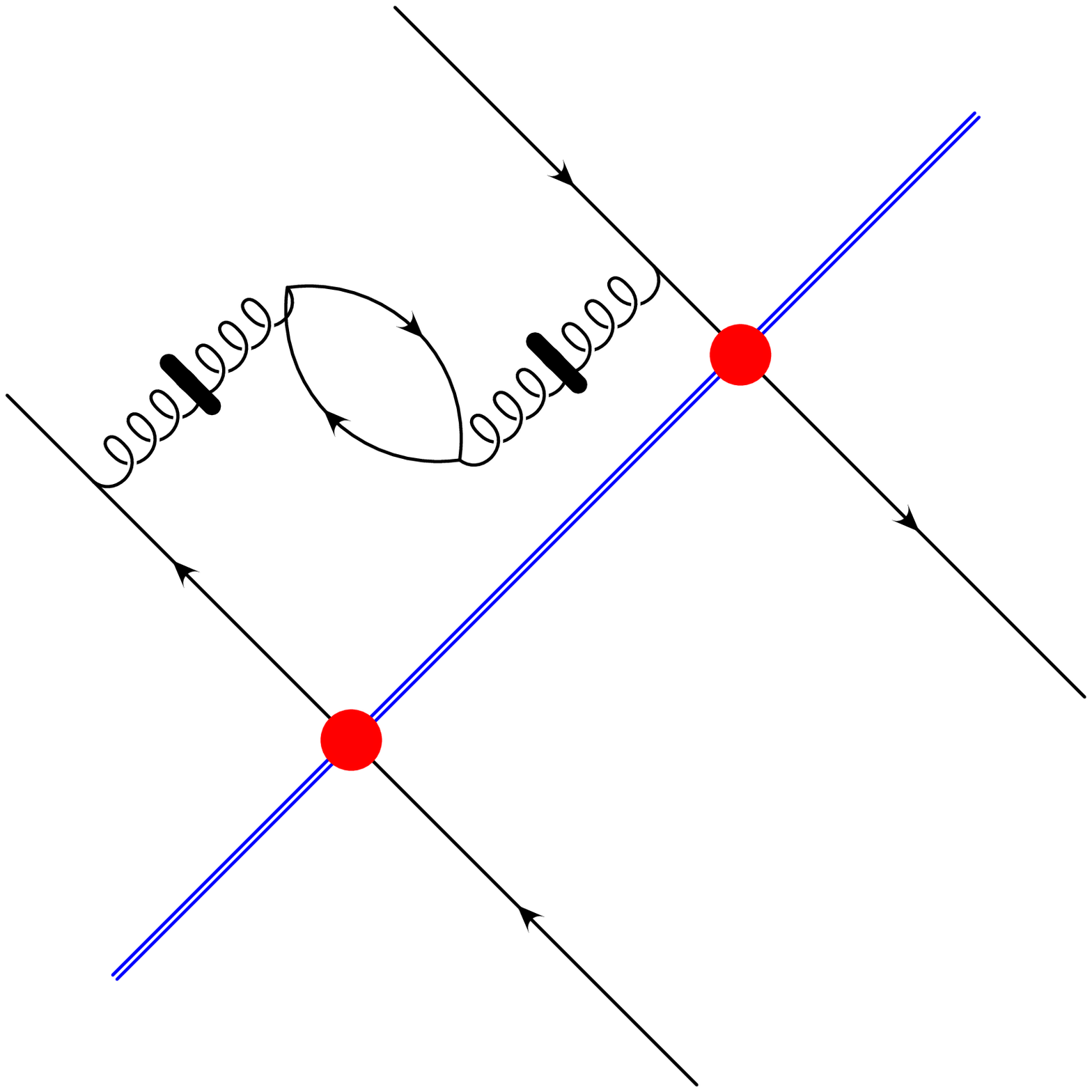}
\\ C$^\prime$
\end{minipage} 
  \caption{\em Diagrams with instantaneous parts of gluon propagators 
    giving the leading $\as N_f$ correction to the kernels of JIMWLK
    and BK small-$x$ evolution equations. There is no analog of
    Fig.~\ref{fig:NLO1} B. All the lines are implied to
    be labeled in the same way as in \fig{fig:NLO1}.}
  \label{fig:NLO1_inst}
\end{figure}

To resum $\as N_f$ corrections we begin by considering the lowest order
diagrams for one step of small-$x$ evolution containing a single quark bubble.
These diagrams give the lowest order $\as N_f$ correction to the JIMWLK and/or
BK evolution kernels and are shown in~\fig{fig:NLO1}. The diagrams are
time-ordered as they are drawn according to the rules of LCPT
\cite{Lepage:1980fj,Brodsky:1997de}.  Gluon lines in Figs.~\ref{fig:NLO1} A
and C also have instantaneous/longitudinal
counterparts~\cite{Lepage:1980fj,Brodsky:1997de}, shown in diagrams A$^\prime$
and C$^\prime$ in \fig{fig:NLO1_inst}.  (The virtual gluon on the left side of
\fig{fig:NLO1}B can not be instantaneous, since the produced gluon on the
right of \fig{fig:NLO1}B can only be transverse and a longitudinal gluon can
not interfere with a transverse gluon, as will be seen in the calculations
done below.)

%%%%%%%%%%%%%%%%%%%%%%%%%%%%%%%%%%%%%%%%%%%%%%%%%%%%%%%%%%%%%%%%%%%%%%%

\subsection{Diagrams A and A$^\prime$}

To calculate the forward scattering amplitude in Figs.  \ref{fig:NLO1}A and
\ref{fig:NLO1_inst}A$^\prime$ we first need to calculate the wave function of
a dipole emitting a gluon which then, in turn, splits into a quark--anti-quark
pair, i.e., the part of the diagrams A and A$^\prime$ located on one side of
the interaction with the target. The calculation is similar to what is
presented in~\cite{Kovchegov:2006qn}. We will work in the $A_+ =0$ light cone
gauge in the framework of the light cone perturbation theory
\cite{Lepage:1980fj,Brodsky:1997de,Perry:1992sw,Mustaki:1990im}. The momentum
space wave function of a dipole (or a single (anti-)quark) splitting into a
gluon which in turn splits into a $q\bar q$ pair with the transverse momenta
${\bm k}_1$ and ${\bm k}_2$ of the quark and the anti-quark with the quark
carrying a fraction $\alpha$ of the gluon's longitudinal (``plus'') momentum
is \cite{Kovchegov:2006qn}
\begin{align}\label{eq:psi_1k}
  \Psi^{(1)}_{\sigma_1, \, \sigma_2} (\bm k_1, \bm k_2, \alpha) \, =
  \, [t^a]_{\text{em}}\otimes [t^a]_{\text{f}} \,
  \psi^{(1)}_{\sigma_1, \, \sigma_2} (\bm k_1, \bm k_2, \alpha) = - 2
  \, g^2 [t^a]_{\text{em}}\otimes [t^a]_{\text{f}}
  \sum\limits_{\lambda = \pm 1} \frac{{\bm
      \epsilon}^{*\lambda}\cdot({\bm k}_1+{\bm k}_2)}{
    ({\bm k}_1+{\bm k}_2)^2} \nonumber \\
  \times \, \frac{{\bm \epsilon}^{\lambda}\cdot [{\bm
      k}_1(1-\alpha)-{\bm k}_2\alpha] (1-2\alpha+\lambda\sigma_1) \,
    \delta_{\sigma_1\sigma_2}}{ {\bm k}_1^2(1-\alpha)+{\bm
      k}_2^2\alpha} - 4 \, g^2 [t^a]_{\text{em}}\otimes
  [t^a]_{\text{f}} \, \frac{\alpha \, (1-\alpha) \,
    \delta_{\sigma_1\sigma_2}}{ {\bm k}_1^2(1-\alpha)+{\bm
      k}_2^2\alpha}.
\end{align}
Here $\lambda = \pm 1$ is the internal gluon's polarization: the gluon
polarization vector for transverse gluons is given by
$\epsilon_\mu^\lambda = (0,0, {\bm \epsilon}^\lambda)$ with $\bm
\epsilon^\lambda = (1 + i \, \lambda) / \sqrt{2}$. The instantaneous
diagram from \fig{fig:NLO1_inst}A$^\prime$ gives the second term on
the right hand side of \eq{eq:psi_1k}.  The produced quark and
anti-quark are massless, which is sufficient for our purposes of
determining the scale of the running coupling.  $\sigma_1 = \pm 1$ and
$\sigma_2 = \pm1$ are quark and anti-quark helicities correspondingly
(defined as in \cite{Kovchegov:2006qn}).  The fraction the of gluon's
``plus'' momentum carried by the quark is denoted by $\alpha \equiv
k_{1+} / (k_{1+} + k_{2+})$.  The wave function also contains a color
factor $[t^a]_{\text{em}}\otimes [t^a]_{\text{f}}$ consisting of two
color matrices originating in the quark-gluon vertices at the points
of emission of the gluon and its splitting into a $q\bar q$ pair.

It is convenient to rewrite \eq{eq:psi_1k} in terms of a different set
of transverse momenta. Defining the momentum of the gluon ${\bm q} =
{\bm k}_1+{\bm k}_2$ and ${\bm k}={\bm k}_1(1-\alpha)-{\bm
  k}_2\alpha$, and noting that ${\bm k}_1^2(1-\alpha)+ {\bm
  k}_2^2\alpha={\bm k}^2+{\bm q}^2\alpha(1-\alpha)$, we write
\begin{align}\label{eq:psi_1kq}
  \psi^{(1)}_{\sigma_1, \, \sigma_2} (\bm k, \bm q, \alpha) = - 2 \,
  g^2 \, \sum\limits_{\lambda = \pm 1} \frac{{\bm
      \epsilon}^{*\lambda}\cdot{\bm q}}{ {\bm q}^2} \frac{{\bm
      \epsilon}^{\lambda}\cdot{\bm k}\, (1-2\alpha+\lambda\sigma_1) \,
    \delta_{\sigma_1\sigma_2}}{{\bm k}^2+{\bm q}^2\alpha(1-\alpha)} -
  4 \, g^2 \, \frac{\alpha \, (1-\alpha) \,
    \delta_{\sigma_1\sigma_2}}{ {\bm k}^2+{\bm q}^2\alpha(1-\alpha)}.
\end{align}
Performing the summation over gluon polarizations $\lambda$ yields
\begin{align}\label{eq:psi_1kq2}
  \psi^{(1)}_{\sigma_1, \, \sigma_2} (\bm k, \bm q, \alpha) = - 2 \,
  g^2 \, \frac{{\bm q}_i}{ {\bm q}^2} \, \left[(1-2\alpha)\delta_{i j}
    +i \sigma_1 \epsilon_{i j}\right] \, \delta_{\sigma_1\sigma_2} \,
  \frac{{\bm k}_j}{{\bm k}^2+{\bm q}^2\alpha(1-\alpha)} - 4 \, g^2 \,
  \frac{\alpha \, (1-\alpha) \, \delta_{\sigma_1\sigma_2}}{ {\bm
      k}^2+{\bm q}^2\alpha(1-\alpha)}
\end{align}
where $q_i$ denotes the $i$th component of vector $\bm q$ and the sum
over repeated indices $i,j = 1,2$ is implied. Here $\epsilon_{12} = 1
= - \epsilon_{21}$, $\epsilon_{11} = \epsilon_{22} = 0$, and, assuming
summation over repeating indices, $\epsilon_{ij} \, q_i \, k_j = q_x
\, k_y - q_y \, k_x$.

To find the contribution of the diagrams in Figs. \ref{fig:NLO1}A and
\ref{fig:NLO1_inst}A$^\prime$ to the next-to-leading order (NLO)
evolution kernel we first have to transform the wave function from
\eq{eq:psi_1kq2} into transverse coordinate space
\begin{align}\label{ft}
  \psi^{(1)}_{\sigma_1, \, \sigma_2} ({\bm z}_1 - {\bm x}_m, {\bm z}_2
  - {\bm x}_m, \alpha) \, = \, \int \frac{d^2 k_1}{(2 \, \pi)^2} \,
  \frac{d^2 k_2}{(2 \, \pi)^2} \, e^{- i {\bm k}_1 \cdot ({\bm z}_1 -
    {\bm x}_m) - i {\bm k}_2 \cdot ({\bm z}_2 - {\bm x}_m)} \,
  \psi^{(1)}_{\sigma_1, \, \sigma_2} (\bm k_1, \bm k_2, \alpha).
\end{align}
Here the transverse coordinates of the quark and the anti-quark are
taken to be ${\bm z}_1$ and ${\bm z}_2$ correspondingly. The gluon in
\fig{fig:NLO1} can be emitted either off the quark or off the
anti-quark in the incoming ``parent'' dipole. The transverse
coordinates of the quark and the anti-quark in the ``parent'' dipole
are ${\bm x}_0$ and ${\bm x}_1$. In \eq{ft} we labeled them ${\bm
  x}_m$ with $m = 0,1$ depending on whether the gluon was emitted off
the quark or off the anti-quark.

In terms of transverse momenta $\bm k$ and $\bm q$ \eq{ft} can be
written as
\begin{align}\label{ft_kq}
  \psi^{(1)}_{\sigma_1, \, \sigma_2} ({\bm z}_1 - {\bm x}_m, {\bm z}_2
  - {\bm x}_m, \alpha) \, = \, \int \frac{d^2 k}{(2 \, \pi)^2} \,
  \frac{d^2 q}{(2 \, \pi)^2} \, e^{- i {\bm k} \cdot {\bm z}_{12} - i
    {\bm q} \cdot ({\bm z} - {\bm x}_m)} \, \psi^{(1)}_{\sigma_1, \,
    \sigma_2} (\bm k, \bm q, \alpha),
\end{align}
where 
\begin{align}\label{z12}
{\bm z}_{12} = {\bm z}_1 -{\bm z}_2
\end{align}
and 
\begin{align}\label{z}
  {\bm z} = \alpha \, {\bm z}_1+(1-\alpha) \, {\bm z}_2
\end{align}
is the transverse position of the gluon. 

Substituting the wave function from \eq{eq:psi_1kq2} into \eq{ft_kq}
and performing the integrations over $\bm k$ and $\bm q$ yields
\begin{align}\label{psi1_xy}
  \psi^{(1)}_{\sigma_1, \, \sigma_2} ({\bm z}_1 - {\bm x}_m, {\bm z}_2
  - {\bm x}_m, \alpha) \, = \, 2 \, g^2 \, \frac{1}{(2 \pi)^2} \,
  \frac{({\bm z} - {\bm x}_m)_i }{ ({\bm z} - {\bm x}_m)^2 + \alpha \,
    (1-\alpha) \,
    z_{12}^2} \nonumber \\
  \times \, \left[(1-2\alpha)\delta_{i j} +i \sigma_1 \epsilon_{i
      j}\right] \, \delta_{\sigma_1\sigma_2} \, \frac{({\bm
      z}_{12})_j}{{\bm z}_{12}^2} - 4 \, g^2 \, \frac{1}{(2 \pi)^2} \,
  \frac{\alpha \, (1-\alpha) \, \delta_{\sigma_1\sigma_2}}{ ({\bm z} -
    {\bm x}_m)^2 + \alpha \, (1-\alpha) \, z_{12}^2}.
\end{align}

To calculate the diagram in Figs.  \ref{fig:NLO1}A and
\ref{fig:NLO1_inst}A$^\prime$ using the wave function from
\eq{psi1_xy} in a general case we have to include the interaction with
the target by defining path-ordered exponential factors in the
fundamental representation
\begin{align}\label{U}
  U_{\bm x} \, = \, {\sf P} \exp \left[ - i g
    \int\limits_{-\infty}^\infty \, d x_+ \, A_- (x_+, x_- =0, {\bm
      x}) \right].
\end{align}
With the help of \eq{U} we can write down
\begin{align}\label{diag_1}
  \parbox{2.3cm}{\includegraphics[width=2.3cm]{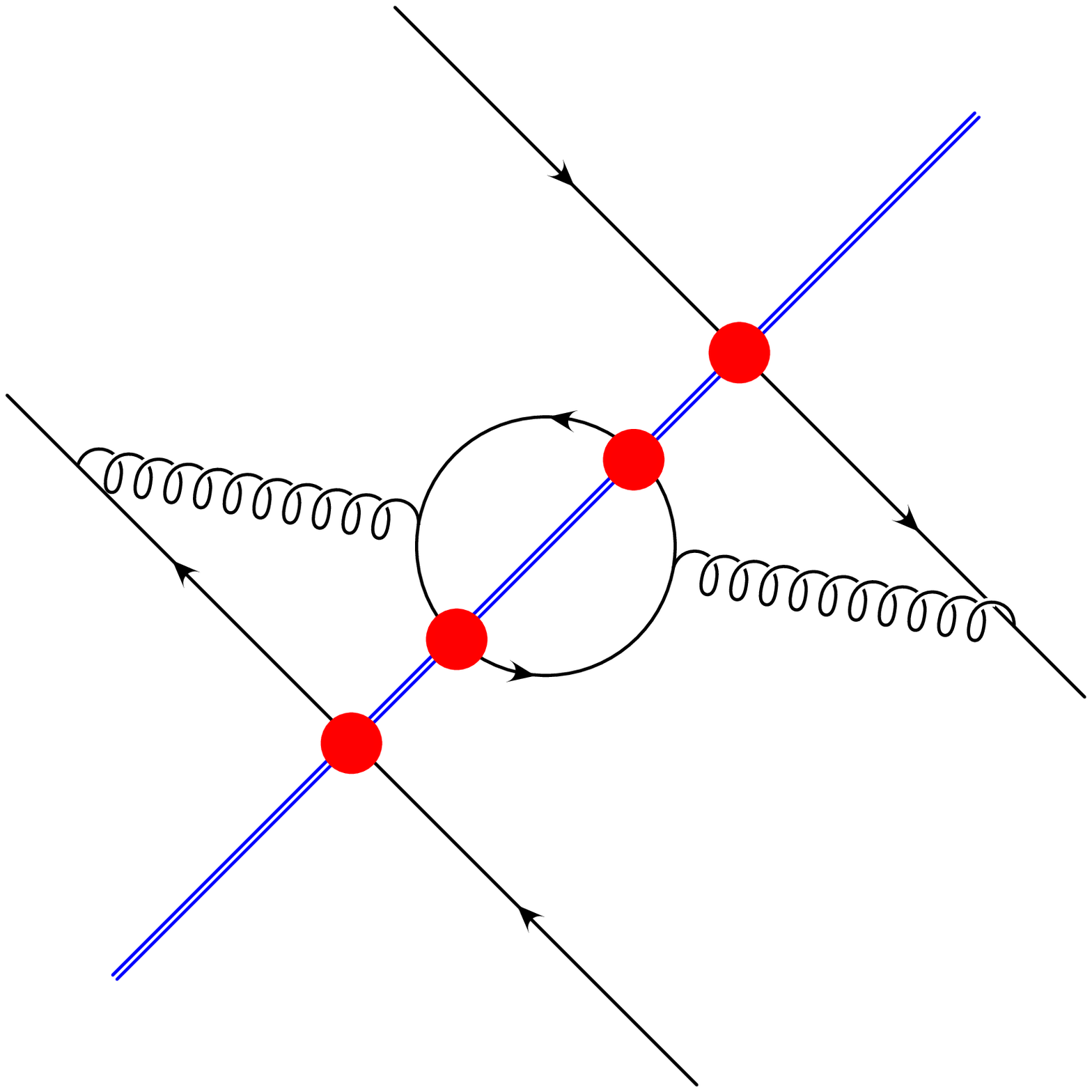}}
  +
  \parbox{2.3cm}{\includegraphics[width=2.3cm]{LoopInt-quarks-inst-1-firsto}}
  \, = & \notag \\ = \, \int d^2z_1 d^2z_2 \, & \am^2\, {\cal
    K}_1^{\text{NLO}} ({\bm x}_0, {\bm x}_1 ; {\bm z}_1, {\bm z}_2) \,
  U_{{\bm x}_0}t^a \otimes U_{{\bm x}_1}^\dagger t^b \ 2 \, \text{tr}(
  t^b U_{{\bm z}_1} t^a U_{{\bm z}_2}^\dagger) \,
  \ln(1/x_{\text{Bj}}),
\end{align}
where $x_{Bj}$ is the Bjorken $x$ variable. In arriving at \eq{diag_1}
we have defined the NLO contribution to the JIMWLK kernel coming from
the diagrams in Figs.  \ref{fig:NLO1}A and
\ref{fig:NLO1_inst}A$^\prime$, labeled ${\cal K}_1^{\text{NLO}}$, by
multiplying the wave function in \eq{psi1_xy} by its complex
conjugate, summing the obtained expression over the helicities of the
quark and the anti-quark in the produced pair and over $N_f$ quark
flavors, and integrating over $\alpha$:
\begin{align}\label{K1JIMWLKdef}
\am^2\,{\cal K}_1^{\text{NLO}} ({\bm x}_0, {\bm x}_1 ; {\bm z}_1,
  {\bm z}_2) \, =& \, \frac{N_f}{2 \, (4 \, \pi)^2} \, \int\limits_0^1 d
  \alpha  \sum_{\sigma_1, \sigma_2 =-1}^1 \nonumber \\
  & \times \, \psi^{(1)}_{\sigma_1, \, \sigma_2} ({\bm z}_1
  - {\bm x}_0, {\bm z}_2 - {\bm x}_0, \alpha) \,
  \psi^{(1)*}_{\sigma_1, \, \sigma_2} ({\bm z}_1 - {\bm x}_1, {\bm
    z}_2 - {\bm x}_1, \alpha).
\end{align}
In this definition of ${\cal K}_1^{\text{NLO}}$ we use the wave function
$\psi^{(1)}_{\sigma_1, \, \sigma_2}$, which is different from the full
wave function $\Psi^{(1)}_{\sigma_1, \, \sigma_2}$ from \eq{eq:psi_1k}
by the fact that the color matrices are included in
$\Psi^{(1)}_{\sigma_1, \, \sigma_2}$ and are not included in
$\psi^{(1)}_{\sigma_1, \, \sigma_2}$. We have used
$\psi^{(1)}_{\sigma_1, \, \sigma_2}$ to define the JIMWLK kernel
${\cal K}_1^{\text{NLO}}$ because the color matrices were already included in
the forward amplitude in \eq{diag_1}. A factor of $1/2$ was inserted
in \eq{K1JIMWLKdef} to account for the factor of $2$ introduced in the
definition of ${\cal K}_1^{\text{NLO}}$ in \eq{diag_1}.

Substituting $\psi^{(1)}$ from \eq{psi1_xy} into \eq{K1JIMWLKdef} and
summing over quark helicities yields
\begin{align}\label{K1JIMWLK}
  {\cal K}_1^{\text{NLO}} & ({\bm x}_0, {\bm x}_1 ; {\bm z}_1, {\bm
    z}_2) \, = \, \frac{N_f}{4 \, \pi^4} \, \int\limits_0^1 d \alpha \notag \\
  & \times \, \Bigg[ \frac{(1- 2 \alpha)^2 {\bm z}_{12} \cdot ({\bm z}
    - {\bm x}_0) \ {\bm z}_{12} \cdot ({\bm z} - {\bm x}_1) +
    \epsilon_{ij} ({\bm z} - {\bm x}_0)_i ({\bm z}_{12})_j \,
    \epsilon_{kl} ({\bm z} - {\bm x}_1)_k ({\bm z}_{12})_l}{({\bm
      z}_{12}^2)^2 \, [({\bm z} - {\bm x}_0)^2 + \alpha \, (1-\alpha)
    \, z_{12}^2] \, [({\bm z} - {\bm x}_1)^2 + \alpha \, (1-\alpha) \,
    z_{12}^2]} \notag \\ & - 2 \, \alpha \, (1-\alpha) \, (1- 2
  \alpha) \, \frac{{\bm z}_{12} \cdot ({\bm z} - {\bm x}_0) + {\bm
      z}_{12} \cdot ({\bm z} - {\bm x}_1)}{{\bm z}_{12}^2 \, [({\bm z}
    - {\bm x}_0)^2 + \alpha \, (1-\alpha) \, z_{12}^2] \, [({\bm z} -
    {\bm x}_1)^2 + \alpha \, (1-\alpha) \, z_{12}^2]} \notag \\ & +
  \frac{4 \, \alpha^2 \, (1-\alpha)^2}{ [({\bm z} - {\bm x}_0)^2 +
    \alpha \, (1-\alpha) \, z_{12}^2] \, [({\bm z} - {\bm x}_1)^2 +
    \alpha \, (1-\alpha) \, z_{12}^2]} \Bigg].
\end{align}
The integral over longitudinal momentum fraction $\alpha$, while
straightforward to perform, would not make the above expression any
more transparent. When squaring $\psi^{(1)}$ from \eq{psi1_xy} one
gets a cross-product between the first and the second terms on the
right hand side of \eq{psi1_xy}, given by the second term in the
square brackets of \eq{K1JIMWLK}. Terms like that are also present in
other physical quantities, such as the $q\bar q$ production cross
section calculated in \cite{Kovchegov:2006qn}.

To obtain the contribution of the diagrams in Figs.  \ref{fig:NLO1}A
and \ref{fig:NLO1_inst}A$^\prime$ to the BK evolution kernel
$K_1^{\text{NLO}} $ we have to sum the wave function
$\Psi^{(1)}_{\sigma_1, \, \sigma_2}$ over all possible emissions of
the gluon off the quark and off the anti-quark, multiply the result by
its complex conjugate, sum over quark and anti-quark helicities and
$N_f$ quark flavors, take a trace over color indices averaging over
$N_c$ colors of the incoming dipole and integrate over $\alpha$
\begin{align}\label{K1BKdef}
  \am^2 \, K_1^{\text{NLO}} ({\bm x}_0, {\bm x}_1 ; {\bm z}_1, {\bm
    z}_2) \, = & \, \frac{N_f}{(4 \, \pi)^2} \, \int\limits_0^1 d
  \alpha \sum_{\sigma_1, \sigma_2 =-1}^1 \sum_{m,n = 0}^1 \,
  (-1)^{m+n} \notag \\ \times \, & \frac{1}{N_c} \, \ \text{tr} \left[
    \Psi^{(1)}_{\sigma_1, \, \sigma_2} ({\bm z}_1 - {\bm x}_0, {\bm
      z}_2 - {\bm x}_0, \alpha) \, \Psi^{(1)*}_{\sigma_1, \, \sigma_2}
    ({\bm z}_1 - {\bm x}_1, {\bm z}_2 - {\bm x}_1, \alpha) \right].
\end{align}
(Note that the capital $K$ denotes the kernel of the BK evolution
equation, while the calligraphic $\cal K$ is reserved for the JIMWLK
evolution kernel.) Using the first line of \eq{eq:psi_1k} along with
\eq{K1JIMWLKdef} in \eq{K1BKdef} one can show that
\begin{align}\label{K1BK}
  K_1^{\text{NLO}} ({\bm x}_0, {\bm x}_1 ; {\bm z}_1, {\bm z}_2) \, = \,
  C_F \, \sum_{m,n = 0}^1 \, (-1)^{m+n} \, {\cal K}_1^{\text{NLO}}
  & ({\bm x}_m, {\bm x}_n ; {\bm z}_1, {\bm z}_2).
\end{align}

For the reasons which will become apparent momentarily, it is more
convenient to leave ${\cal K}^{\text{NLO}}_1$ written in terms of integrals
in transverse momentum space. Using Eqs. (\ref{eq:psi_1kq2}),
(\ref{ft_kq}) in \eq{K1JIMWLKdef} and summing over quark helicities
yields
\begin{align}\label{K1JIMWLKmom}
  {\cal K}_1^{\text{NLO}} ({\bm x}_0, {\bm x}_1 ; {\bm z}_1, {\bm z}_2) \,
  = & \, 4 \, %\am^2 \, 
  N_f \, \int\limits_0^1 d \alpha \, \int \frac{d^2
    k}{(2\pi)^2}\frac{d^2 k'}{(2\pi)^2}\frac{d^2 q}{(2\pi)^2}
  \frac{d^2 q'}{(2\pi)^2} \ e^{ -i {\bm q}\cdot ({\bm z}-{\bm x}_0) +i
    {\bm q}' \cdot ({\bm z}-{\bm x}_1) -i({\bm k}-{\bm k}') \cdot {\bm
      z}_{12}} \notag \\ & \times \left[ \frac{1}{{\bm q}^2{\bm
        q}'^{2}} \frac{(1-2\alpha)^2 {\bm q}\cdot{\bm k}\ {\bm k}'
      \cdot {\bm q}' + {\bm q} \cdot {\bm q}' \ {\bm k}\cdot{\bm k}' -
      {\bm q}\cdot{\bm k}' \ {\bm k}\cdot{\bm q}'}{\Big[{\bm k}^2+{\bm
        q}^2\alpha(1-\alpha)\Big]\Big[{\bm k}'^2+{\bm
        q}'^2\alpha(1-\alpha)\Big]} \right. \notag \\ & + \frac{2 \,
    \alpha \, (1-\alpha) \, (1-2 \alpha)}{\Big[{\bm k}^2+{\bm
      q}^2\alpha(1-\alpha)\Big]\Big[{\bm k}'^2+{\bm
      q}'^2\alpha(1-\alpha)\Big]} \, \left( \frac{{\bm k} \cdot {\bm
        q}}{{\bm q}^2} + \frac{{\bm k}' \cdot {\bm q}'}{{\bm q}'^2}
  \right)\notag \\ & \left. + \, \frac{4 \, \alpha^2 \,
      (1-\alpha)^2}{\Big[{\bm k}^2+{\bm
        q}^2\alpha(1-\alpha)\Big]\Big[{\bm k}'^2+{\bm
        q}'^2\alpha(1-\alpha)\Big]} \right],
\end{align}
where we have used the identity
\begin{align}
  \epsilon_{i j} \, {\bm q}_i \, {\bm k}_j \ \epsilon_{kl} \, {\bm
    q}'_k \, {\bm k}'_l = {\bm q}\cdot{\bm q}' \ {\bm k}\cdot{\bm
    k}' - {\bm q}\cdot{\bm k}' \ {\bm k}\cdot{\bm q}'.
\end{align}

%%%%%%%%%%%%%%%%%%%%%%%%%%%%%%%%%%%%%%%%%%%%%%%%%%%%%%%%%%%%%%%%%%%%%%%

\subsection{Diagram B}

Unlike the diagram A, the diagram B in \fig{fig:NLO1} looks more like
a ``typical'' running coupling correction to the leading order
JIMWLK/BK kernels. The contribution of the diagram B along with its
mirror-reflection with respect to the line denoting the interaction
with the target can be written as
\begin{align}\label{diag_2}
    \parbox{2.3cm}{\includegraphics[width=2.3cm]{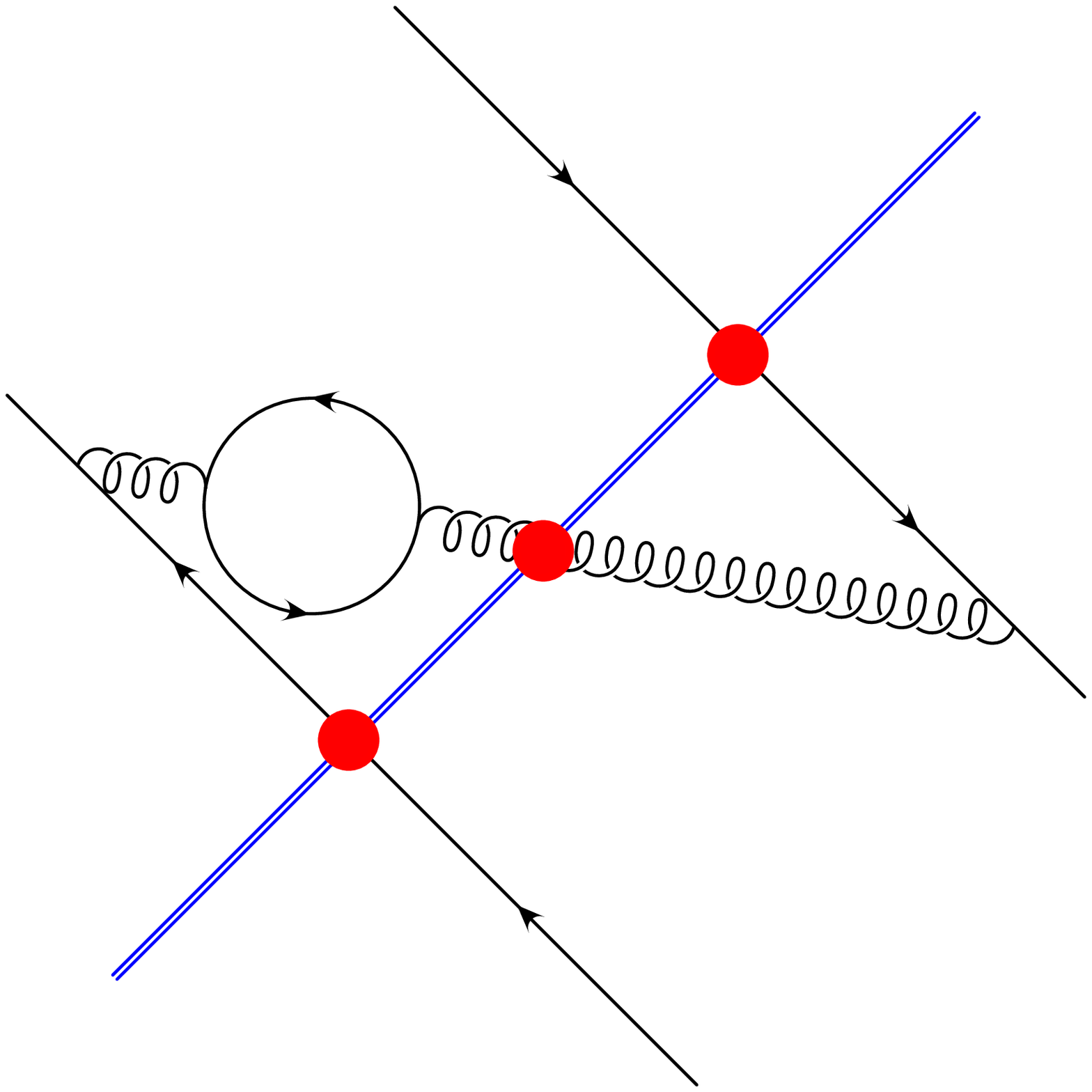}}
  +
    \parbox{2.3cm}{\includegraphics[width=2.3cm]{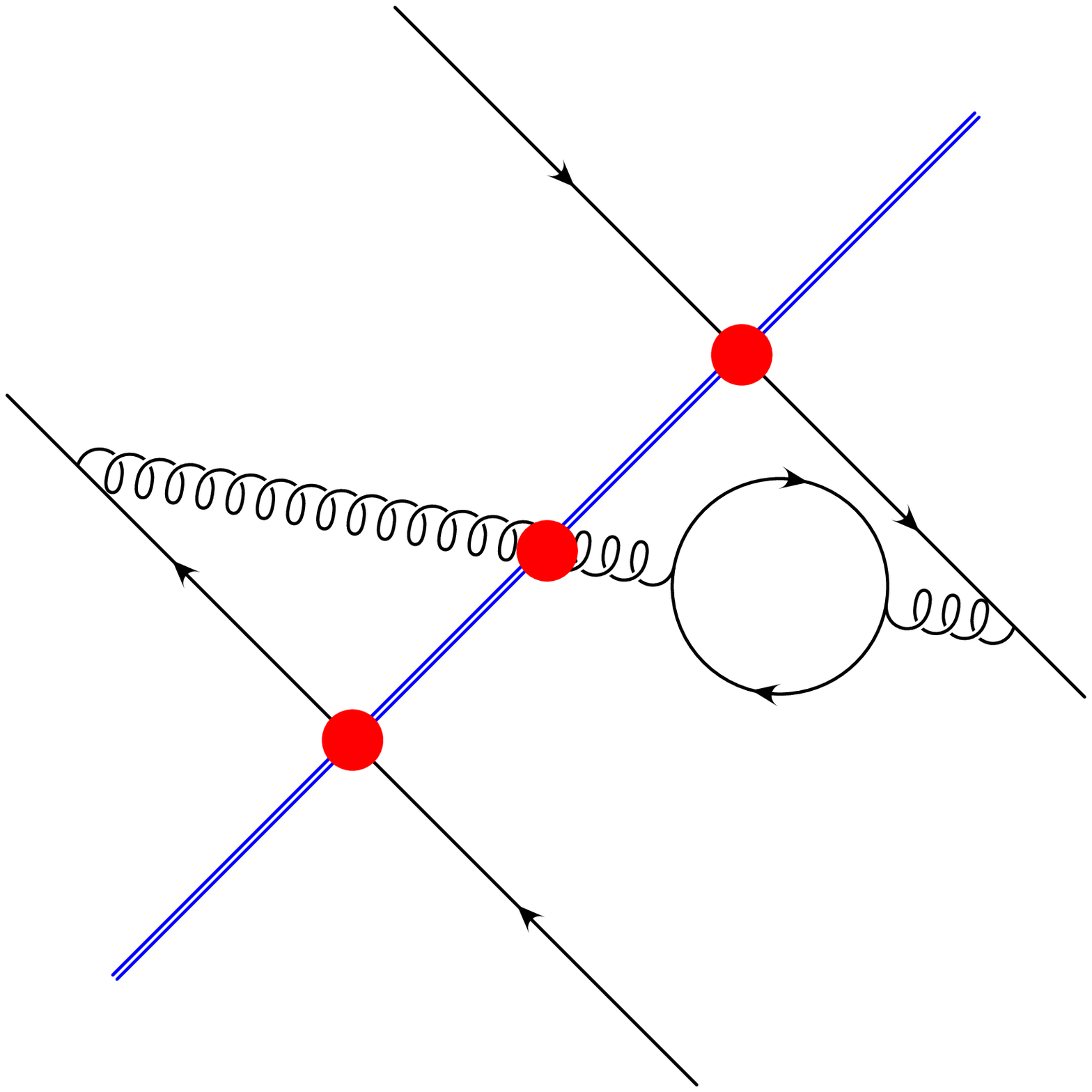}}
  \, = \, \int d^2z \ \am^2\, {\cal K}_2^{\text{NLO}} ({\bm x}_0, {\bm x}_1
  ; {\bm z}) \ U_{{\bm x}_0}t^a \otimes U_{{\bm
      x}_1}^\dagger t^b \ U^{ab}_{\bm z} \ \ln(1/x_{\text{Bj}})
\end{align}
with the corresponding NLO contribution ${\cal K}_2^{\text{NLO}}$ to the
JIMWLK kernel calculated using the rules of the light-cone
perturbation theory \cite{Lepage:1980fj,Brodsky:1997de}. We first
decompose the kernel ${\cal K}_2^{\text{NLO}}$ into a sum of the
contributions of the diagram \fig{fig:NLO1}B (denoted ${\cal K}_{2\,
  \text{left}}^{\text{NLO}}$) and its mirror-image (denoted ${\cal K}_{2\,
  \text{right}}^{\text{NLO}}$):
\begin{align}\label{K2dec}
  {\cal K}_2^{\text{NLO}} ({\bm x}_0, {\bm x}_1 ; {\bm z}) \, = \, {\cal
    K}_{2 \, \text{left}}^{\text{NLO}} ({\bm x}_0, {\bm x}_1 ; {\bm z}) + {\cal
    K}_{2 \, \text{right}}^{\text{NLO}} ({\bm x}_0, {\bm x}_1 ; {\bm z}).
\end{align}
Below we will only calculate ${\cal K}_{2\, \text{left}}^{\text{NLO}}$: to
construct ${\cal K}_{2\, \text{right}}^{\text{NLO}}$ one only has to replace
${\bm x}_0 \leftrightarrow {\bm x}_1$ in its argument. A simple calculation
along the same lines as the calculation of the diagram A done above yields
\begin{align}\label{K21}
  \am^2 \, {\cal K}_{2 \, \text{left}}^{\text{NLO}} ({\bm x}_0, {\bm
    x}_1 ; {\bm z}) \, = \, \frac{N_f}{(4 \, \pi)^2} \, 2 \, g^4 \,
  \int\limits_0^1 \frac{d \alpha}{\alpha (1 - \alpha)} \, \int
  \frac{d^2 k}{(2\pi)^2}\frac{d^2 q}{(2\pi)^2} \frac{d^2 q'}{(2\pi)^2}
  \ e^{ -i {\bm q}\cdot ({\bm z}-{\bm x}_0) +i {\bm q}' \cdot ({\bm
      z}-{\bm x}_1)} \notag \\ \sum_{\sigma_1, \sigma_2 =-1}^1 \,
  \sum_{\lambda, \lambda' =-1}^1 \, \frac{{\bm
      \epsilon}^{*\lambda}\cdot{\bm q}}{ {\bm q}^2} \frac{{\bm
      \epsilon}^{\lambda}\cdot{\bm k}\, (1-2\alpha+\lambda\sigma_1)
    \delta_{\sigma_1\sigma_2}}{{\bm k}^2+{\bm q}^2\alpha(1-\alpha)} \,
  \frac{{\bm \epsilon}^{*\lambda'}\cdot{\bm k}\, (1-2\alpha+\lambda'
    \sigma_1)}{{\bm q}^2} \, \frac{{\bm \epsilon}^{\lambda'}\cdot{\bm
      q}'}{ {\bm q}'^2}
\end{align}
with all the notation being the same as in the case of the diagram
\ref{fig:NLO1}A and $\lambda' = \pm 1$ the polarization of the gluon
interacting with the target. Different from ${\cal K}_1^{\text{NLO}}$,
the kernel ${\cal K}_{2 \, \text{left}}^{\text{NLO}}$ in \eq{K21} has
a part of the color factor included in it: it includes of $1/2$ coming
from the color trace of the quark loop, which is required by the
definition of ${\cal K}_2^{\text{NLO}}$ in \eq{diag_2}. Similar to the
above, to obtain the corresponding correction to the BK evolution
kernel, we use
\begin{align}\label{K2BK}
  K_2^{\text{NLO}} ({\bm x}_0, {\bm x}_1 ; {\bm z}) \, = \, C_F \, \sum_{m,n
    = 0}^1 \, (-1)^{m+n} \, {\cal K}_2^{\text{NLO}} & ({\bm x}_m, {\bm x}_n ;
  {\bm z})
\end{align} [Similar relationships holds for for both ${\cal K}_{2\,
  \text{left}}^{\text{NLO}}$ and ${\cal K}_{2\, \text{right}}^{\text{NLO}}$
separately.]

We can simplify \eq{K21}. First we sum over the quark helicities and
gluon polarizations to obtain
\begin{align}\label{K22}
  {\cal K}_{2 \, \text{left}}^{\text{NLO}} ({\bm x}_0, {\bm x}_1 ; {\bm z}) \, = \,
  4 \, N_f \, %\alpha^2_\mu \, 
  \int\limits_0^1 \frac{d \alpha}{\alpha
    (1 - \alpha)} \, \int \frac{d^d k}{(2\pi)^d}\frac{d^2 q}{(2\pi)^2}
  \frac{d^2 q'}{(2\pi)^2} \ e^{ -i {\bm q}\cdot ({\bm z}-{\bm x}_0) +i
    {\bm q}' \cdot ({\bm z}-{\bm x}_1)} \notag \\ \times \,
  \frac{(1-2\alpha)^2 {\bm q}\cdot{\bm k}\ {\bm k} \cdot {\bm q}' +
    {\bm q} \cdot {\bm q}' \ {\bm k}^2 - {\bm q}\cdot{\bm k} \ {\bm
      k}\cdot{\bm q}'}{({\bm q}^2)^2 \, {\bm q}'^2 \, \Big[{\bm
      k}^2+{\bm q}^2\alpha(1-\alpha)\Big]}.
\end{align}
The integral over $\bm k$ is UV-divergent, as expected. We will
regularize it by using dimensional regularization, for which purpose
we have replaced $d^2 k /(2 \pi)^2 \rightarrow d^d k /(2 \pi)^d$ in
\eq{K2BK} with $d$ the number of dimensions.  Anticipating the
integration over the angles of the vector $\bm k$ we also replace
\begin{align}\label{angles}
  {\bm k}_i \, {\bm k}_j \, \rightarrow \, \frac{{\bm k}^2}{d} \,
  \delta_{ij}
\end{align}
in \eq{K22}, obtaining
\begin{align}\label{K23}
  {\cal K}_{2 \, \text{left}}^{\text{NLO}} ({\bm x}_0, {\bm x}_1 ;
  {\bm z}) \, = \,
  4 \, N_f \, %\alpha^2_\mu \, 
  \int\limits_0^1 \frac{d \alpha}{\alpha
    (1 - \alpha)} \, \int \frac{d^2 q}{(2\pi)^2} \frac{d^2
    q'}{(2\pi)^2} \ e^{ -i {\bm q}\cdot ({\bm z}-{\bm x}_0) +i {\bm
      q}' \cdot ({\bm z}-{\bm x}_1)} \,
  \frac{{\bm q} \cdot {\bm q}'}{({\bm q}^2)^2 \, {\bm q}'^2} \notag \\
  \times \, \frac{1}{d} \, \int \frac{d^d k}{(2\pi)^d} \frac{{\bm
      k}^2}{{\bm k}^2+{\bm q}^2\alpha(1-\alpha)} \, \left[
    (1-2\alpha)^2 + d - 1\right].
\end{align}
Now the $\bm k$-integral is easily doable (see, e.g.,
\cite{Peskin:1995ev}) yielding
\begin{align}\label{K24}
  {\cal K}_{2 \, \text{left}}^{\text{NLO}} ({\bm x}_0, {\bm x}_1 ;
  {\bm z}) \, = \, 4 \, N_f
  \, %\alpha^2_\mu \, 
  \int\limits_0^1 \frac{d \alpha}{\alpha (1 -
    \alpha)} \, \int \frac{d^2 q}{(2\pi)^2} \frac{d^2 q'}{(2\pi)^2} \ 
  e^{ -i {\bm q}\cdot ({\bm z}-{\bm x}_0) +i {\bm q}' \cdot ({\bm
      z}-{\bm x}_1)} \,
  \frac{{\bm q} \cdot {\bm q}'}{({\bm q}^2)^2 \, {\bm q}'^2} \notag \\
  \times \, \frac{1}{2 \, (4 \pi)^{d/2}} \, \Gamma \left( -
    \frac{d}{2} \right) \, \left[ {\bm q}^2\alpha(1-\alpha)
  \right]^{d/2} \, \left[ (1-2\alpha)^2 + d - 1\right].
\end{align}
Writing $d = 2 - \epsilon$ and expanding around $\epsilon =0$ we get
\begin{align}\label{K25}
  {\cal K}_{2 \, \text{left}}^{\text{NLO}} ({\bm x}_0, {\bm x}_1 ; {\bm z}) \, = \, 
  \frac{N_f}{2 \, \pi} \, 
  %N_f \, \frac{\alpha^2_\mu}{2 \, \pi} \, 
    \int\limits_0^1 d \alpha \, \int
  \frac{d^2 q}{(2\pi)^2} \frac{d^2 q'}{(2\pi)^2} \ e^{ -i {\bm q}\cdot
    ({\bm z}-{\bm x}_0) +i {\bm q}' \cdot ({\bm z}-{\bm x}_1)} \,
  \frac{{\bm q} \cdot {\bm q}'}{{\bm q}^2 \, {\bm q}'^2} \notag \\
  \times \, \left\{ \left[ (1-2\alpha)^2 + 1 \right] \, \left[ \ln
      \frac{{\bm q}^2 \, \alpha \, (1-\alpha)}{\mu_{{\text{MS}}}^2} + \gamma -
      \ln 4 \pi - 1 \right] + 2 \right\},
\end{align}
where we replaced $1/\epsilon$ with $\ln \mu_{{\text{MS}}}$. Integrating over
$\alpha$ we obtain
\begin{align}\label{K26}
  {\cal K}_{2 \, \text{left}}^{\text{NLO}} ({\bm x}_0, {\bm x}_1 ; {\bm z}) \, = \, 
  %\frac{2 \,  N_f}{3} \, \frac{\alpha^2_\mu}{\pi} \, 
  \frac{2 \,  N_f}{3\, \pi} \,  
  \int \frac{d^2 q}{(2\pi)^2}
  \frac{d^2 q'}{(2\pi)^2} \ e^{ -i {\bm q}\cdot ({\bm z}-{\bm x}_0) +i
    {\bm q}' \cdot ({\bm z}-{\bm x}_1)} \, \frac{{\bm q} \cdot {\bm
      q}'}{{\bm q}^2 \, {\bm q}'^2} \, \left\{ \ln \frac{{\bm
        q}^2}{\mu_{\overline{{\text{MS}}}}^2} - \frac{5}{3} \right\}
\end{align}
with $\mu_{\overline{{\text{MS}}}}^2 =\mu_{{\text{MS}}}^2 \, 4 \pi \, e^{-\gamma}$.

While \eq{K26} is sufficiently simple for our later purposes, we can
further simplify it by Fourier-transforming it into transverse
coordinate space. A straightforward integration yields the NLO
contribution to the JIMWLK kernel coming from the diagram B
\begin{align}\label{K27}
  {\cal K}_{2 \, \text{left}}^{\text{NLO}} ({\bm x}_0, {\bm x}_1 ; {\bm z}) \, = \,
  \frac{%\alpha^2_\mu \, 
    N_f}{6 \, \pi^3} \, \frac{{\bm z}-{\bm
      x}_0}{|{\bm z}-{\bm x}_0|^2} \cdot \frac{{\bm z}-{\bm
      x}_1}{|{\bm z}-{\bm x}_1|^2} \, \left\{ \ln \frac{4}{|{\bm
        z}-{\bm x}_0|^2 \, \mu_{\overline{{\text{MS}}}}^2} - \frac{5}{3} 
    %+ 2\ln 2 
    - 2 \gamma \right\}.
\end{align}
Similarly one can show that
\begin{align}\label{K28}
  {\cal K}_{2 \, \text{right}}^{\text{NLO}} ({\bm x}_0, {\bm x}_1 ; {\bm z}) \, = \,
  \frac{%\alpha^2_\mu \, 
    N_f}{6 \, \pi^3} \, \frac{{\bm z}-{\bm
      x}_0}{|{\bm z}-{\bm x}_0|^2} \cdot \frac{{\bm z}-{\bm
      x}_1}{|{\bm z}-{\bm x}_1|^2} \, \left\{ \ln \frac{4}{|{\bm
        z}-{\bm x}_1|^2 \, \mu_{\overline{{\text{MS}}}}^2} - \frac{5}{3} 
    %+ 2\ln 2 
    - 2 \gamma \right\}
\end{align}
and
\begin{align}\label{K29}
  {\cal K}_2^{\text{NLO}} ({\bm x}_0, {\bm x}_1 ; {\bm z}) \, =& \, {\cal
    K}_{2 \, \text{left}}^{\text{NLO}} ({\bm x}_0, {\bm x}_1 ; {\bm z}) + {\cal
    K}_{2 \, \text{right}}^{\text{NLO}} ({\bm x}_0, {\bm x}_1 ; {\bm z}) \nonumber
  \\ \, =& \, \frac{%\alpha^2_\mu \, 
    N_f}{6 \, \pi^3} \, \frac{{\bm
      z}-{\bm x}_0}{|{\bm z}-{\bm x}_0|^2} \cdot \frac{{\bm z}-{\bm
      x}_1}{|{\bm z}-{\bm x}_1|^2} \, \left\{ \ln \frac{4 \,
      e^{-\frac{5}{3} - 2 \gamma}}{|{\bm z}-{\bm x}_0|^2 \,
      \mu_{\overline{{\text{MS}}}}^2} + \ln \frac{4 \, e^{-\frac{5}{3} - 2
        \gamma}}{|{\bm z}-{\bm x}_1|^2 \, \mu_{\overline{{\text{MS}}}}^2}
  \right\}.
\end{align}
The corresponding contribution to the NLO BK kernel can be easily
obtained from \eq{K29} using \eq{K2BK}.

Recalling that the leading order (LO) JIMWLK kernel is given by
\begin{align}\label{eq:LO-JIMWLK-kernel}
  {\cal K}^{\text{LO}} ({\bm x}_0, {\bm x}_1 ; {\bm z}) \, = \,
  \frac{1%\alpha_\mu
  }{\pi^2} \, \frac{{\bm z}-{\bm x}_0}{|{\bm z}-{\bm
      x}_0|^2} \cdot \frac{{\bm z}-{\bm x}_1}{|{\bm z}-{\bm x}_1|^2}
\end{align}
we immediately see that adding ${\cal K}_{2}^{\text{NLO}}$ from \eq{K29} to
it yields
\begin{align}\label{LO+NLO21}
  \am \, {\cal K}^{LO} ({\bm x}_0, {\bm x}_1 ; {\bm z}) + \am^2\,
  {\cal K}_2^{\text{NLO}} ({\bm x}_0, {\bm x}_1 ; {\bm z}) \, = \,
  \frac{\alpha_\mu}{\pi^2} \, \frac{{\bm z}-{\bm x}_0}{|{\bm z}-{\bm
      x}_0|^2} \cdot \frac{{\bm z}-{\bm x}_1}{|{\bm z}-{\bm x}_1|^2}
  \notag \\ \times \, \left\{ 1 + \frac{\alpha_\mu \, N_f}{6 \, \pi}
    \, \left[ \ln \frac{4 \, e^{-\frac{5}{3} - 2 \gamma}}{|{\bm
          z}-{\bm x}_0|^2 \, \mu_{\overline{{\text{MS}}}}^2} + \ln \frac{4 \,
        e^{-\frac{5}{3} - 2 \gamma}}{|{\bm z}-{\bm x}_1|^2 \,
        \mu_{\overline{{\text{MS}}}}^2} \right] \right\}.
\end{align}
Anticipating the appearance of the full QCD beta-function we perform
the replacement of \eq{repl} in \eq{LO+NLO21} to obtain
\begin{align}\label{LO+NLO22}
  \am\, {\cal K}^{LO} ({\bm x}_0, {\bm x}_1 ; {\bm z}) 
  + \am^2\, {\cal K}_2^{\text{NLO}}
  ({\bm x}_0, {\bm x}_1 ; {\bm z}) \, = \, \frac{\alpha_\mu}{\pi^2} \,
  \frac{{\bm z}-{\bm x}_0}{|{\bm z}-{\bm x}_0|^2} \cdot \frac{{\bm
      z}-{\bm x}_1}{|{\bm z}-{\bm x}_1|^2} \notag \\ \times \, \left\{
    1 - \alpha_\mu \, \beta_2 \, \left[ \ln \frac{4 \, e^{-\frac{5}{3}
          - 2 \gamma}}{|{\bm z}-{\bm x}_0|^2 \, \mu_{\overline{{\text{MS}}}}^2}
      + \ln \frac{4 \, e^{-\frac{5}{3} - 2 \gamma}}{|{\bm z}-{\bm
          x}_1|^2 \, \mu_{\overline{{\text{MS}}}}^2} \right] \right\}.
\end{align}
Now one can readily see that the diagram B in \fig{fig:NLO1} gives a
contribution to the one-loop running coupling correction to the LO
JIMWLK and BK kernels, as expected.

%%%%%%%%%%%%%%%%%%%%%%%%%%%%%%%%%%%%%%%%%%%%%%%%%%%%%%%%%%%%%%%%%%%%%%%

\subsection{Diagrams C and C$^\prime$} 
\label{sec:diagram-c}

The contribution of the diagrams in Figs.  \ref{fig:NLO1}C and
\ref{fig:NLO1_inst}C$^\prime$ along with their mirror-reflections can
be written as
\begin{align}\label{diag_3}
\parbox{2.3cm}{\includegraphics[width=2.3cm]{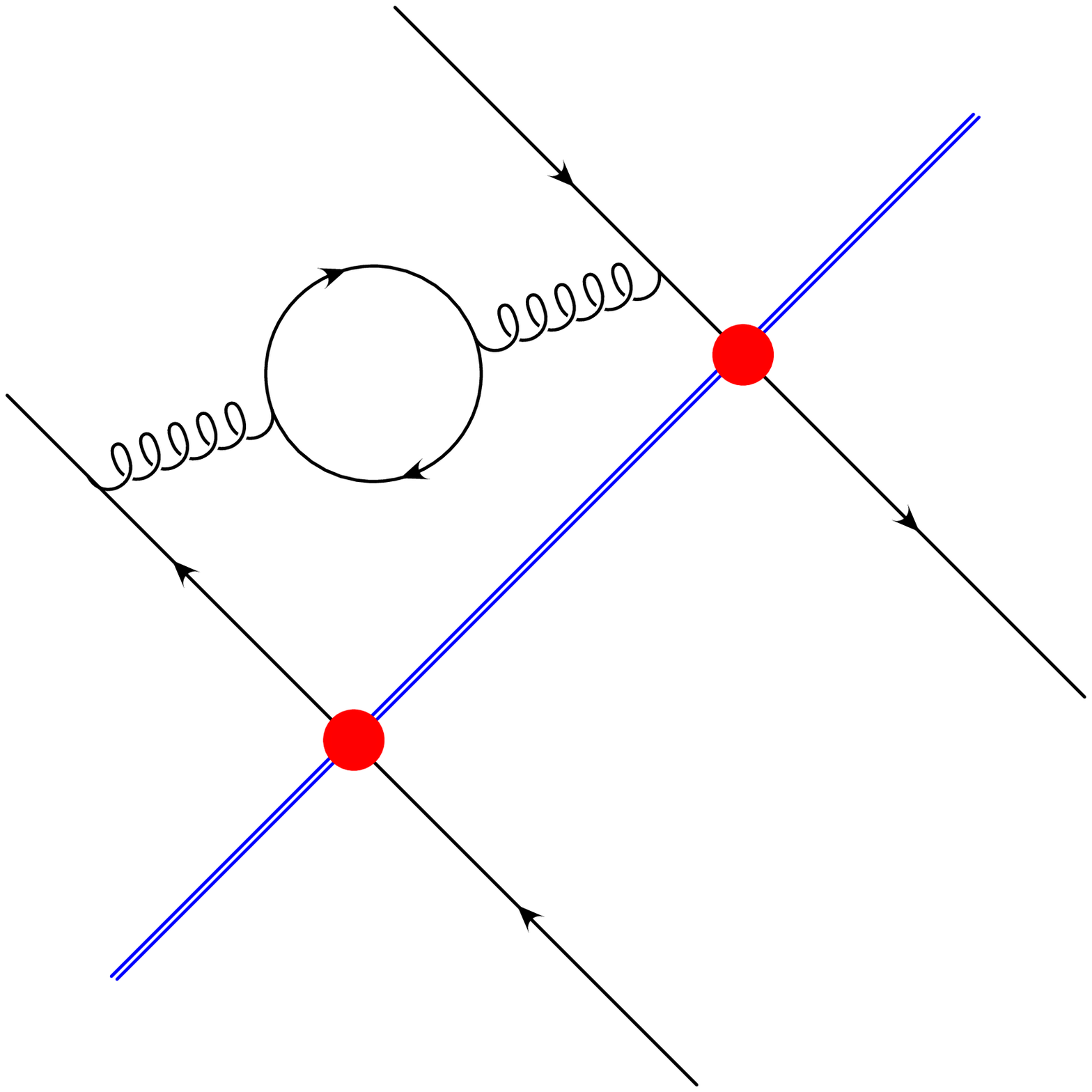}}
+
\parbox{2.3cm}{\includegraphics[width=2.3cm]{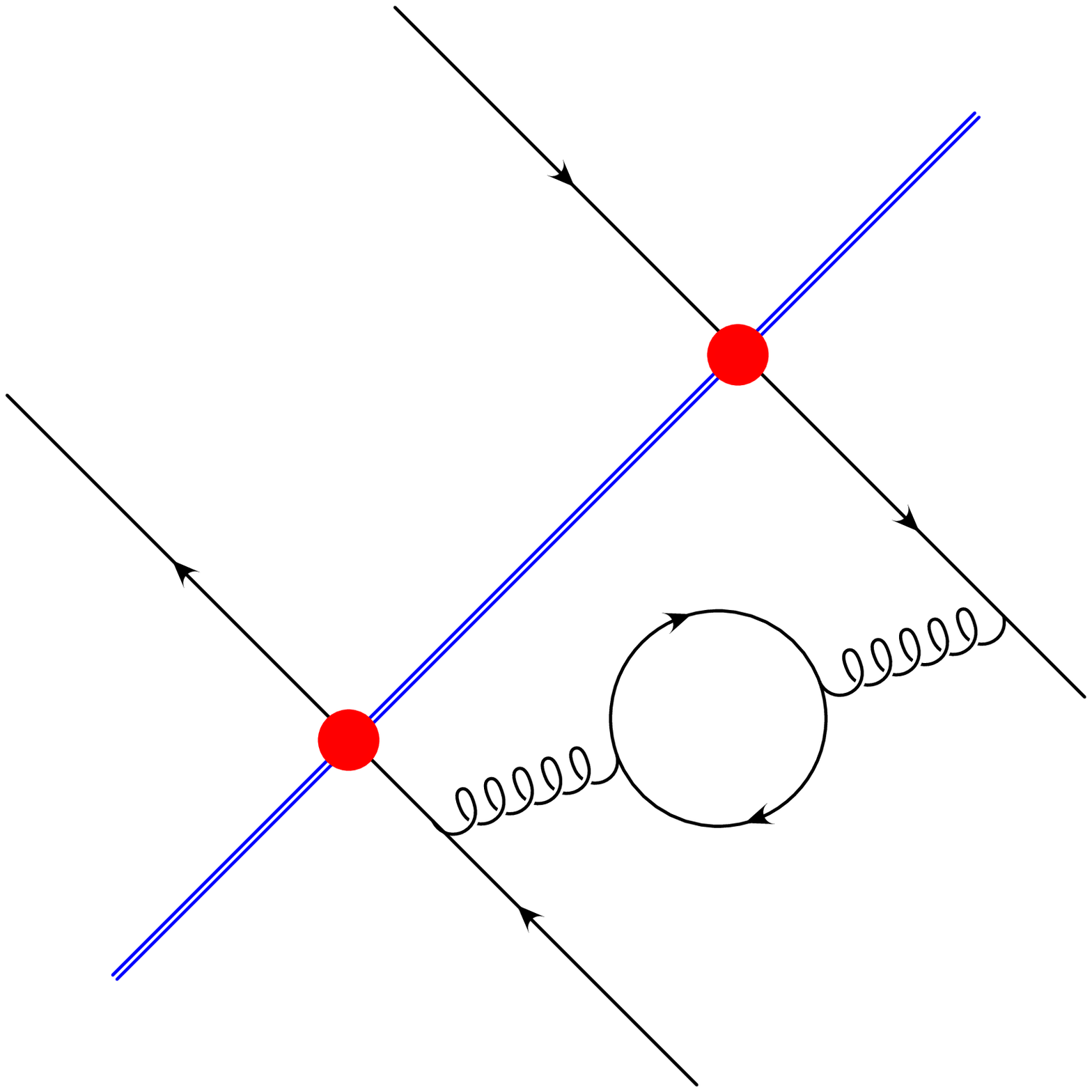}}
+
\parbox{2.3cm}{\includegraphics[width=2.3cm]{LoopNonInt-up-to-inst-1-firsto}}
+
\parbox{2.3cm}{\includegraphics[width=2.3cm]{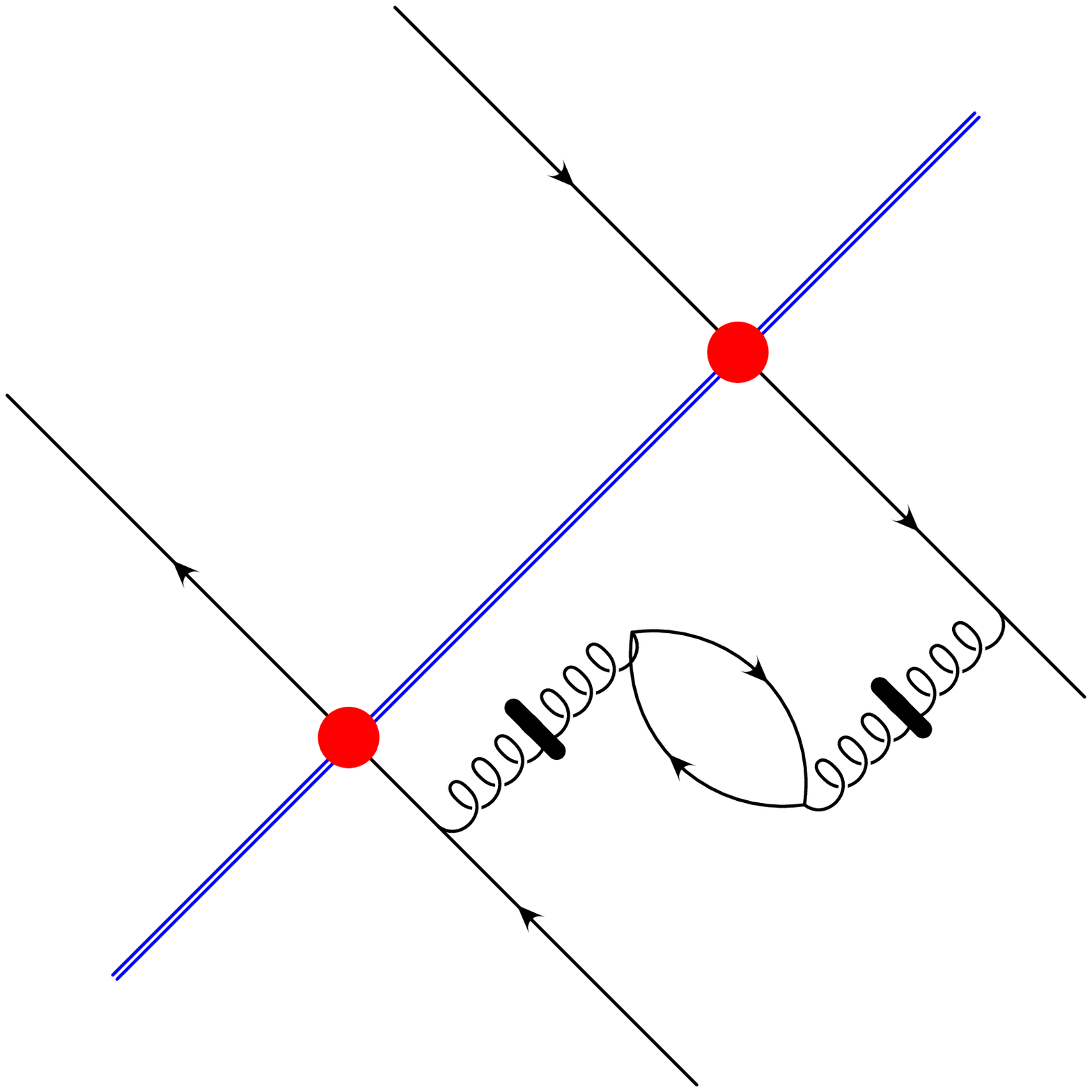}}
  \, = \notag \\ = \, \int d^2z \ \am^2\, {\cal K}_3^{\text{NLO}} ({\bm x}_0, {\bm x}_1 ; {\bm
    z}) \ U_{{\bm x}_0}t^a \otimes U_{{\bm x}_1}^\dagger t^a \ 
  \ln(1/x_{\text{Bj}})
\end{align}
where $\bm z$ is still the gluon's transverse coordinate which we
choose to keep explicitly even though, since both gluon lines are now
completely virtual, they do not interact with the target. Instead of
calculating the NLO correction to the JIMWLK kernel ${\cal
  K}_3^{\text{NLO}}$ coming from the diagrams C and C$^\prime$
explicitly we will use the conservation of probability condition,
which states that, in the absence of interaction with the target, the
sum of all three diagrams in \fig{fig:NLO1} (along with the mirror
images of the diagrams B and C reflected with respect to the line
representing the interaction with the target) gives zero. Intuitively
this condition is clear: in the absence of interactions there will be
no contribution to the evolution kernel. In the diagrammatic sense,
adding up all the graphs in \fig{fig:NLO1} corresponds to summing over
all the cuts for the diagram of the gluon emission with a quark bubble
correction. Similarly, if the interactions are absent, the sum of the
diagrams in \fig{fig:NLO1_inst} along with the mirror-reflection of
the diagram C$^\prime$ also gives zero. This probability conservation
condition was originally used by Mueller to calculate the virtual
correction to the leading order gluon emission in the dipole evolution
kernel in \cite{Mueller:1994rr}. In our case it formally reads
\begin{align}\label{prob}
  \int d^2z_1 d^2 z_2 \ {\cal K}_1^{\text{NLO}} ({\bm x}_0, {\bm x}_1 ; {\bm
    z}_1, {\bm z}_2) + \int d^2z \ {\cal K}_2^{\text{NLO}} ({\bm x}_0, {\bm
    x}_1 ; {\bm z}) + \int d^2z \ {\cal K}_3^{\text{NLO}} ({\bm x}_0, {\bm
    x}_1 ; {\bm z}) \, = \, 0.
\end{align}
Noting that
\begin{align}
  d^2 z_1 \, d^2 z_2 \, = \, d^2 z \, d^2 z_{12}
\end{align}
with ${\bm z}_{12}$ and $\bm z$ defined in Eqs. (\ref{z12}) and
(\ref{z}) above, an explicit diagram calculation (keeping all the
transverse momenta fixed in momentum space) yields an even stronger
identity than \eq{prob}:
\begin{align}\label{prob2}
  {\cal K}_3^{\text{NLO}} ({\bm x}_0, {\bm x}_1 ; {\bm z}) \, = \, - \int d^2
  z_{12} \ {\cal K}_1^{\text{NLO}} ({\bm x}_0, {\bm x}_1 ; {\bm z}_1, {\bm
    z}_2) - \ {\cal K}_2^{\text{NLO}} ({\bm x}_0, {\bm x}_1 ; {\bm z}).
\end{align}

${\cal K}_2^{\text{NLO}}$ in \eq{prob2} is given explicitly in
\eq{K29}.  Using the momentum-space expression (\ref{K1JIMWLKmom}) for
${\cal K}_1^{\text{NLO}}$ we get
\begin{align}\label{K1int1}
  \int d^2 z_{12} \, {\cal K}_1^{\text{NLO}} ({\bm x}_0, {\bm x}_1 ; {\bm
    z}_1, {\bm z}_2) \, = & \, 4 \, 
  N_f \, \int\limits_0^1 d \alpha \, \int \frac{d^d
    k}{(2\pi)^d}\frac{d^2 q}{(2\pi)^2} \frac{d^2 q'}{(2\pi)^2} \ e^{
    -i {\bm q}\cdot ({\bm z}-{\bm x}_0) +i {\bm q}' \cdot ({\bm
      z}-{\bm x}_1) } \notag \\ & \times \left[ \frac{1}{{\bm q}^2{\bm
        q}'^{2}} \frac{(1-2\alpha)^2 {\bm q}\cdot{\bm k}\ {\bm k}
      \cdot {\bm q}' + {\bm q} \cdot {\bm q}' \ {\bm k}^2 - {\bm
        q}\cdot{\bm k} \ {\bm k}\cdot{\bm q}'}{\Big[{\bm k}^2+{\bm
        q}^2\alpha(1-\alpha)\Big]\Big[{\bm k}^2+{\bm
        q}'^2\alpha(1-\alpha)\Big]} \right. \notag \\ & + \,
  \left.\frac{4 \, \alpha^2 \, (1-\alpha)^2}{\Big[{\bm k}^2+{\bm
        q}^2\alpha(1-\alpha)\Big]\Big[{\bm k}^2+{\bm
        q}'^2\alpha(1-\alpha)\Big]} \right],
\end{align}
where the $\bm k$-integral is UV-divergent, which we regularize using
dimensional regularization. With the help of \eq{angles} we rewrite
\eq{K1int1} as
\begin{align}\label{K1int2}
  \int d^2 z_{12} \, {\cal K}_1^{\text{NLO}} ({\bm x}_0, {\bm x}_1 ;
  {\bm z}_1, {\bm z}_2) \, = \, 4 \, N_f \, \int\limits_0^1 d \alpha
  \, \int \frac{d^2 q}{(2\pi)^2} \frac{d^2 q'}{(2\pi)^2} \ e^{ -i {\bm
      q}\cdot ({\bm z}-{\bm x}_0) +i {\bm q}' \cdot ({\bm z}-{\bm
      x}_1) } \notag \\ \times \, \Bigg\{ \frac{{\bm q} \cdot {\bm
      q}'}{{\bm q}^2{\bm q}'^{2}} \, \frac{1}{d} \, \int \frac{d^d
    k}{(2\pi)^d} \, \frac{{\bm k}^2}{\Big[{\bm k}^2+{\bm
      q}^2\alpha(1-\alpha)\Big]\Big[{\bm k}^2+{\bm
      q}'^2\alpha(1-\alpha)\Big]} \, \left[ (1-2\alpha)^2 + d -
    1\right] \notag \\ + \, 4 \, \alpha^2 \, (1-\alpha)^2 \, \int
  \frac{d^2 k}{(2\pi)^2} \, \frac{1}{\Big[{\bm k}^2+{\bm
      q}^2\alpha(1-\alpha)\Big]\Big[{\bm k}^2+{\bm
      q}'^2\alpha(1-\alpha)\Big]} \Bigg\},
\end{align}
where we put $d=2$ in the second term in the curly brackets since the
integral in that term is not divergent.  Performing the $\bm
k$-integrals yields
\begin{align}\label{K1int3}
  \int d^2 z_{12} \, {\cal K}_1^{\text{NLO}} ({\bm x}_0, {\bm x}_1 ; {\bm
    z}_1, {\bm z}_2) \, = \, 4 \, %\am^2 \, 
  N_f \, \int\limits_0^1 d \alpha \, \int \frac{d^2 q}{(2\pi)^2}
  \frac{d^2 q'}{(2\pi)^2} \ e^{ -i {\bm q}\cdot ({\bm z}-{\bm x}_0) +i
    {\bm q}' \cdot ({\bm z}-{\bm x}_1) } \notag \\ \times \, \Bigg\{
  \frac{{\bm q} \cdot {\bm q}'}{{\bm q}^2{\bm q}'^{2}} \frac{1}{d} \,
  \frac{1}{(4 \, \pi)^{d/2}} \, \Gamma \left( 1 - \frac{d}{2} \right)
  \, [\alpha \, (1 - \alpha)]^{\frac{d}{2} -1} \, \left[(1-2\alpha)^2
    + d - 1\right] \, \frac{[{\bm q}^2]^{d/2} - [{\bm
      q}'^2]^{d/2}}{{\bm q}^{2} - {\bm q}'^{2}} \notag \\ + \,
  \frac{1}{\pi} \, \alpha \, (1 - \alpha) \, \frac{\ln ({\bm
      q}^{2}/{\bm q}'^{2})}{{\bm q}^{2} - {\bm q}'^{2}} \Bigg\}.
\end{align}
Writing $d = 2 - \epsilon$, expanding around $\epsilon =0$, replacing
$1/\epsilon$ with $\ln \mu_{{\text{MS}}}$ and integrating over $\alpha$ we
obtain
\begin{align}\label{K1int4}
  \int d^2 z_{12} \, {\cal K}_1^{\text{NLO}} ({\bm x}_0, {\bm x}_1 ;
  {\bm z}_1, {\bm z}_2) \, = \,- \frac{2 \, N_f}{3 \, \pi} \, \int
  \frac{d^2 q}{(2\pi)^2} \frac{d^2 q'}{(2\pi)^2} \ e^{ -i {\bm q}\cdot
    ({\bm z}-{\bm x}_0) +i {\bm q}' \cdot ({\bm z}-{\bm x}_1) } \notag
  \\ \times \, \Bigg\{ \frac{{\bm q} \cdot {\bm q}'}{{\bm q}^2{\bm
      q}'^{2}} \, \frac{{\bm q}^{2} \, \left( \ln \frac{{\bm
          q}^2}{\mu_{\overline{{\text{MS}}}}^2} - \frac{5}{3} \right) - {\bm
      q}'^{2} \, \left( \ln \frac{{\bm q}'^2}{\mu_{\overline{{\text{MS}}}}^2} -
      \frac{5}{3} \right) }{{\bm q}^{2} - {\bm q}'^{2}} - \frac{\ln
    ({\bm q}^{2}/{\bm q}'^{2})}{{\bm q}^{2} - {\bm q}'^{2}} \Bigg\}.
\end{align}
The details of integrations over $\bm q$ and ${\bm q}'$ are shown in
Appendix \ref{K1FT}. The result reads
\begin{align}\label{K1int5}
  \int d^2 z_{12} \, {\cal K}_1^{\text{NLO}} ({\bm x}_0, {\bm x}_1 ;
  {\bm z}_1, {\bm z}_2) \, = \,- \frac{ N_f}{6 \, \pi^3} \, \frac{{\bm
      z}-{\bm x}_0}{|{\bm z}-{\bm x}_0|^2} \cdot \frac{{\bm z}-{\bm
      x}_1}{|{\bm z}-{\bm x}_1|^2} \, \ln \frac{4 \, e^{-\frac{5}{3}
      -2 \gamma}}{R^2 ({\bm x}_0, {\bm x}_1 ; {\bm z}) \,
    \mu_{\overline{{\text{MS}}}}^2},
\end{align}
where we have defined a transverse coordinate scale $R ({\bm x}_0,
{\bm x}_1 ; {\bm z})$ such that
\begin{align}\label{Rln}
  \ln R^2 ({\bm x}_0, {\bm x}_1 ; {\bm z}) \,
  \mu_{\overline{{\text{MS}}}}^2 \, \equiv \, \frac{|{\bm z} -{\bm
      x}_0|^2 \, \ln \left[ |{\bm z}-{\bm x}_1|^2 \,
      \mu_{\overline{{\text{MS}}}}^2 \right] - |{\bm z}-{\bm x}_1|^2
    \, \ln \left[ |{\bm z} -{\bm x}_0|^2 \,
      \mu_{\overline{{\text{MS}}}}^2
    \right]}{|{\bm z} -{\bm x}_0|^2 - |{\bm z}-{\bm x}_1|^2} \notag \\
  + \, \frac{|{\bm z}-{\bm x}_0|^2 \, |{\bm z}-{\bm x}_1|^2}{({\bm
      z}-{\bm x}_0) \cdot ({\bm z}-{\bm x}_1)} \, \frac{\ln (|{\bm z}
    -{\bm x}_0|^2 / |{\bm z}-{\bm x}_1|^2)}{|{\bm z} -{\bm x}_0|^2 -
    |{\bm z}-{\bm x}_1|^2}
\end{align}
or, equivalently, 
\begin{align}\label{Rexp}
  R^2 ({\bm x}_0, {\bm x}_1 ; {\bm z}) \, = \, |{\bm z} -{\bm x}_0| \,
  |{\bm z}-{\bm x}_1| \, \left( \frac{|{\bm z}-{\bm x}_1|}{|{\bm z}
      -{\bm x}_0|} \right)^{\frac{({\bm z} -{\bm x}_0)^2 + ({\bm
        z}-{\bm x}_1)^2}{({\bm z} - {\bm x}_0)^2 - ({\bm z}-{\bm
        x}_1)^2} - 2 \, \frac{|{\bm z}-{\bm x}_0|^2 \, |{\bm z}-{\bm
        x}_1|^2}{({\bm z}-{\bm x}_0) \cdot ({\bm z}-{\bm x}_1)} \,
    \frac{1}{|{\bm z} -{\bm x}_0|^2 - |{\bm z}-{\bm x}_1|^2}}.
\end{align}

Employing Eqs. (\ref{K1int5}) and (\ref{K29}) in \eq{prob2} we obtain
\begin{align}\label{K3}
  {\cal K}_3^{\text{NLO}} ({\bm x}_0, {\bm x}_1 ; {\bm z}) & \, = \, 
  \frac{%\am^2 \, 
    N_f}{6 \, \pi^3} \, \frac{{\bm z}-{\bm x}_0}{|{\bm z}-{\bm
      x}_0|^2} \cdot \frac{{\bm z}-{\bm x}_1}{|{\bm z}-{\bm x}_1|^2}
  \notag \\ & \times \, \left\{ \ln \frac{4 \, e^{-\frac{5}{3} -2
        \gamma}}{R^2 ({\bm x}_0, {\bm x}_1 ; {\bm z}) \,
      \mu_{\overline{{\text{MS}}}}^2} - \ln \frac{4 \, e^{-\frac{5}{3}
        - 2 \gamma}}{|{\bm z}-{\bm x}_0|^2 \,
      \mu_{\overline{{\text{MS}}}}^2} - \ln \frac{4 \, e^{-\frac{5}{3}
        - 2 \gamma}}{|{\bm z}-{\bm x}_1|^2 \,
      \mu_{\overline{{\text{MS}}}}^2} \right\}.
\end{align}
This is the contribution of the diagrams C in \fig{fig:NLO1} and
C$^\prime$ in \fig{fig:NLO1_inst} (along with their mirror
reflections) to the NLO JIMWLK kernel. To obtain the corresponding
contribution to the NLO BK kernel one again should use the following
formula
\begin{align}\label{K3BK}
  K_3^{\text{NLO}} ({\bm x}_0, {\bm x}_1 ; {\bm z}) \, = \, C_F \, \sum_{m,n
    = 0}^1 \, (-1)^{m+n} \, {\cal K}_3^{\text{NLO}} & ({\bm x}_m, {\bm x}_n ;
  {\bm z}).
\end{align}

Finally one may substitute the scale $R ({\bm x}_0, {\bm x}_1 ; {\bm
  z})$ from \eq{Rln} explicitly into \eq{K3} to obtain
\begin{align}\label{K3old}
  {\cal K}_3^{\text{NLO}} ({\bm x}_0, {\bm x}_1 ; {\bm z})  \, = \, -
  \frac{ N_f}{6 \, \pi^3} \, \Bigg[ \frac{{\bm z}-{\bm x}_0}{|{\bm
      z}-{\bm x}_0|^2} \cdot \frac{{\bm z}-{\bm x}_1}{|{\bm z}-{\bm
      x}_1|^2} & \notag \\  \times \, \frac{|{\bm z} -{\bm x}_0|^2 \,
    \ln \frac{4 \, e^{-\frac{5}{3} -2 \gamma}}{|{\bm z}-{\bm x}_0|^2
      \, \mu_{\overline{{\text{MS}}}}^2} - |{\bm z}-{\bm x}_1|^2 \, \ln \frac{4
      \, e^{-\frac{5}{3} -2 \gamma}}{|{\bm z}-{\bm x}_1|^2 \,
      \mu_{\overline{{\text{MS}}}}^2}}{|{\bm z} -{\bm x}_0|^2 - |{\bm z}-{\bm
      x}_1|^2} - & \frac{\ln (|{\bm z} -{\bm x}_0|^2 / |{\bm z}-{\bm
      x}_1|^2)}{|{\bm z} -{\bm x}_0|^2 - |{\bm z}-{\bm x}_1|^2}
  \Bigg].
\end{align}
The result in \eq{K3old} agrees with the NLO correction extracted from the
calculation performed in~\cite{Gardi:2006} where the dispersion method was
used in calculating the virtual part of the evolution kernel to determine the
scale of the running coupling for small-$x$ evolution.

%%%%%%%%%%%%%%%%%%%%%%%%%%%%%%%%%%%%%%%%%%%%%%%%%%%%%%%%%%%%%%%%%%%%%%%%%%%%%%%%%

\section{Ultraviolet Subtraction and Scheme Dependence}
\label{Subtr}

\subsection{Subtraction for the JIMWLK Equation}
\label{sec:subtr-JIMWLK}

To understand how the diagrams calculated above translate into
corrections to the JIMWLK equation, let us recall how the JIMWLK
Hamiltonian relates to the leading order diagrams.

The leading order JIMWLK Hamiltonian is a sum of real and virtual
contributions defined by
\begin{equation}
  \label{eq:JIMWLK-Hamiltonian-LO-0}
  \begin{split}
 {\cal H}^{\text{LO}}[U] = & {\cal H}^{\text{LO}}_{\text{real}}[U] + {\cal
   H}^{\text{LO}}_{\text{virtual}}[U]
\\
{\cal H}^{\text{LO}}_{\text{real}}[U] := & \frac{\am}{2} \int d^2x\, d^2y\, d^2z\
   {\cal K}^{\text{LO}}({\bm x},{\bm y};{\bm z})\ 
  U_{\bm z}^{a b}(i\Bar\nabla^a_{\bm x}i\nabla^b_{\bm y}   
  +i\nabla^a_{\bm x} i\Bar\nabla^b_{\bm y})
\\
{\cal H}^{\text{LO}}_{\text{virtual}}[U] := &
\frac{\am}{2}
\int d^2x\, d^2y\, d^2z\
   {\cal K}^{\text{LO}}({\bm x},{\bm y};{\bm z})\ 
  ( i\nabla^a_{\bm x} i\nabla^a_{\bm y}+i\Bar\nabla^a_{\bm x} i\Bar\nabla^a_{\bm y})
\ .    
  \end{split}
\end{equation}
Alternatively we will employ a notation in which an integration convention
over repeated transverse coordinates is implied and write more compactly
\begin{align}
  \label{eq:JIMWLK-Hamiltonian-LO}
  {\cal H}^{\text{LO}}[U] =\frac{\am}{2} 
   {\cal K}^{\text{LO}}_{{\bm x}, {\bm y}; {\bm z}}
   \left[ 
   U_{\bm z}^{a b}(i\Bar\nabla^a_{\bm x}i\nabla^b_{\bm y}   
  +i\nabla^a_{\bm x} i\Bar\nabla^b_{\bm y})
+
  ( i\nabla^a_{\bm x} i\nabla^a_{\bm y}+i\Bar\nabla^a_{\bm x} i\Bar\nabla^a_{\bm y})
  \right].
\end{align}
Integration conventions will be implied throughout when we employ
subscripts to list the transverse arguments of the kernels.

In the above, $\nabla^a_{\bm x}$ and $\Bar\nabla^a_{\bm x}$ are
functional derivatives with respect to the path ordered exponentials
(corresponding to the left and right-invariant vector fields on the
$SU(N_c)$ group) defined operationally via
\begin{subequations}
  \label{Lie-der}
\begin{align}
  i\nabla^a_{\bm{x}} U_{\bm{y}} := & 
  -U_{\bm{x}} t^a \delta^{(2)}_{\bm{x y}}
  \ ,
  \hspace{1cm}
   i\nabla^a_{\bm{x}} U^\dagger_{\bm{y}} := 
    t^a U^\dagger_{\bm{x}} \delta^{(2)}_{\bm{x y}} \ 
\intertext{and}
   i\Bar\nabla^a_{\bm{x}} U_{\bm{y}} := & 
   t^a U_{\bm{x}} \delta^{(2)}_{\bm{x y}}
   \ ,
  \hspace{1cm}
   i\Bar\nabla^a_{\bm{x}} U^\dagger_{\bm{y}} := 
   - U^\dagger_{\bm{x}} t^a \delta^{(2)}_{\bm{x y}}
   \ .
\end{align}
\end{subequations}
${\cal K}^{\text{LO}}$ was already given in~(\ref{eq:LO-JIMWLK-kernel}).
Our notation here is somewhat different from the usual in that we absorb a
factor $1/\pi^2$ into the leading oder kernel. 

The JIMWLK Hamiltonian determines the $Y$ dependence of  expectation values
of arbitrary functionals $O[U]$ of Wilson lines $U_{\bm x}$ 
\begin{equation}
  \label{eq:corrs}
  \langle O[U] \rangle(Y) := \int\! \Hat{D}[U]  O[U] Z_Y[U]
\end{equation}
via the $Y$ dependence of the functional weight $Z_Y[U]$. The evolution
equation for $Z_Y[U]$ is known as the JIMWLK equation:
\begin{equation}
  \label{eq:JIMWLK-eq}
  \partial_Y \ \Hat Z_Y[U]=
  - {\cal H}^{\text{LO}}[U]\  
  Z_Y[U] 
\ .
\end{equation}

The leading order JIMWLK Hamiltonian in \eq{eq:JIMWLK-Hamiltonian-LO} is {\em
  constructed} such that it adds the leading order real and virtual
corrections to, say, an interacting $q\Bar q$ pair, represented by its Wilson
line bilinear $U_{{\bm x}_0}\otimes U_{{\bm x}_1}^\dagger$:
\begin{align}
  \label{eq:JIMWLK-LO-diagram-cont}
  \ln(1/x_{\text{Bj}})\, 
{\cal H}^{\text{LO}}[U] \ U_{{\bm x}_0}\otimes U_{{\bm x}_1}^\dagger = & 
\parbox{2cm}{\includegraphics[width=2cm]{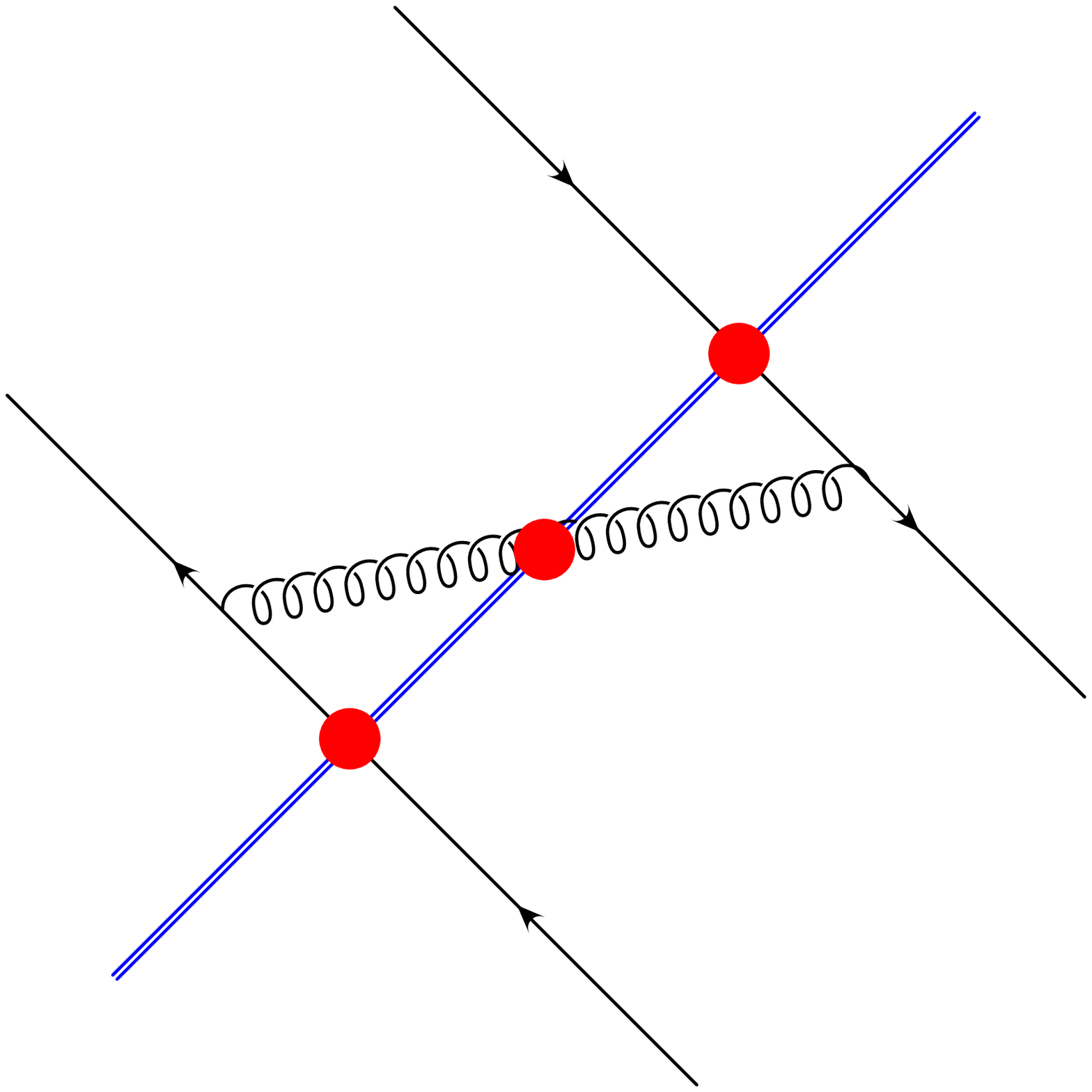}}
+
\parbox{2cm}{\includegraphics[width=2cm]{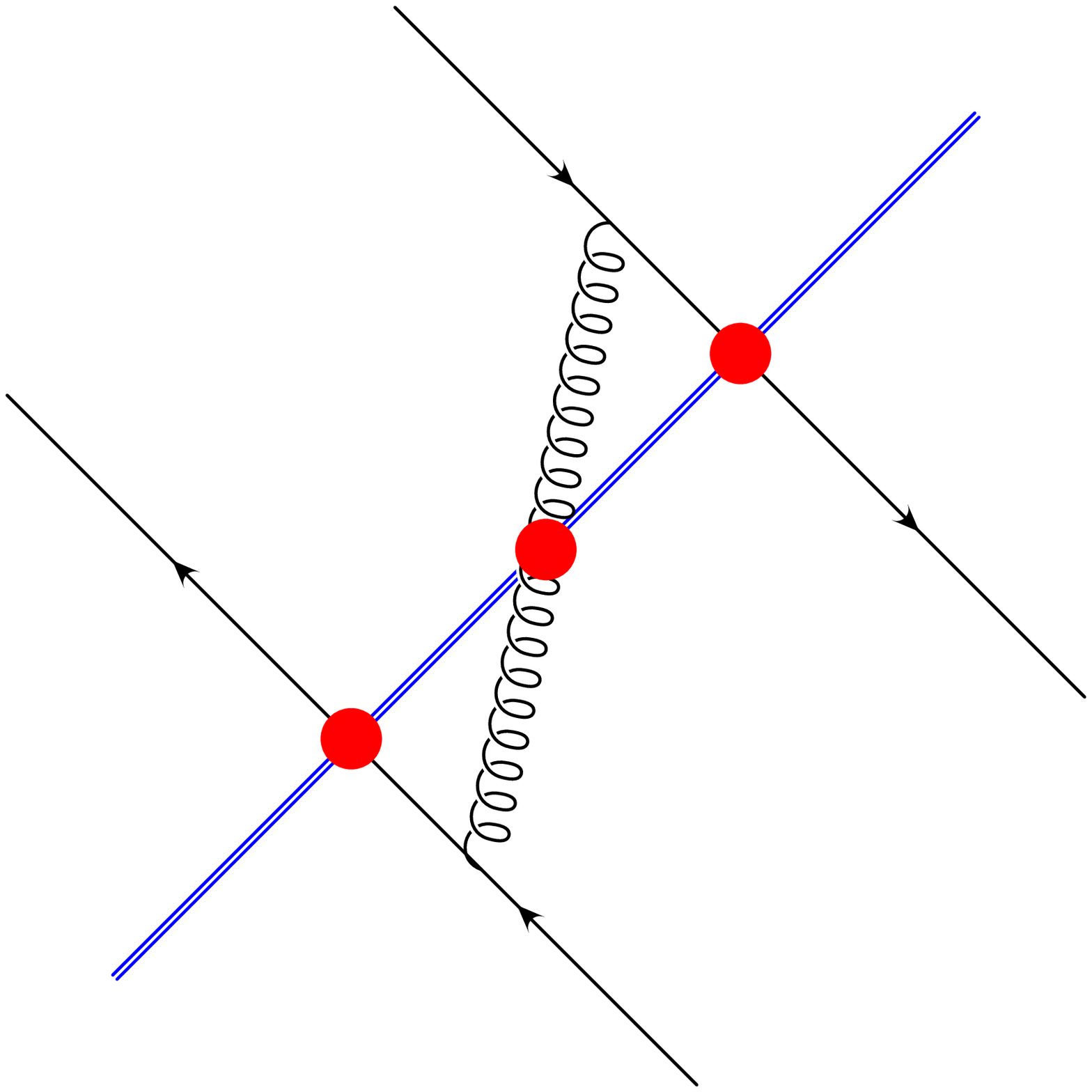}}
+%\parbox{2cm}{\includegraphics[width=2cm]{chiqqb-tup}}
\parbox{2cm}{\includegraphics[width=2cm]{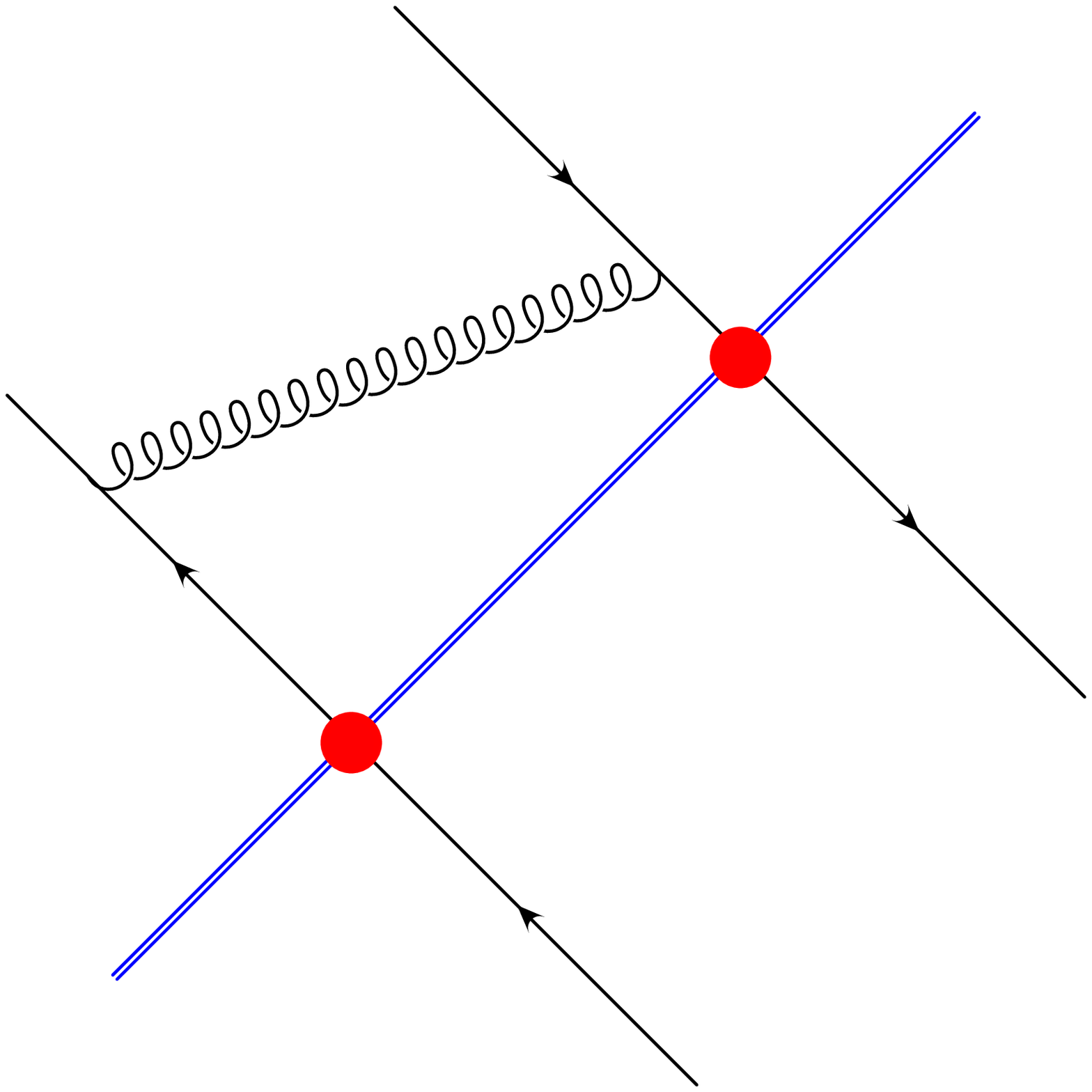}}
+\parbox{2cm}{\includegraphics[width=2cm]{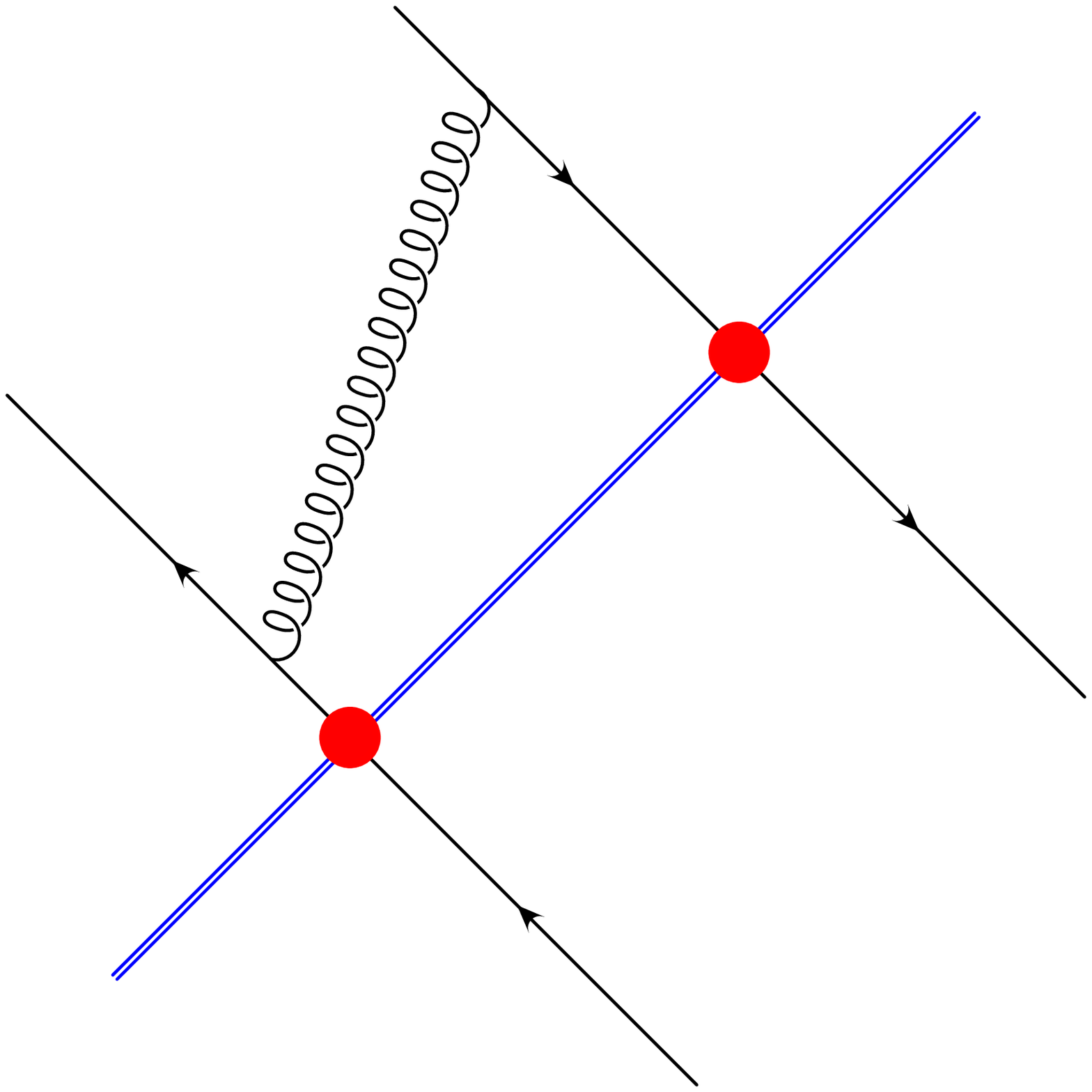}}
\notag \\ & \phantom{+}\hspace{4cm}
+%\parbox{2cm}{\includegraphics[width=2cm]{chiqqb-tdown}}
\parbox{2cm}{\includegraphics[width=2cm]{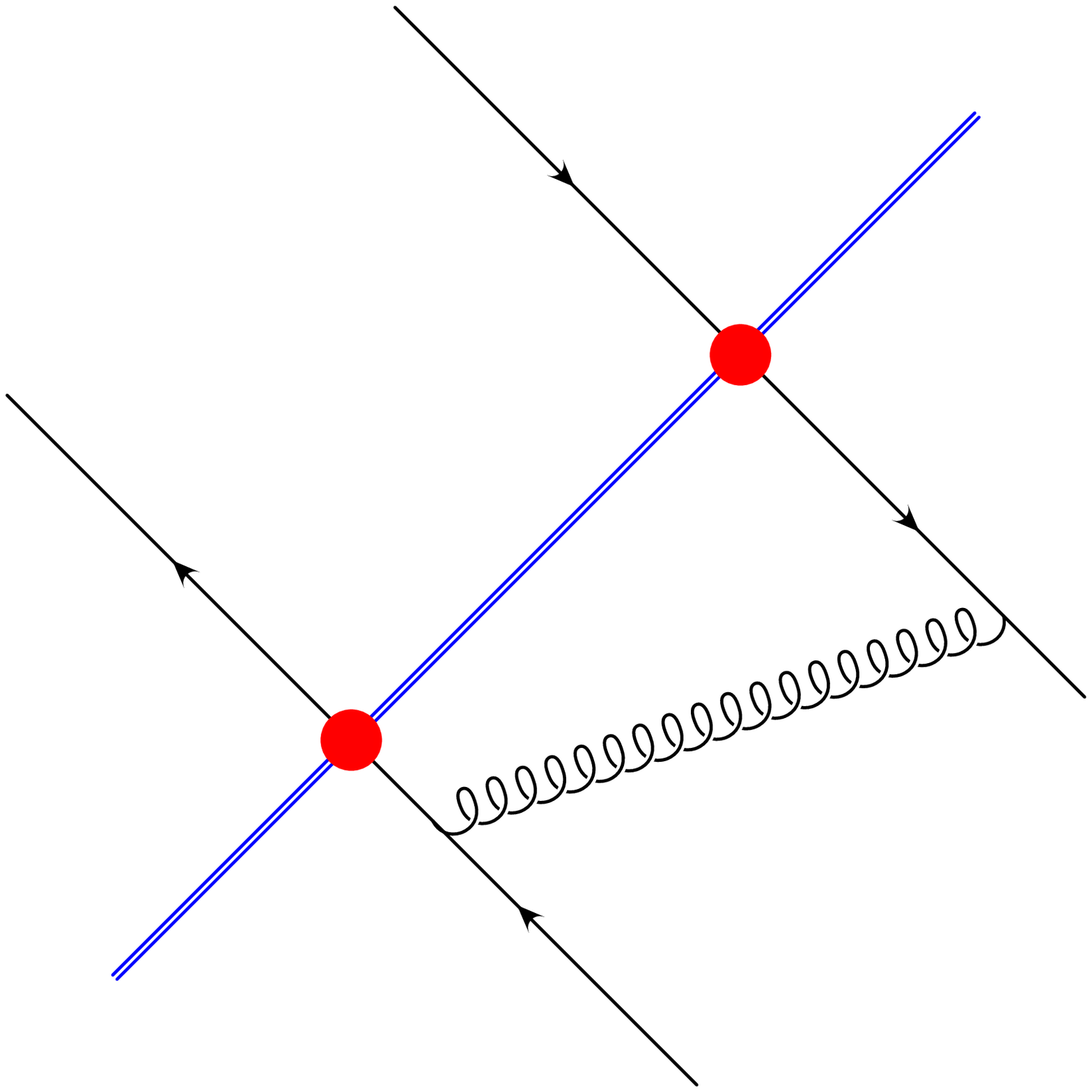}}
+\parbox{2cm}{\includegraphics[width=2cm]{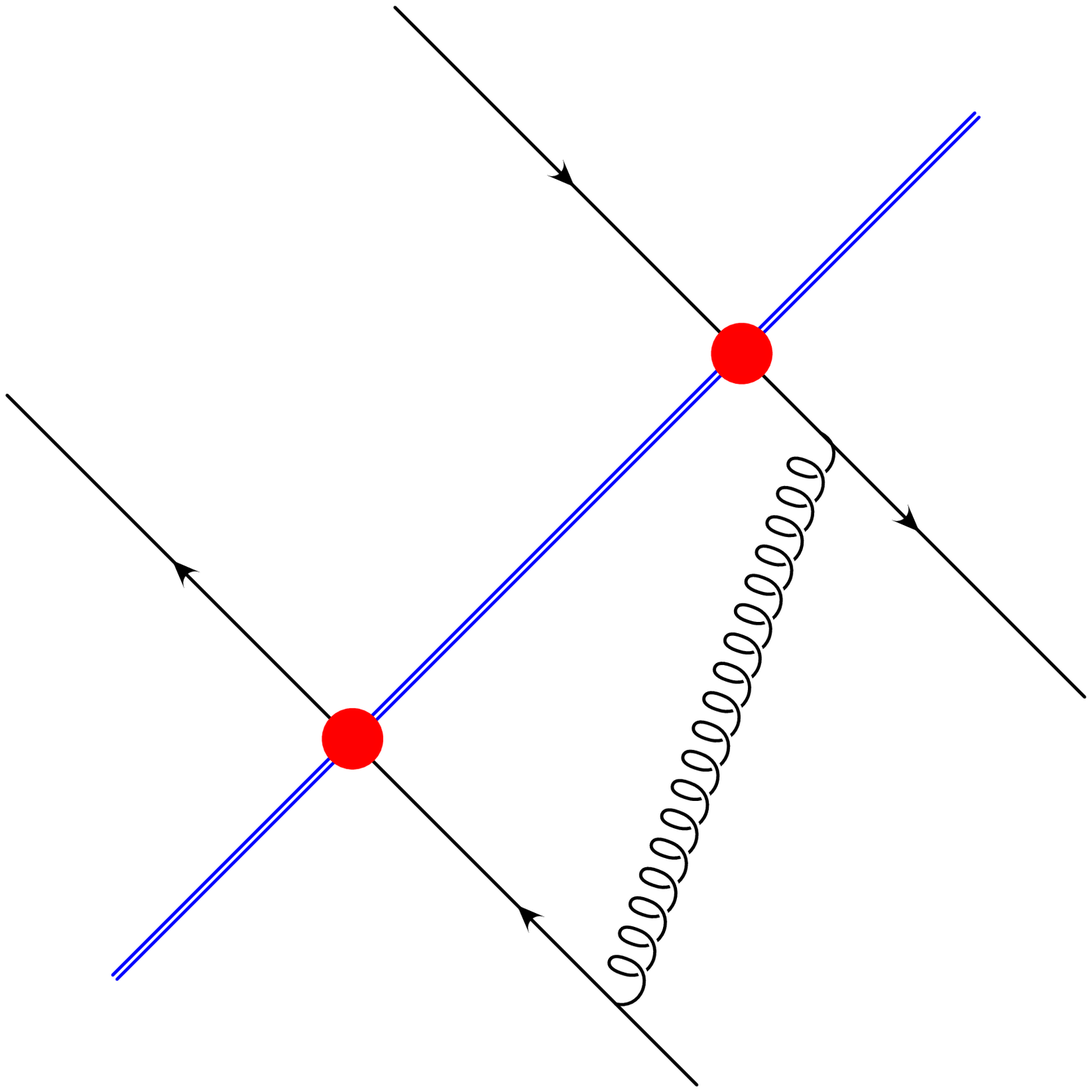}}
\notag \\ &
+\text{real and virtual self-energy-like terms.}
\end{align}
Taking a trace of \eq{eq:JIMWLK-LO-diagram-cont} and normalizing by the number
of colors turns the above into the (unfactorized) right hand side of the BK
equation for $S ({{\bm x}_0, {\bm x}_1},Y):=\left< \frac{\tr U_{{\bm x}_0}
    U_{{\bm x}_1}^\dagger}{N_c}\right> (Y)$ shown in \eq{eqS} below.
$\frac{\tr U_{{\bm x}_0} U_{{\bm x}_1}^\dagger}{N_c}$ is but the most generic
of the operators $O[U]$ referred to in~\eqref{eq:corrs}.

As in the BK case, real-virtual cancellation and thus UV finiteness
follow from the appearance of the same kernel ${\cal K}^{\text{LO}}$
in both real and virtual contributions. This ensures that the limits
${\bm z}\to {\bm x}$ and ${\bm z}\to {\bm y}$ cancel between the two
terms under the integral. Probability conservation at leading order
manifests itself more globally in the absence of interaction with the
target, i.e., in the limit $U\to 1$: There~\eqref{Lie-der} ensures
that ${\cal H}^{\text{LO}}_{\text{virtual}}\to -{\cal
  H}^{\text{LO}}_{\text{real}}$ so that there is no evolution without
interaction with the target.

${\cal H}^{\text{LO}}$ may in fact be used to act on any tensor
product of quarks, antiquarks and gluons in a projectile's
wavefunction that interact with the target via corresponding Wilson
lines to produce a sum of leading order $\alpha_s
\ln(1/x_{\text{Bj}})$ corrections to this eikonal interaction. This is
the technical mechanism by which the JIMWLK equation translates into
the Balitsky hierarchy.

Note the efficiency with which the JIMWLK Hamiltonian encodes the
contributions: due to the symmetry properties of the kernels, the
self-energy-like diagrams arise from the same terms that create the
exchange diagrams. The only distinctions left in the Hamiltonian are:
\begin{itemize}
\item The order of the vertices w.r.t. the target interaction, i.e. the Wilson
  lines. This is encoded in the use of the $\nabla$ and $\Bar\nabla$.
\item The interaction (or lack thereof) encoded in presence or absence
  of an adjoint Wilson line at the transverse position at which the
  newly created gluon interacts with the target, as shown in the
  second and third lines of \eq{eq:JIMWLK-Hamiltonian-LO-0}.
\end{itemize}

This pattern extends itself to the NLO contributions studied here.
Only the variants of diagram A differ slightly in structure from the
contributions already encountered at leading order: they depend on two
new transverse coordinates and contain a factor $2 \text{tr}( t^b
U_{{\bm z}_1} t^a U_{{\bm z}_2}^\dagger)$ instead of the $U^{a b}_{\bm
  z}$ of the real emissions at leading order.  This leads to the
following correspondence of diagrams and terms in the NLO corrections
in the Hamiltonian
\begin{subequations}
 \label{eq:JIMWLK-NLO-terms-from-diags}
\begin{align}
\label{eq:JIMWLK-NLO-from-A}
\ln(1/x_{\text{Bj}}) &\, {\cal H}^{\text{NLO}}_1 
\hspace{.2cm} U_{{\bm x}_0}\otimes U_{{\bm x}_1}^\dagger =
\am\, \ln(1/x_{\text{Bj}})
\int d^2x d^2y d^2z_1 d^2z_2 \, {\cal K}_1^{\text{NLO}} ({\bm x}, {\bm y}
  ; {\bm z}_1, {\bm z}_2) \,
\notag \\ & \times  2 \text{tr}( t^b U_{{\bm z}_1} t^a U_{{\bm
      z}_2}^\dagger) \, (i\Bar\nabla^a_{\bm x}i\nabla^b_{\bm y}
  +i\nabla^a_{\bm x} i\Bar\nabla^b_{\bm y}) 
  \hspace{.2cm} U_{{\bm x}_0}\otimes U_{{\bm x}_1}^\dagger
= \parbox{2cm}{\includegraphics[width=2cm]{LoopInt-quarks-1-firsto}} 
 + \parbox{2cm}{\includegraphics[width=2cm]{LoopInt-quarks-inst-1-firsto}} + \ldots \ .
\intertext{All other corrections have the same $U$ structure already
  encountered at the leading order. We have corrections to real emission}
\label{eq:JIMWLK-NLO-from-B}  
\ln(1/x_{\text{Bj}})  & \, {\cal H}^{\text{NLO}}_2 
\hspace{.2cm} U_{{\bm x}_0}\otimes U_{{\bm x}_1}^\dagger =
\am\, \ln(1/x_{\text{Bj}})
\int d^2x d^2y d^2z \ {\cal K}_2^{\text{NLO}} ({\bm x}, {\bm y}
  ; {\bm z}) \ 
\notag \\ & \times U^{ab}_{\bm z}  \, (i\Bar\nabla^a_{\bm x}i\nabla^b_{\bm y}
  +i\nabla^a_{\bm x} i\Bar\nabla^b_{\bm y})
  \hspace{.2cm} U_{{\bm x}_0}\otimes U_{{\bm x}_1}^\dagger
%\notag \\ = &
=
\parbox{2cm}{\includegraphics[width=2cm]{LoopInt-glue-1-firsto-left}} 
  +
\parbox{2cm}{\includegraphics[width=2cm]{LoopInt-glue-1-firsto-right}}
 +\ldots
\intertext{and virtual terms}
\label{eq:JIMWLK-NLO-from-C} 
\ln(1/x_{\text{Bj}}) &\, {\cal H}^{\text{NLO}}_3 
\hspace{.2cm} U_{{\bm x}_0}\otimes U_{{\bm x}_1}^\dagger =
-\am\, \ln(1/x_{\text{Bj}})
 \int d^2x d^2 y d^2z \ {\cal K}_3^{\text{NLO}} ({\bm x}, {\bm y}; {\bm
    z}) 
\notag \\ &  \hspace{5cm} \times
 ( i\nabla^a_{\bm x} i\nabla^a_{\bm y}+i\Bar\nabla^a_{\bm x}
 i\Bar\nabla^a_{\bm y})
  \hspace{.2cm} U_{{\bm x}_0}\otimes U_{{\bm x}_1}^\dagger
\notag \\ &
=
\parbox{2cm}{\includegraphics[width=2cm]{LoopNonInt-up-to-1-firsto}} 
+
\parbox{2cm}{\includegraphics[width=2cm]{LoopNonInt-down-to-1-firsto}}
+
\parbox{2cm}{\includegraphics[width=2cm]{LoopNonInt-up-to-inst-1-firsto}} 
+
\parbox{2cm}{\includegraphics[width=2cm]{LoopNonInt-down-to-inst-1-firsto}}
+ \ldots \ .
\end{align}
\end{subequations}
The minus sign in the last term is due to the different $\nabla$
structures in real and virtual terms and is important for the
real-virtual cancellations. The dots represent both symmetrization in
external coordinates ${\bm x}_0$ and ${\bm x}_1$ as well as the
inclusion of ``self energy like diagrams'' in which the gluon line
connects back to the quark (or antiquark) it originates from.

We group the contributions accordingly (again
employing an integration convention for all repeated transverse
coordinates ${\bm x}, {\bm y}, {\bm z}, {\bm z}_i$)
\begin{align}
  \label{eq:JIMWLK-Nfcontribs}
\text{new:} & & {\cal H}^{\text{NLO}}_1 = &
%\int d^2x d^2y d^2z_1 d^2z_2
\ \frac{\am^2}2 \
 {\cal K}^{\text{NLO}}_{1\ {\bm x}, {\bm y}  ; {\bm z}_1, {\bm z}_2} \
 2\, \text{tr}( t^b U_{{\bm z}_1} t^a U_{{\bm
      z}_2}^\dagger) \, 
(i\Bar\nabla^a_{\bm x}i\nabla^b_{\bm y}
  +i\nabla^a_{\bm x} i\Bar\nabla^b_{\bm y})
%(i\Bar\nabla^a_{\bm x}i\nabla^b_{\bm y}
%  +i\nabla^a_{\bm x} i\Bar\nabla^b_{\bm y}) 
\\
\label{eq:JIMWLK-unsubtr-real}
\text{real:} & &
{\cal H}^{\text{LO}}_{\text{real}} + {\cal H}^{\text{NLO}}_2
= &\ \frac12 \left(\am\,{\cal K}^{\text{LO}}_{{\bm x}, {\bm y}; {\bm z}}
   +\am^2\, {\cal K}^{\text{NLO}}_{2\ {\bm x}, {\bm y}; {\bm z}}\right)
 \  U^{ab}_{\bm z}  \, 
(i\Bar\nabla^a_{\bm x}i\nabla^b_{\bm y}
  +i\nabla^a_{\bm x} i\Bar\nabla^b_{\bm y})
%(i\Bar\nabla^a_{\bm x}i\nabla^b_{\bm y}
%  +i\nabla^a_{\bm x} i\Bar\nabla^b_{\bm y})
\\
\text{virtual:} & &
{\cal H}^{\text{LO}}_{\text{virtual}} + {\cal H}^{\text{NLO}}_3 = &
\ \frac12\left(\am\,{\cal K}^{\text{LO}}_{{\bm x}, {\bm y}; {\bm z}} -
\am^2{\cal K}^{\text{NLO}}_{3\ {\bm x}, {\bm y}; {\bm z}}\right) 
 ( i\nabla^a_{\bm x} i\nabla^a_{\bm y}
   +i\Bar\nabla^a_{\bm x} i\Bar\nabla^a_{\bm y})
\end{align}
and observe also at this order, that, due to probability conservation as
expressed by Eq.~\eqref{prob}, the limit $U\to 1$ leads to a cancellation of
the sum of {\em all} these contributions. We note: probability
conservation connects all of the above contributions.

The above separation of terms is quite unsatisfactory also if we wish to
extract the running coupling contributions to the leading order
Hamiltonian. Two complementary issues emerge:
\begin{enumerate}
\item Any running coupling correction should come as a uniform modification in
  both real and virtual terms of the leading order kernel, i.e. as a
  replacement
  \begin{align}
    \label{eq:running-coupling-replacement-NLO}
   \am {\cal K}^{\text{LO}}({\bm x}, {\bm y}; {\bm z}) \to &\,
   \am {\cal K}^{\text{LO}}({\bm x}, {\bm y}; {\bm z}) +\am^2  {\cal
     K}^{\text{NLO}}({\bm x}, {\bm y}; {\bm z})
\intertext{with a yet unspecified NLO kernel. This is required if 
an all orders resummation of quark bubbles is to take the form}
\label{eq:eq:running-coupling-replacement-resummed}
    \am {\cal K}^{\text{LO}}({\bm x}, {\bm y}; {\bm z}) \to &\,
    \alpha_s(f({\bm x}, {\bm y}; {\bm z})) {\cal K}^{\text{LO}}({\bm x}, {\bm y}; {\bm z})
  \end{align}
  inside the coordinate integrals of the JIMWLK Hamiltonian and if the pattern
  of real virtual cancellation (and thus probability conservation) be
  maintained beyond the leading order. The sum of real and virtual
  contributions in the above is not of this form; there is no common NLO
  kernel in both terms. Not even the divergent contributions (traceable by the
  $\mu$-dependence of the transverse logarithms) in Eqs.~\eqref{K29}
  and~\eqref{K3old} coincide. This is related to the second issue:
\item The new term, Eq.~(\ref{eq:JIMWLK-NLO-from-A}), contains UV
  divergent contributions where $|{\bm z}_{12}|$, the separation of
  quark and antiquark, reaches the UV cutoff. To extract the UV
  divergence, which is driven by scales much larger than the
  saturation scale $Q_s$, the $U$-dependent part of the quark loop,
  the factor $2\,\text{tr}( t^b U_{{\bm z}_1} t^a U_{{\bm
      z}_2}^\dagger)$, may be expanded in ${\bm z}_{12}$ around some
  fixed base point $\Bar{\bm z}$
  \begin{align}
    \label{eq:quarkloop-local-expansion}
    2\, \text{tr}( t^b & U_{{\bm z}_1}  t^a U_{{\bm
      z}_2}^\dagger) =  U^{a b}_{\Bar{\bm z}}
  \notag \\ &\, + 
  \left[(\Bar{\bm z}-{\bm z}_1)
  \partial_{{\bm z}_1}+(\Bar{\bm z}-{\bm z}_2)
  \partial_{{\bm z}_2}\right]
  2\, \text{tr}( t^b U_{{\bm z}_1} t^a U_{{\bm z}_2}^\dagger)\Big\vert_{{\bm
      z}_{1,2}
    =\Bar{\bm z}}
  + \ldots
  \end{align}
  so that to leading order in this expansion (i.e., keeping the $U^{a
    b}_{\Bar{\bm z}}$ term only) the resulting $U$-dependence
  of~(\ref{eq:JIMWLK-NLO-from-A}) takes a form similar to that
  in~(\ref{eq:JIMWLK-unsubtr-real}). The integral over ${\bm z}_{12}$
  may then be carried out and its divergence exposed as illustrated in
  Fig.~\ref{fig:finite-div-separation}.
%%%%%%%%%%%%%%%%%%%%%%%%%%%%%%%%%%%%%%%%%%%%%%%%%%%%%%%%%%%%%%%%%%%%%%%%%%%
  \begin{figure}[ht]
    \centering
    \begin{equation*}
      \parbox{3cm}{\includegraphics[width=3cm]{LoopInt-quarks-1-firsto}}
      =
\underbrace{
\left[
\parbox{3cm}{\includegraphics[width=3cm]{LoopInt-quarks-1-firsto}}-
\parbox{3cm}{\includegraphics[width=3cm]{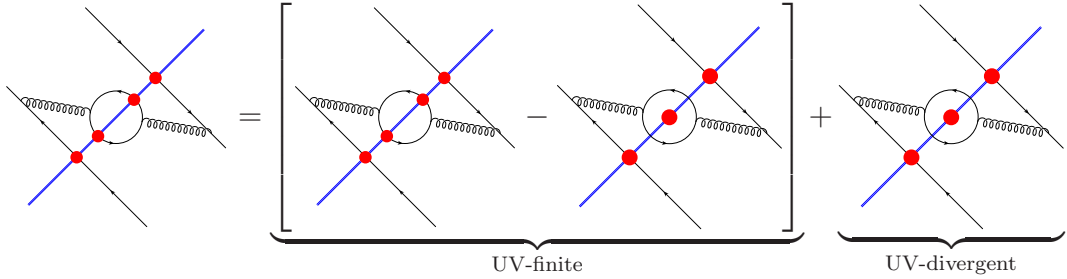}}
\right]}_{\text{UV-finite}} 
+\underbrace{
\parbox{3cm}{\includegraphics[width=3cm]{LoopInt-quarks-1-firsto-subtr}}
}_{\text{UV-divergent}}
    \end{equation*}
    \caption{\em Separating UV-finite and UV-divergent parts of
      Fig.~\ref{fig:NLO1}A}
    \label{fig:finite-div-separation}
  \end{figure}
%%%%%%%%%%%%%%%%%%%%%%%%%%%%%%%%%%%%%%%%%%%%%%%%%%%%%%%%%%%%%%%%%%%%%%%%%%%
  [For this to be sufficient it is of course mandatory that only the
  leading order in this Taylor expansion contains a UV divergence.]
  This divergent contribution must patch up the mismatch between the
  real and virtual terms discussed previously. While the divergence is
  independent of the choice of base point, the finite terms associated with
  the separation shown in Fig.~\ref{fig:finite-div-separation} will
  depend on this choice. This will lead to a scheme dependence to be
  discussed below.
\end{enumerate}
For the JIMWLK Hamiltonian, we are thus led to consider a term of the form
\begin{align}
  \label{eq:JIMWLK-subtraction}
 \am^2 \ln(1/x_{\text{Bj}}) \int d^2z_\xi\, d^2z_{12}\ & 
{\cal K}_1^{\text{NLO}} ({\bm x}, {\bm y}
  ; {\bm z}_1, {\bm z}_2) \  U^{ab}_{\Bar{\bm z}}  \, 
(i\Bar\nabla^a_{\bm x}i\nabla^b_{\bm y}
  +i\nabla^a_{\bm x} i\Bar\nabla^b_{\bm y})
  \hspace{.2cm} U_{{\bm x}_0} \otimes U_{{\bm x}_1}^\dagger
\notag \\ = & 
 \parbox{2cm}{\includegraphics[width=2cm]{LoopInt-quarks-1-firsto-subtr}}
+ \parbox{2cm}{\includegraphics[width=2cm]{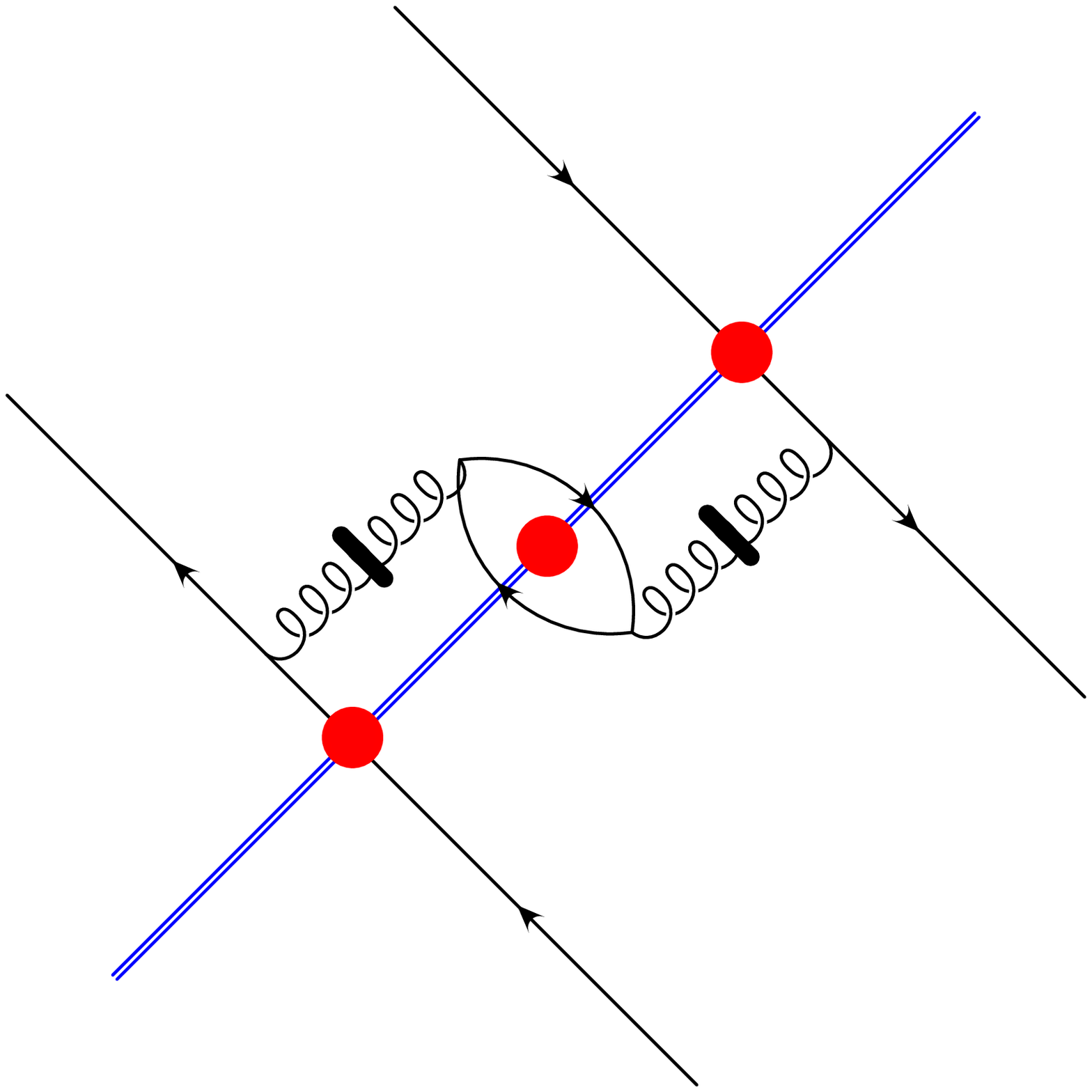}}
+ \ldots
%  = \int d^2z_\xi\, d^2z_{12}\, {\cal K}_1^{\text{NLO}} ({\bm x}, {\bm y}
%  ; {\bm z}_1, {\bm z}_2) 
% U_{{\bm x}_0}t^a \otimes U_{{\bm
%      x}_1}^\dagger t^b \ U^{a b}_{\Bar{\bm z}} \, \ln(1/x_{\text{Bj}}),
\end{align}
which carries the UV divergence of~(\ref{eq:JIMWLK-NLO-from-A}).

By subtracting this contribution from (\ref{eq:JIMWLK-Nfcontribs}) and
adding it to (\ref{eq:JIMWLK-unsubtr-real}), we shift the UV
divergence from a genuinely and physically new contribution in which a
distinguishable, well separated $q\Bar q$ pair interacts with the
target, to the contribution that is not distinguishable from the
single interacting gluon already present at leading order. While the
logarithmically UV divergent term is uniquely defined, the finite
scale dependent terms under the logarithm are not constrained. This is
the origin of our scheme dependence.

To be explicit, we make use of~(\ref{prob2}) for our choice of
$\Bar{\bm z}$ and define what we will call the subtraction term in the
gluon scheme\footnote{The concept of an explicit UV subtraction was first
  introduced by Balitsky in the calculation of transverse coordinate
  space version of NLO BFKL in \cite{Bal:2006}. We thank Ian Balitsky
  for communicating it to us in private.}
\begin{align}
  \label{eq:H-subtr}
&  \Tilde {\cal H}_1^{\text{NLO}} =  
\frac{\am^2}{2} \
\Tilde {\cal K}^{\text{NLO}}_{1\ {\bm x}, {\bm y}  ; {\bm z}}  \  
U^{ab}_{{\bm z}}  \, 
(i\Bar\nabla^a_{\bm x}i\nabla^b_{\bm y}
  +i\nabla^a_{\bm x} i\Bar\nabla^b_{\bm y})
\intertext{where the kernel in the subtraction is calculated with $\Bar{\bm
    z}$ placed at the gluon position ${\bm z}$:}
& \Tilde {\cal K}_1^{\text{NLO}}({\bm x}, {\bm y}
  ; {\bm z}) = \int d^2z_{12}\, {\cal K}_1^{\text{NLO}} ({\bm x}, {\bm y}
  ; {\bm z}_1, {\bm z}_2).\label{K1K}
\end{align}
The explicit form of the right hand side of \eq{K1K} was already
obtained in our calculation of the fully virtual corrections in
Sect.~\ref{sec:diagram-c} with the answer given by Eq.~\eqref{K1int5}.
We use it first to define a genuinely UV finite $q\Bar q$ contribution
of the form
\begin{align}
  \label{eq:qqbar-subtr-JIMWLK}
%  {\cal H}^{\text{NLO}}_{q\Bar q} := 
%  {\cal H}_1^{\text{NLO}}-\Tilde {\cal H}_1^{\text{NLO}}
\frac{\am^2}{2} 
   {\cal K}^{\text{NLO}}_{1\ {\bm x}, {\bm y}; {\bm z}_1, {\bm z}_2}
   \ 2\, \tr(t^a U_{{\bm z}_1} t^bU_{{\bm z}_1}^\dagger)
   (i\Bar\nabla^a_{\bm x}i\nabla^b_{\bm y}   
  +i\nabla^a_{\bm x} i\Bar\nabla^b_{\bm y})
-   \frac{\am^2}{2} 
   \Tilde {\cal K}^{\text{NLO}}_{1\ {\bm x}, {\bm y}; {\bm z}}
  \ U_{\bm z}^{a b}
(i\Bar\nabla^a_{\bm x}i\nabla^b_{\bm y}   
  +i\nabla^a_{\bm x} i\Bar\nabla^b_{\bm y})\, .
\end{align}
This contribution only is of interest if we wish to go beyond the inclusion of
running coupling corrections to include genuine NLO contributions. While such
a calculation would be interesting and important, it remains beyond the scope
of this paper. Here we note that the term in \eq{eq:qqbar-subtr-JIMWLK} is
UV-finite and vanishes {\em by itself} in the no-interaction limit of $U \to
1$: This term no longer mixes with the remaining contributions under
probability conservation and --contrary to the unsubtracted contributions-- we
may neglect if we are only interested in the the scale of the running
coupling.

The remaining contributions now assemble directly into a form that
fulfills all requirements of a running coupling contribution.
Using~\eqref{prob2}, we find that adding~\eqref{eq:H-subtr}
to~\eqref{eq:JIMWLK-unsubtr-real} leaves us with identical kernels
both for real and virtual contributions
\begin{align}
  \label{eq:NLO-JIMWLK-kernel}
  \am\, {\cal K}^{\text{LO}}({\bm x}, {\bm y}; {\bm z})
-\am^2\,{\cal K}^{\text{NLO}}_3({\bm x}, {\bm y}; {\bm z})
=
  \am\, {\cal K}^{\text{LO}}({\bm x}, {\bm y}; {\bm z})
+\am^2\,\left(\Tilde {\cal K}^{\text{NLO}}_1({\bm x}, {\bm y}; {\bm z})
  + {\cal K}^{\text{LO}}_2({\bm x}, {\bm y}; {\bm z})\right).
\end{align}
The leading $N_f$ contributions to the running coupling corrections for the
JIMWLK Hamiltonian take a from analogous to~\eqref{eq:JIMWLK-Hamiltonian-LO}
\begin{align}
  \label{eq:JIMWLK-Hamiltonian-running-NLO}
  \frac{\am}{2} 
   \left[{\cal K}^{\text{LO}}_{{\bm x}, {\bm y}; {\bm z}}
     +\am\left(\Tilde {\cal K}^{\text{LO}}_{1\ {\bm x}, {\bm y}; {\bm z}} 
       +{\cal K}^{\text{LO}}_{2\ {\bm x}, {\bm y}; {\bm z}}\right)\right]
   \left[ 
  \Tilde U_z^{a b}(i\Bar\nabla^a_{\bm x}i\nabla^b_{\bm y}   
  +i\nabla^a_{\bm x} i\Bar\nabla^b_{\bm y})
+
  ( i\nabla^a_{\bm x} i\nabla^a_{\bm y}+i\Bar\nabla^a_{\bm x} i\Bar\nabla^a_{\bm y})
  \right]
\, .
\end{align}
The equations in the Balitsky hierarchy created with this operator are
finite and unitary for fixed projectile configurations. The
subtraction fully decouples conformal
contributions~\eqref{eq:qqbar-subtr-JIMWLK} and non-conformal
contributions~\eqref{eq:JIMWLK-Hamiltonian-running-NLO} up to the
order $\am^2$ making it feasible to discuss running coupling
corrections independently of the conformal contributions.

\subsection{Subtraction for the BK Equation}
\label{sec:subtr-BK}

The UV subtraction described above for JIMWLK evolution equation can be
translated to the BK framework~\cite{Balitsky:1996ub, Kovchegov:1999yj,
  Kovchegov:1999ua} simply by repeating the steps that allow to identify BK as
a limiting case of JIMWLK at the leading order. Here we will instead formulate
the argument again entirely within the BK framework to provide a self contained
discussion. We begin by writing the standard LO BK evolution equation for the
forward amplitude of a quark dipole scattering on a nucleus
\begin{align}\label{N}
  N ({\bm x}_0,{\bm x}_1, Y) \, \equiv \, 1 - \frac{1}{N_c} \, \left<
    \tr{ \, \left[ U_{{\bm x}_0} \, U_{{\bm x}_1}^\dagger \right]}
  \right> (Y),
\end{align}
where $U$'s are from \eq{U}, the transverse coordinates of the quark
and the anti-quark are ${\bm x}_0$ and ${\bm x}_1$, and the dipole's
rapidity is $Y$. The LO BK equation reads
\begin{align}\label{eqN}
  \frac{\partial N ({\bm x}_{0}, {\bm x}_1, Y)}{\partial Y} \, = \,
  \frac{\am \, C_F}{\pi^2} \, \int d^2 x_2 \, \frac{x_{01}^2}{x_{20}^2
    \, x_{21}^2} \, \left[ N ({\bm x}_{0}, {\bm x}_2, Y) + N ({\bm
      x}_{2}, {\bm x}_1, Y) - N ({\bm x}_{0}, {\bm x}_1, Y) \right.
  \notag \\ \left. - N ({\bm x}_{0}, {\bm x}_2, Y) \, N ({\bm x}_{2},
    {\bm x}_1, Y) \right],
\end{align}
where $x_{mn} = |{\bm x}_m - {\bm x}_n|$ and the large-$N_c$ limit is
assumed. Using the LO JIMWLK kernel from \eq{eq:LO-JIMWLK-kernel} we
can define the LO dipole kernel by
\begin{align}\label{KLOBK}
  K^{\text{LO}} ({\bm x}_0, {\bm x}_1 ; {\bm z}) \, = \, C_F \, \sum_{m,n
    = 0}^1 \, (-1)^{m+n} \, {\cal K}^{\text{LO}} & ({\bm x}_m, {\bm x}_n ;
  {\bm z}).
\end{align}
The dipole kernel (\ref{KLOBK}) sums up the same diagrams as shown in
\eq{eq:JIMWLK-LO-diagram-cont} for the LO JIMWLK Hamiltonian
\cite{Mueller:1994rr}.  Using \eq{KLOBK} we rewrite \eq{eqN} as
\begin{align}\label{eqN2}
  \frac{\partial N ({\bm x}_{0}, {\bm x}_1, Y)}{\partial Y} \, = \,
  \am \, \int d^2 x_2 \, K^{\text{LO}} ({\bm x}_0, {\bm x}_1 ; {\bm
    x}_2) \, \left[ N ({\bm x}_{0}, {\bm x}_2, Y) + N ({\bm x}_{2},
    {\bm x}_1, Y) - N ({\bm x}_{0}, {\bm x}_1, Y) \right.  \notag \\ 
  \left. - N ({\bm x}_{0}, {\bm x}_2, Y) \, N ({\bm x}_{2}, {\bm x}_1,
    Y) \right].
\end{align}
For the purpose of performing the UV subtraction, it is more
convenient to rewrite \eq{eqN2} in terms of the $S$-matrix
\begin{align}\label{S}
  S ({\bm x}_0,{\bm x}_1, Y) \, \equiv \, \frac{1}{N_c} \, \left< \tr{
      \, \left[ U_{{\bm x}_0} \, U_{{\bm x}_1}^\dagger \right]}
  \right> (Y) \, = \, 1 - N ({\bm x}_{0}, {\bm x}_1, Y)
\end{align}
obtaining
\begin{align}\label{eqS}
  \frac{\partial S ({\bm x}_{0}, {\bm x}_1, Y)}{\partial Y} \, = \,
  \am \, \int d^2 x_2 \, K^{\text{LO}} ({\bm x}_0, {\bm x}_1 ; {\bm
    x}_2) \, \left[ S ({\bm x}_{0}, {\bm x}_2, Y) \, S ({\bm x}_{2},
    {\bm x}_1, Y) - S ({\bm x}_{0}, {\bm x}_1, Y) \right].
\end{align}

Now we are ready to include the NLO corrections calculated in Section
\ref{LObb}. Adding the diagrams in Figs. \ref{fig:NLO1} and
\ref{fig:NLO1_inst} to the LO dipole kernel yields the following
evolution equation
\begin{align}\label{eqS_NLO}
  \frac{\partial S ({\bm x}_{0}, {\bm x}_1, Y)}{\partial Y} \, = \,
  \am \, \int d^2 x_2 \, K^{\text{LO}} ({\bm x}_0, {\bm x}_1 ; {\bm
    x}_2) \, \left[ S ({\bm x}_{0}, {\bm x}_2, Y) \, S ({\bm x}_{2},
    {\bm x}_1, Y) - S ({\bm x}_{0}, {\bm x}_1, Y) \right] \notag \\ +
  \am^2 \, \int d^2 z_1 \, d^2 z_2 \, K_1^{\text{NLO}} ({\bm x}_0,
  {\bm x}_1 ; {\bm z}_1, {\bm z}_2) \, S ({\bm x}_{0}, {\bm z}_1, Y)
  \, S ({\bm z}_{2}, {\bm x}_1, Y) \notag \\ + \, \am^2 \, \int d^2
  x_2 \, K_2^{\text{NLO}} ({\bm x}_0, {\bm x}_1 ; {\bm x}_2) \, S
  ({\bm x}_{0}, {\bm x}_2, Y) \, S ({\bm x}_{2}, {\bm x}_1, Y) \notag
  \\ + \, \am^2 \, \int d^2 x_2 \, K_3^{\text{NLO}} ({\bm x}_0, {\bm
    x}_1 ; {\bm x}_2) \, S ({\bm x}_{0}, {\bm x}_1, Y).
\end{align}
Similar to the JIMWLK case we notice that while kernels
$K_2^{\text{NLO}}$ and $K_3^{\text{NLO}}$ appear as higher order
corrections to the leading order BK kernel $K^{\text{LO}}$ having the
same transverse coordinate dependence, the kernel $K_1^{\text{NLO}}$
stands out. It includes integrals over two transverse vectors, ${\bm
  z}_1$ and ${\bm z}_2$, instead of one. Since the shape of the
leading kernel $K^{\text{LO}}$ in not preserved in $K_1^{\text{NLO}}$
and as we are looking for running coupling corrections to
$K^{\text{LO}}$, one may naively discard $K_1^{\text{NLO}}$ as not
giving any running coupling contribution. However, before we embark on
extracting the running coupling corrections, let us formulate general
rules for such corrections. Similar to the JIMWLK case we require the
following:
\begin{itemize}
\item {\sl Unitarity}: as \eq{eqS} gives an explicitly unitary
  solution for $S$, i.e., as rapidity $Y \rightarrow \infty$ then $S
  \rightarrow 0$, we require the running coupling corrections to
  preserve this unitarity property. This requirement is satisfied as
  long as the right hand side of the evolution equation has only terms
  containing powers of $S$.
  
\item {\sl No interaction --- no evolution condition}: we require that
  in the absence of interaction the right hand side of the resulting
  evolution equation should become $0$ when $S=1$ is inserted there.
  This condition is easily satisfied in the standard Feynman
  perturbation theory where the non-interacting graphs are zero. In
  the light cone perturbation theory (LCPT) the non-interacting
  diagrams are not zero, which allows us to define and calculate light
  cone wave functions. Because of that it is a little harder to show
  in LCPT that in the absence of interactions all diagrams for the
  amplitude (not to be confused with the wave function) cancel. For
  instance, the {\sl no interaction --- no evolution} condition is
  satisfied by \eq{eqS_NLO}: if we put $S=1$ on its right hand side we
  will get zero due to the condition in \eq{prob2}. We want this
  property to be preserved after running coupling corrections are
  included.
\end{itemize}

\begin{figure}[ht]
    \centering
    \begin{equation*}
      \parbox{3cm}{\includegraphics[width=3cm]{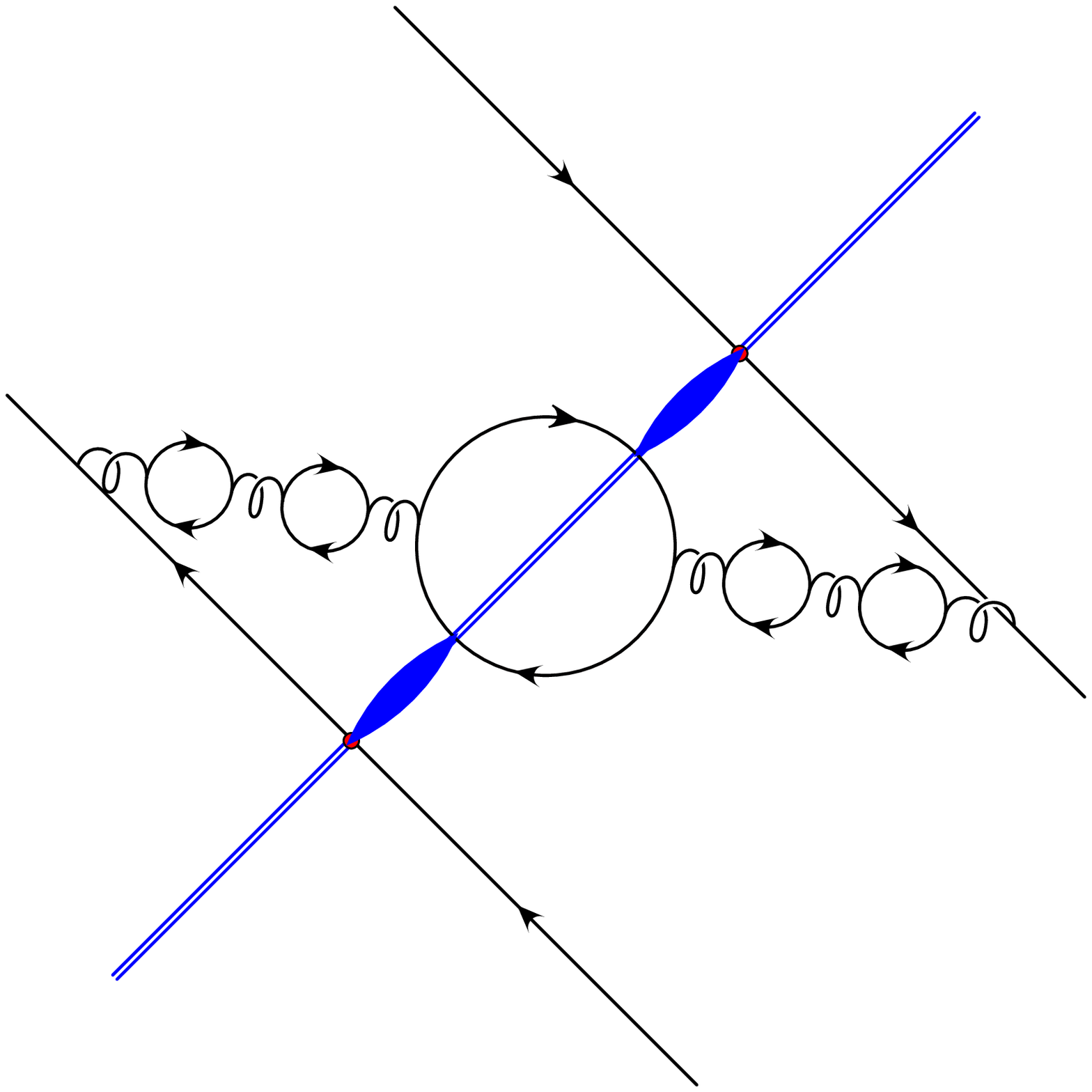}}
      =
\underbrace{
\left[
\parbox{3cm}{\includegraphics[width=3cm]{LoopInt-quarks-1-dipole}}-
\parbox{3cm}{\includegraphics[width=3cm]{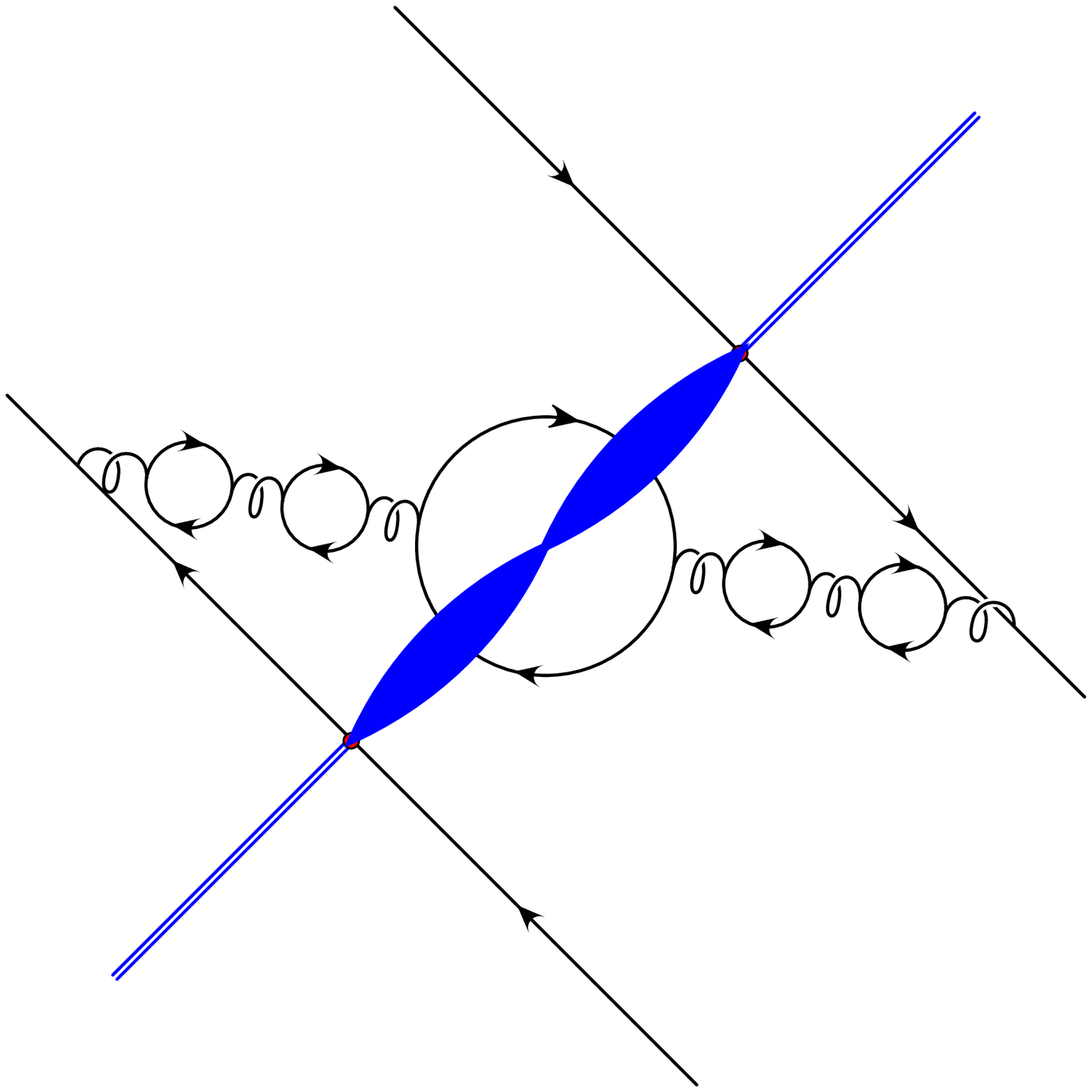}}
\right]}_{\text{UV-finite}} 
+\underbrace{
\parbox{3cm}{\includegraphics[width=3cm]{LoopInt-quarks-1-dipole-subtr}}
}_{\text{UV-divergent}}
    \end{equation*}
    \caption{\em Separating UV-finite and UV-divergent parts of
      the kernel $K_1^{\text{NLO}}$ in the NLO BK evolution. The ovals
      denote color dipoles. }
    \label{fig:BKsubtr}
  \end{figure}

From the above conditions one can see that simply discarding
$K_1^{\text{NLO}}$ from the right hand side of \eq{eqS_NLO} would not
work: while the equation obtained this way would satisfy the unitarity
condition, it would not satisfy the second condition stated above,
since, for $S=1$ we will not get zero on the right hand side anymore.
What strengthens the case for keeping a part of $K_1^{\text{NLO}}$ is
that it contains a UV divergence, as can be seen from \eq{K1int5},
which may contribute to the running of the coupling constant. Kernels
$K_2^{\text{NLO}}$ and $K_3^{\text{NLO}}$ also contain UV divergences,
which need to be canceled by the divergence in $K_1^{\text{NLO}}$ as
follows from \eq{prob2}. Without $K_1^{\text{NLO}}$ the right hand
side of \eq{eqS_NLO} would become infinite. Therefore, to keep the
right hand side of the resulting evolution equation finite, and in
order to satisfy the second one of the above conditions, we propose
the subtraction illustrated in \fig{fig:BKsubtr}.
%%%%%%%%%%%%%%%%%%%%%%%%%%%%%%%%%%%%%%%%%%%%%%%%%%%%%%%%%%%%%%%%%%%%%%%%%%%
%%%%%%%%%%%%%%%%%%%%%%%%%%%%%%%%%%%%%%%%%%%%%%%%%%%%%%%%%%%%%%%%%%%%%%%%%%%

Formally we write
\begin{align}\label{K1subtr}
  \int d^2 z_1 \, d^2 z_2 \, K_1^{\text{NLO}} ({\bm x}_0, {\bm x}_1 ;
  {\bm z}_1, {\bm z}_2) \, S ({\bm x}_{0}, {\bm z}_1, Y) \, S ({\bm
    z}_{2}, {\bm x}_1, Y) \notag \\ \, = \, \int d^2 z \, d^2 z_{12}
  \, K_1^{\text{NLO}} ({\bm x}_0, {\bm x}_1 ; {\bm z}_{1}, {\bm z}_2)
  \, \left[ S ({\bm x}_{0}, {\bm z}_1, Y) \, S ({\bm z}_{2}, {\bm
      x}_1, Y) - S ({\bm x}_{0}, {\bm z}, Y) \, S ({\bm z}, {\bm x}_1,
    Y) \right. \notag \\ + \, \left.  S ({\bm x}_{0}, {\bm z}, Y) \, S
    ({\bm z}, {\bm x}_1, Y) \right],
\end{align}
where $\bm z$ is the position of the virtual gluon in \fig{fig:NLO1}A
defined in \eq{z}.\footnote{Indeed as $\bm z$ from \eq{z} depends on
  the longitudinal momentum fraction of the quark $\alpha$, switching
  from ${\bm z}_1$ and ${\bm z}_2$ to ${\bm z}_{12}$ and $\bm z$
  implies a change in the $\alpha$-integral in $K_1^{\text{NLO}}$.
  However, since above we have used ${\bm z}_{12}$ and $\bm z$ vectors
  everywhere, no changes apply to our earlier results.} Now the first
two terms in the square brackets on the right hand side of
\eq{K1subtr} give a UV-finite result, as shown in \fig{fig:BKsubtr},
which goes to zero both for $S=0$ and $S=1$. These terms combined do
not have a UV-divergence and do not contribute to the running coupling
constant.  They give a non-running coupling NLO BK evolution piece and
we will discard them here. The last term on the right hand side of
\eq{K1subtr} we will keep. Similar to the JIMWLK case we define the
subtraction kernel by
\begin{align}\label{K1sub}
  {\tilde K}_1^{\text{NLO}} ({\bm x}_0, {\bm x}_1 ; {\bm z}) \, \equiv
  \, \int d^2 z_{12} \, K_1^{\text{NLO}} ({\bm x}_0, {\bm x}_1 ; {\bm
    z}_{1}, {\bm z}_2).
\end{align}
An explicit form of ${\tilde K}_1^{\text{NLO}} ({\bm x}_0, {\bm x}_1 ;
{\bm z})$ can be found using Eqs. (\ref{K1int5}) and (\ref{K1BK}).
With the help of the definition in \eq{K1sub} we rewrite the last term
in \eq{K1subtr} as
\begin{align}
  \int d^2 x_2 \, {\tilde K}_1^{\text{NLO}} ({\bm x}_0, {\bm x}_1 ;
  {\bm x}_2) \, S ({\bm x}_{0}, {\bm x}_2, Y) \, S ({\bm x}_{2}, {\bm
    x}_1, Y).
\end{align}
Keeping only this term in kernel $K_1^{\text{NLO}}$ modifies
\eq{eqS_NLO} to give
\begin{align}\label{eqS_NLO_sub1}
 \frac{\partial S ({\bm x}_{0}, {\bm x}_1, Y)}{\partial Y} \, = \,
  \am \, \int d^2 x_2 \, K^{\text{LO}} ({\bm x}_0, {\bm x}_1 ; {\bm
    x}_2) \, \left[ S ({\bm x}_{0}, {\bm x}_2, Y) \, S ({\bm x}_{2},
    {\bm x}_1, Y) - S ({\bm x}_{0}, {\bm x}_1, Y) \right] %\displaybreak[0] 
\notag \\ +
  \am^2 \, \int d^2 x_2 \, {\tilde K}_1^{\text{NLO}} ({\bm x}_0,
  {\bm x}_1 ; {\bm x}_2) \, S ({\bm x}_{0}, {\bm x}_2, Y)
  \, S ({\bm x}_{2}, {\bm x}_1, Y) \notag \\ + \, \am^2 \, \int d^2
  x_2 \, K_2^{\text{NLO}} ({\bm x}_0, {\bm x}_1 ; {\bm x}_2) \, S
  ({\bm x}_{0}, {\bm x}_2, Y) \, S ({\bm x}_{2}, {\bm x}_1, Y) \notag
  \\ + \, \am^2 \, \int d^2 x_2 \, K_3^{\text{NLO}} ({\bm x}_0, {\bm
    x}_1 ; {\bm x}_2) \, S ({\bm x}_{0}, {\bm x}_1, Y).
\end{align}
Using \eq{prob2} we rewrite \eq{eqS_NLO_sub1} as
\begin{align}\label{eqS_NLO_sub}
  \frac{\partial S ({\bm x}_{0}, {\bm x}_1, Y)}{\partial Y} \, = \,
  \int d^2 x_2 \, \left[ \am \, K^{\text{LO}} ({\bm x}_0, {\bm x}_1 ;
    {\bm x}_2) + \am^2 \, {\tilde K}_1^{\text{NLO}} ({\bm x}_0, {\bm
      x}_1 ; {\bm x}_2) + \am^2 \, K_2^{\text{NLO}} ({\bm x}_0, {\bm
      x}_1 ; {\bm x}_2) \right] \notag \\ \times \, \left[ S ({\bm
      x}_{0}, {\bm x}_2, Y) \, S ({\bm x}_{2}, {\bm x}_1, Y) - S ({\bm
      x}_{0}, {\bm x}_1, Y) \right].
\end{align}
\eq{eqS_NLO_sub} obeys both of the conditions stated above: its right
hand side is zero at both $S=0$ and $S=1$. Moreover the right hand
side of \eq{eqS_NLO_sub} is UV finite, which is essential for
obtaining a meaningful result. The kernels ${\tilde K}_1^{\text{NLO}}$
and $K_2^{\text{NLO}}$ both look like corrections to the LO kernel. In
the following, when we study fermion bubble insertions to all orders,
we will use the format of \eq{eqS_NLO_sub} to systematically include
their contributions into the running of the coupling constant.

The choice of subtracting and adding $S ({\bm x}_{0}, {\bm z}, Y) \, S
({\bm z}, {\bm x}_1, Y)$ depending on gluon's position $\bm z$ in
\eq{K1subtr} is indeed quite arbitrary. For instance, one can use
${\bm z}_1$ or ${\bm z}_2$ (or any other linear combination of the two
vectors ${\bm z}_1$ and ${\bm z}_2$) in place of ${\bm
  z}$.\footnote{We thank Ian Balitsky for helping us to reach this
  conclusion.} We can not find any argument or criterion which would
prefer one choice of the ``subtraction point'' over the other. We
choose $\bm z$ as our ``subtraction point'' since it appears to be
convenient and goes along the lines of calculating $K_3^{\text{NLO}}$
in \eq{K1int5}. This choice appears to also be preferred by the
dispersive method of calculating the running coupling correction to
small-$x$ evolution used in \cite{Gardi:2006}.  Indeed the uncertainty
in selecting the ``subtraction point'' does not affect our ability to
extract the UV divergent part of $K_1^{\text{NLO}}$.  However, it may
change the scale $R$ under the logarithm in \eq{K1int5}, resulting in
a different scale for the running coupling constant. We believe that
the modification of the running coupling scale due to varying the
``subtraction point'' will be numerically insignificant: however, a
detailed study of this question is left for further investigations.
Here we will refer to this dependence of the running coupling scale on
the ``subtraction point'' as of some sort of a scheme dependence for
the running coupling constant.

%%%%%%%%%%%%%%%%%%%%%%%%%%%%%%%%%%%%%%%%%%%%%%%%%%%%%%%%%%%%%%%%%%%%%%%%%%%%%%%%%

%\subsection{Scheme Dependence due to Subtractions}

%Unitarity for fixed configurations manifests itself in the limit of
%vanishing correlators $\langle U^{(\dagger)}_{{\bm x}_1} \otimes
%\cdots \otimes U^{(\dagger)}_{{\bm x}_n}\rangle\to 0$ as the
%requirement that also there evolution ceases and the r.h.s. of the
%evolution equation for any correlator vanishes. {\bf does not seem to
%  pose any restriction on the S level -- this should be an argument on
%  the N-level. What did I have there earlier?}

%%%%%%%%%%%%%%%%%%%%%%%%%%%%%%%%%%%%%%%%%%%%%%%%%%%%%%%%%%%%%%%%%%%%%%%%%%%%%%%%%

\newpage

\section{Resummation of Bubbles to All Orders: Setting the Scale for the 
Running Coupling Constant}
\label{all_orders}

\begin{figure}[htbp]
  \centering
  \begin{minipage}{5.2cm}
\centering
\includegraphics[width=5.2cm]{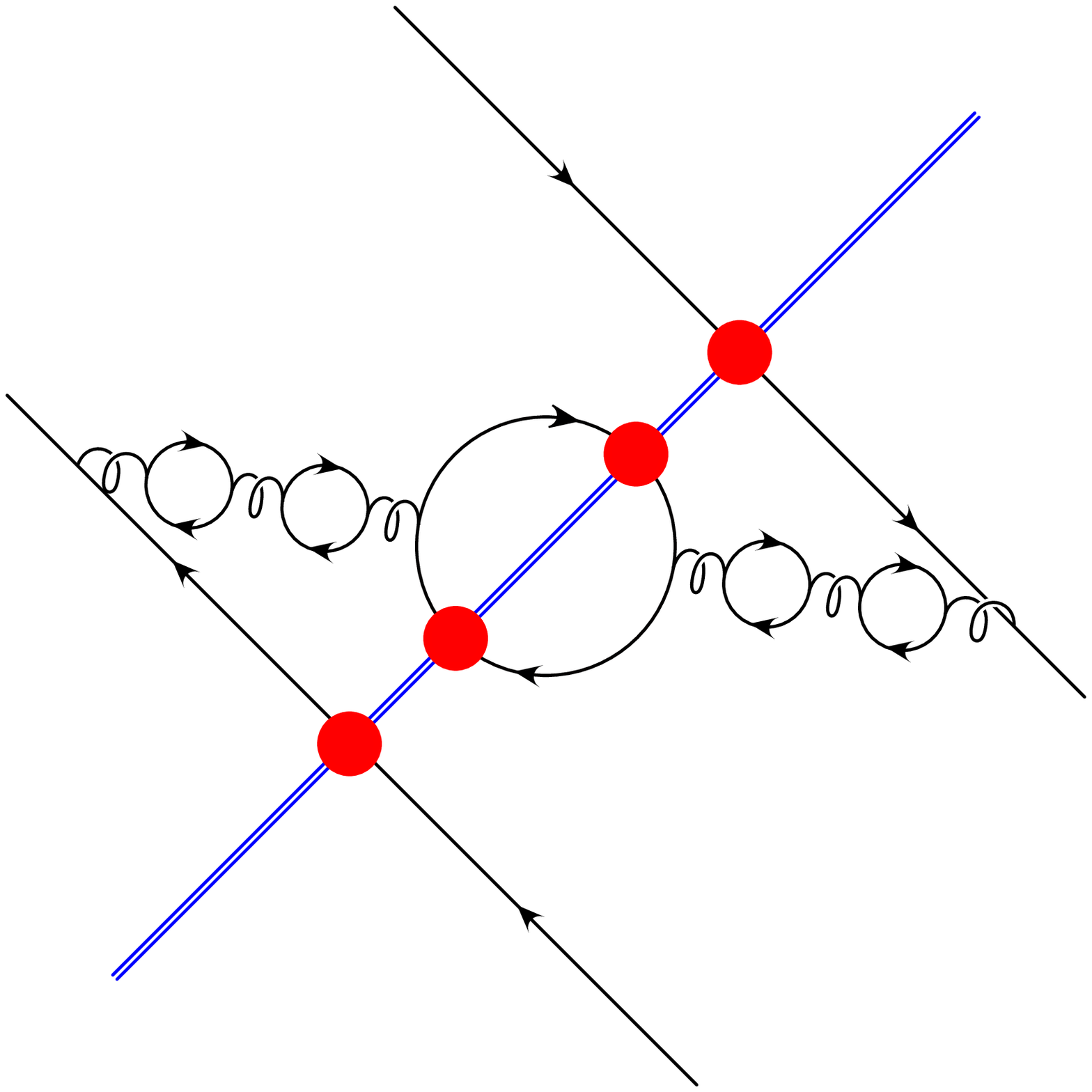}\\
A
  \end{minipage}
  \begin{minipage}{5.2cm}  
\centering
  \includegraphics[width=5.2cm]{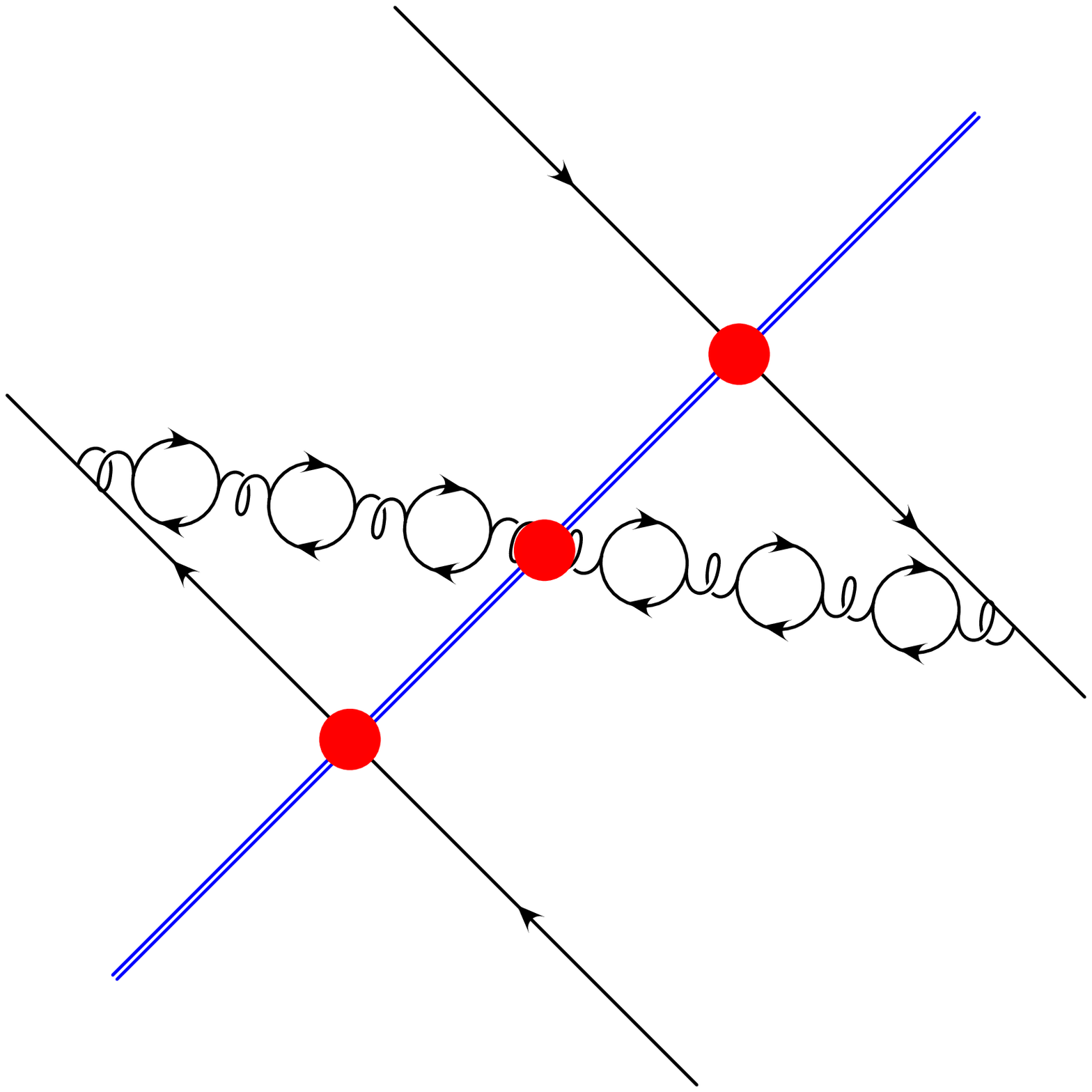}  
\\ B
\end{minipage}  
\begin{minipage}{5.2cm} 
\centering 
  \includegraphics[width=5.2cm]{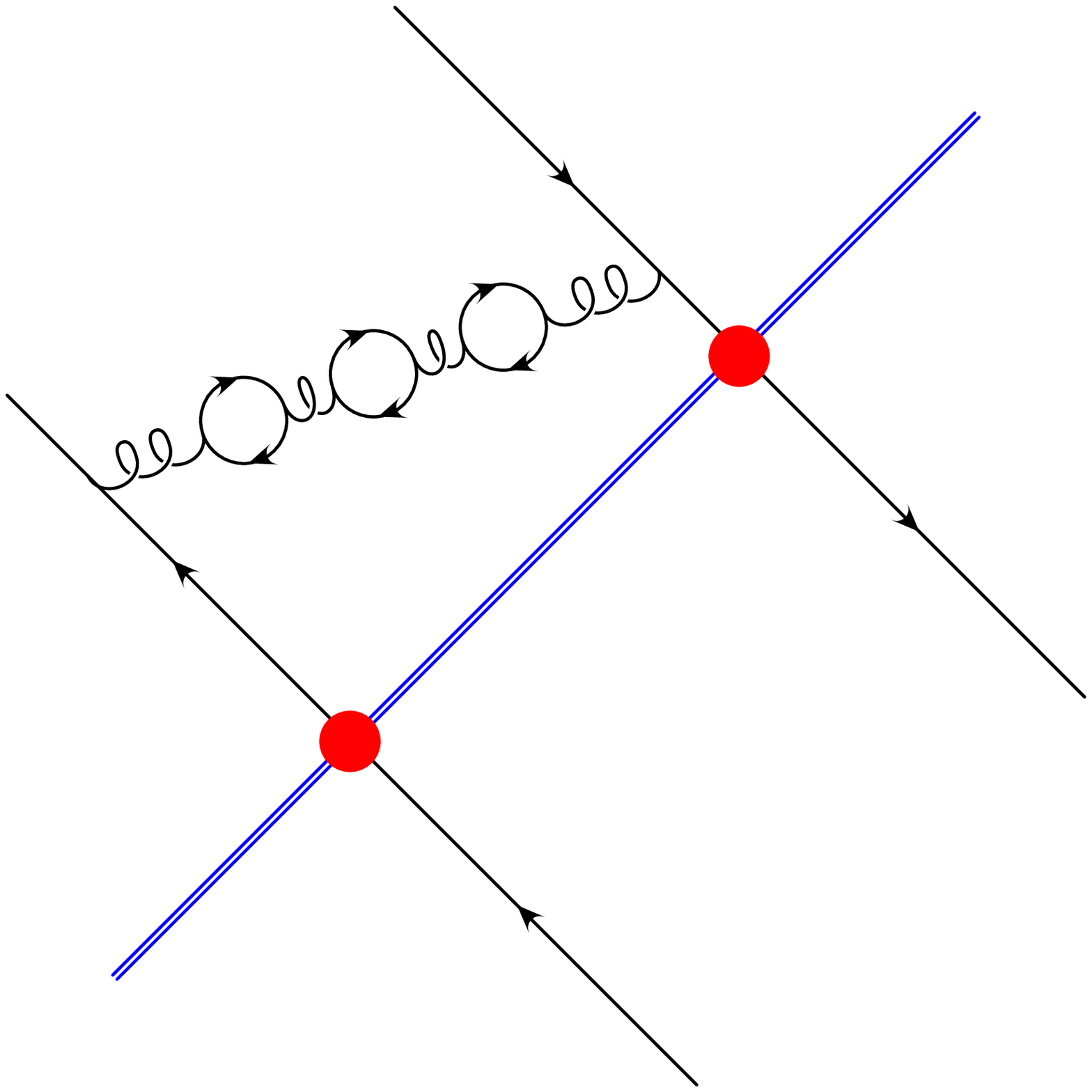}
\\ C
\end{minipage} 
  \caption{\em Diagrams giving the higher order $\as N_f$ corrections 
    to the kernels of JIMWLK and BK small-$x$ evolution equations.  To
    get the all-order $\as N_f$ contribution one has to sum an
    infinite series of quark bubble insertions. }
  \label{fig:NLO_all}
\end{figure}

\begin{figure}[htbp]
  \centering
  \begin{minipage}{5.2cm}
\centering
\includegraphics[width=5.2cm]{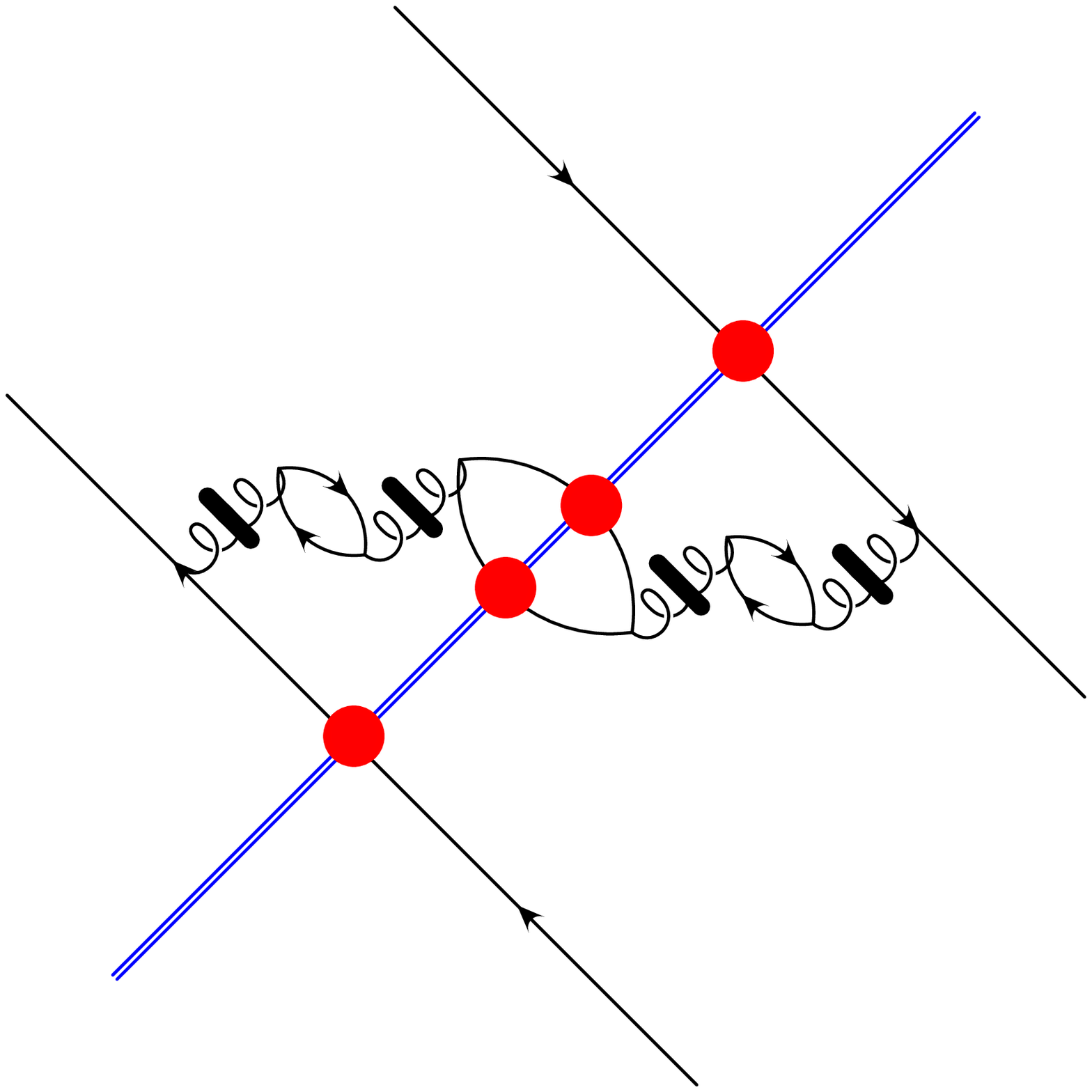}\\
A$^\prime$
  \end{minipage}
\hspace{5.2cm}
\begin{minipage}{5.2cm} 
\centering 
  \includegraphics[width=5.2cm]{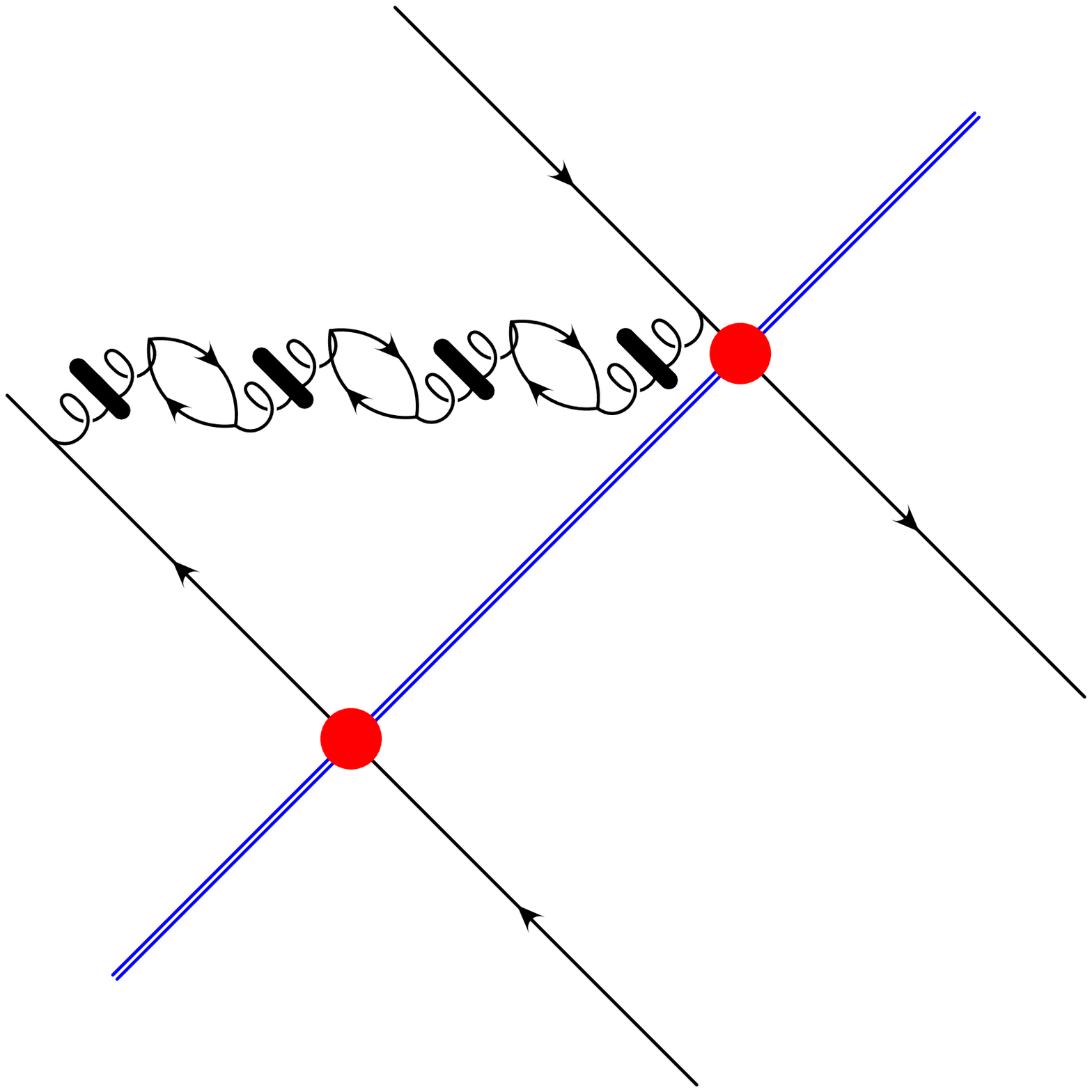}
\\ C$^\prime$
\end{minipage} 
  \caption{\em Diagrams giving the higher order $\as N_f$ corrections 
    to the kernels of JIMWLK and BK small-$x$ evolution equations
    containing instantaneous gluon lines.  Again, to get the all-order
    $\as N_f$ contribution one has to sum an infinite series of quark
    bubble insertions. }
  \label{fig:NLO_inst}
\end{figure}

Now we are ready to resum all powers of $\as \, N_f$ corrections in
the JIMWLK and BK evolution kernels. To accomplish that one has to
insert infinite chains of gluon bubbles onto the gluon lines in Figs.
\ref{fig:NLO1} and \ref{fig:NLO1_inst}. An example of corresponding
higher-order diagrams is shown in Figs. \ref{fig:NLO_all} and
\ref{fig:NLO_inst}.

\begin{figure}[htbp]
  \centering
  \begin{minipage}{5.2cm}
\centering
\includegraphics[width=5.2cm]{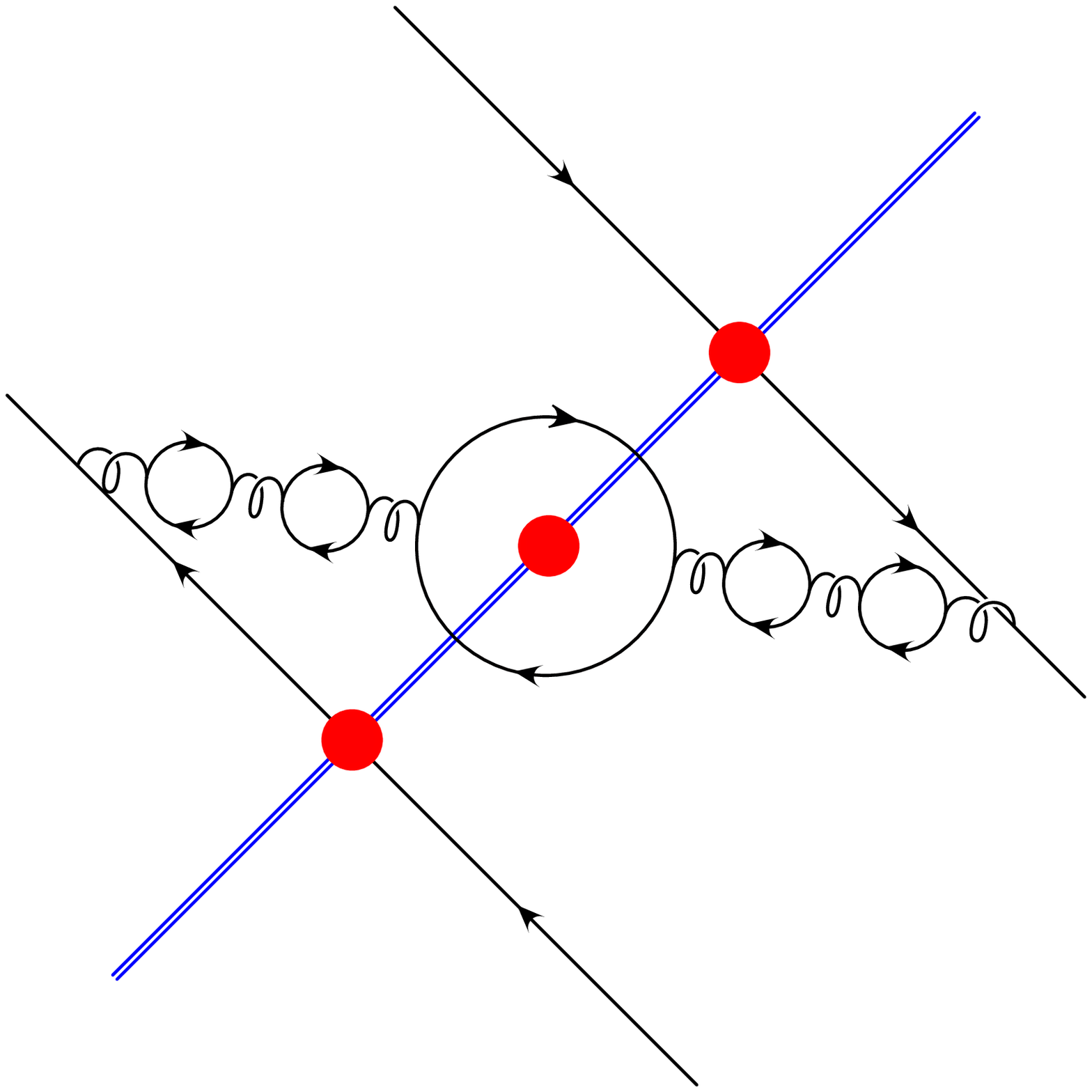} \\
A
 \end{minipage}
\hspace{3cm}
\begin{minipage}{5.2cm} 
\centering 
\includegraphics[width=5.2cm]{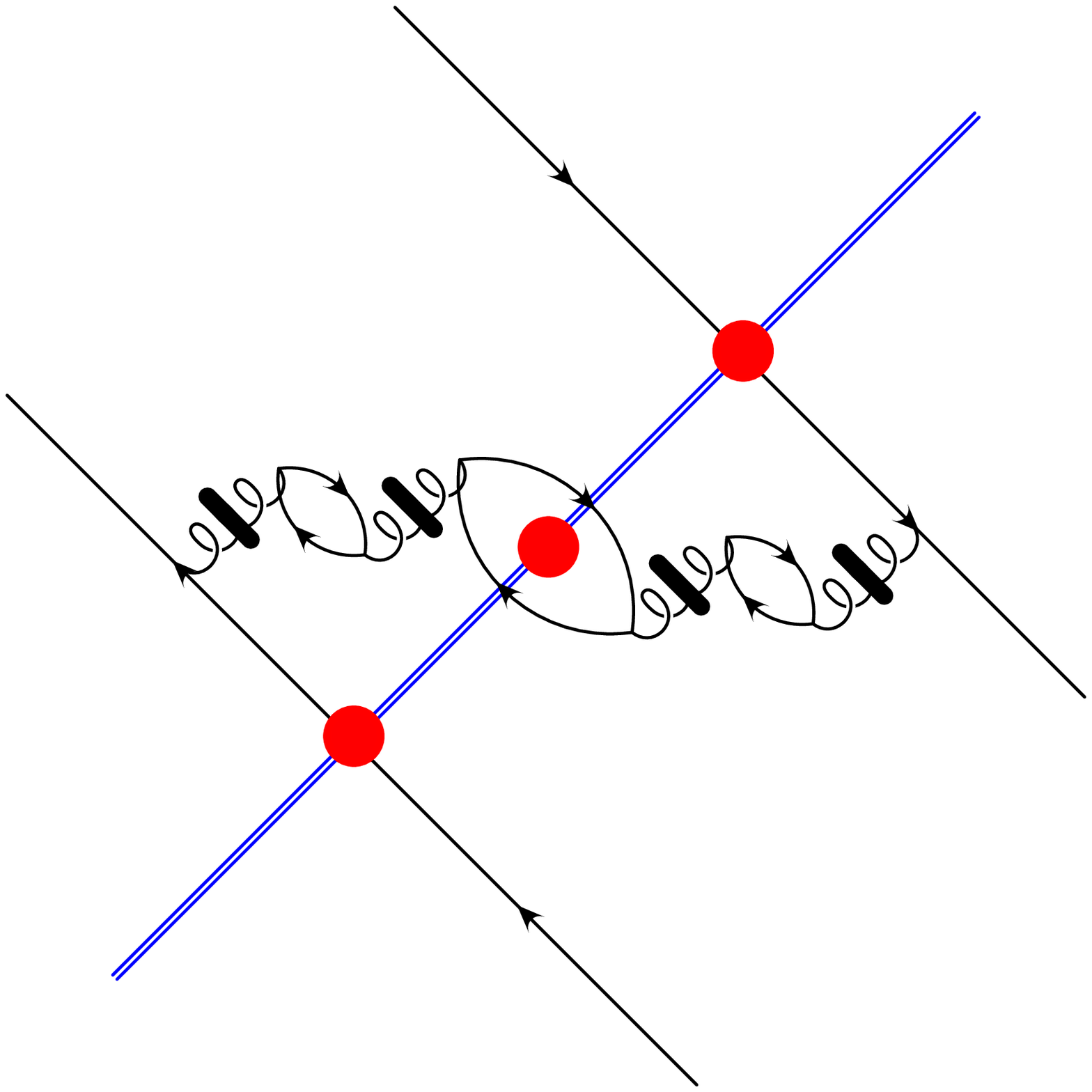} \\
A$^\prime$
  \end{minipage}
   \caption{\em The ``subtraction'' diagrams which should be considered 
     in place of the diagrams in Figs. \ref{fig:NLO_all} A and
     \ref{fig:NLO_inst} A$^\prime$ for the running coupling scale
     calculations.}
 \label{fig:NLO_A}
\end{figure}

An explicit calculation using the rules of light-cone perturbation
theory \cite{Lepage:1980fj,Brodsky:1997de} shows that inserting
all-order quark bubbles on the gluon lines generates geometric series
in momentum space. Before calculating the diagrams in Figs.
\ref{fig:NLO_all} and \ref{fig:NLO_inst} we remember that, as was
discussed above, in order to find the running coupling correction,
instead of the diagram in Figs.  \ref{fig:NLO_all}A and
\ref{fig:NLO_inst}A$^\prime$ we should consider the ``subtraction''
diagrams A and A$^\prime$ shown in \fig{fig:NLO_A}.

By an explicit calculation, similar to the calculation of a single
bubble insertion which led to \eq{K26}, one can show that the
contribution of the ``dressed'' subtraction diagrams A and A$^\prime$
in \fig{fig:NLO_A} to the JIMWLK kernel reads
\begin{align}\label{K1sdr}
  \am^2 \, {\tilde {\cal K}}_{\oone} ({\bm x}_0, {\bm x}_1 ; {\bm z})
  \, = \, 4 \, \am^2 \, \beta_2 \, \int \frac{d^2 q}{(2\pi)^2}
  \frac{d^2 q'}{(2\pi)^2} \ e^{ -i {\bm q}\cdot ({\bm z}-{\bm x}_0) +i
    {\bm q}' \cdot ({\bm z}-{\bm x}_1) } \notag \\ \times \, \Bigg\{
  \frac{{\bm q} \cdot {\bm q}'}{{\bm q}^2{\bm q}'^{2}} \, \frac{{\bm
      q}^{2} \, \ln \frac{{\bm q}^2 \,
      e^{-5/3}}{\mu_{\overline{{\text{MS}}}}^2} - {\bm q}'^{2} \, \ln
    \frac{{\bm q}'^2 \, e^{-5/3}}{\mu_{\overline{{\text{MS}}}}^2} }{{\bm q}^{2}
    - {\bm q}'^{2}} - \frac{\ln ({\bm q}^{2}/{\bm q}'^{2})}{{\bm
      q}^{2} - {\bm q}'^{2}} \Bigg\} \notag \\ \times \,
  \frac{1}{\left( 1 + \am \beta_2 \ln \frac{{\bm q}^2 \,
        e^{-5/3}}{\mu_{\overline{{\text{MS}}}}^2} \right) \, \left( 1 + \am
      \beta_2 \ln \frac{{\bm q}'^2 \, e^{-5/3}}{\mu_{\overline{{\text{MS}}}}^2}
    \right)},
\end{align}
where we have also replaced all factors of $N_f$ by $- 6 \, \pi \,
\beta_2$.  Similarly, the contribution of the ``dressed'' diagram in
\fig{fig:NLO_all} B is obtained by iterating the quark bubbles from
\eq{K26} on both sides of the cut. The series of bubbles on each side
of the cut generates a geometric series. The zeroth-order term in this
series is the leading order JIMWLK kernel. The first-order term in the
series is given by $K_2^{\text{NLO}}$ in \eq{K29}. It is therefore
more convenient to write down the {\sl sum} of the LO JIMWLK kernel
and the contribution of the diagram in \fig{fig:NLO_all} B. The result
is
\begin{align}\label{K2dr}
  \am \, {\cal K}^{\text{LO}} ({\bm x}_0, {\bm x}_1 ; {\bm z}) + \am^2
  \, {\cal K}_{\otwo} ({\bm x}_0, {\bm x}_1 ;
  {\bm z}) \, = \, 4 \, \am \, \int \frac{d^2 q}{(2\pi)^2} \frac{d^2
    q'}{(2\pi)^2} \ e^{ -i {\bm q}\cdot ({\bm z}-{\bm x}_0) +i {\bm
      q}' \cdot ({\bm z}-{\bm x}_1) } \,
  \frac{{\bm q} \cdot {\bm q}'}{{\bm q}^2{\bm q}'^{2}} \notag \\
  \times \, \frac{1}{\left( 1 + \am \beta_2 \ln \frac{{\bm q}^2 \,
        e^{-5/3}}{\mu_{\overline{{\text{MS}}}}^2} \right) \, \left( 1 + \am
      \beta_2 \ln \frac{{\bm q}'^2 \, e^{-5/3}}{\mu_{\overline{{\text{MS}}}}^2}
    \right)}.
\end{align}
The contribution to the virtual part of the JIMWLK or BK kernels is
given by the diagrams in Figs. \ref{fig:NLO_all}C and
\ref{fig:NLO_inst}C$^\prime$. It can be easily extracted from Eqs.
(\ref{K1sdr}) and (\ref{K2dr}) using the condition (\ref{prob}) which
holds to all orders in quark bubbles. The resulting virtual kernel
would be equal to the real kernel as was explained above and shown in
Eqs.  (\ref{eq:JIMWLK-Hamiltonian-running-NLO}) and
(\ref{eqS_NLO_sub}). The dispersive method of \cite{Gardi:2006}, which
was used there to calculate Figs. \ref{fig:NLO_all}C and
\ref{fig:NLO_inst}C$^\prime$, can also be used to obtain \eq{K2dr},
and, by employing the probability conservation condition
(\ref{prob2}), to recover \eq{K1sdr} as well.

Before proceeding to evaluate the kernels in Eqs. (\ref{K1sdr}) and
(\ref{K2dr}) let us first analyze their dependence on the UV cutoff
$\mu_{\overline{{\text{MS}}}}$. This is instructive because the cutoff is
indeed a constant and therefore $\mu_{\overline{{\text{MS}}}}$-dependence does
not get modified by the Fourier transform. Hence the
$\mu_{\overline{{\text{MS}}}}$-dependence of the kernel is the same in
transverse coordinate and momentum spaces. Keeping only $\am$,
$\beta_2$ and $\mu_{\overline{{\text{MS}}}}$ we thus write
\begin{align}\label{K1simpl}
  \am^2 \, {\tilde {\cal K}}_{\oone} \, \propto \, \frac{-
    \am^2 \, \beta_2 \, \ln \mu_{\overline{{\text{MS}}}}^2}{ (1 - \am \beta_2
    \ln {\mu_{\overline{{\text{MS}}}}^2})^2}
\end{align}
and
\begin{align}\label{K2simpl}
  \am \, {\cal K}^{\text{LO}} + \am^2 \, {\cal K}_{\otwo} \,
  \propto \, \frac{\am}{ (1 - \am \beta_2 \ln
    {\mu_{\overline{{\text{MS}}}}^2})^2}.
\end{align}
From \eq{K2simpl} we immediately see that the sum of the LO kernel and
the diagram in \fig{fig:NLO_all}B does not give us a renormalizable
quantity, as it can not be expressed in terms of the renormalized
coupling constant. It lacks a power of $\am$ to give us a square of
the physical coupling $\as^2$. Now the need of extracting the UV
divergence from the graph in \fig{fig:NLO_all}A becomes manifest.
Adding the  ``subtraction'' term (\ref{K1simpl}) to
(\ref{K2simpl}) yields
\begin{align}\label{Ksimpl}
  \am \, {\cal K}^{\text{LO}} + \am^2 \, {\cal K}_{\otwo} +
  \am^2 \, {\tilde {\cal K}}_{\oone}\, \propto \, \frac{\am \,
    (1 - \am \beta_2 \ln {\mu_{\overline{{\text{MS}}}}^2})}{ (1 - \am \beta_2
    \ln {\mu_{\overline{{\text{MS}}}}^2})^2} \, \propto \, \frac{\as \,
    \as}{\as} ,
\end{align}
which is indeed renormalizable as it can be expressed in terms of the
physical coupling $\as$. (We have left a factor of $\as$ both in the
numerator and in the denominator of \eq{Ksimpl} on purpose to
underline the fact that the arguments of all three couplings, which we
did not keep, may be different, which would prohibit the
cancellation.)

The sum of the kernels in Eqs. (\ref{K1sdr}) and (\ref{K2dr}) gives
the all-order in $\as \, N_f$ (or $\as \, \beta_2$) contribution to
the running-coupling part of the real JIMWLK kernel:
\begin{align}\label{Kall1}
  \am \, & {\cal K}_{\text{rc}} ({\bm x}_0, {\bm x}_1 ; {\bm z}) \,
  \equiv \, \am \, {\cal K}^{\text{LO}} ({\bm x}_0, {\bm x}_1 ; {\bm
    z}) + \am^2 \, {\cal K}_{\bigcirc \hspace{-2.2mm} 2} ({\bm x}_0,
  {\bm x}_1 ; {\bm z})+ \am^2 \, {\tilde {\cal K}}_{\bigcirc
    \hspace{-2.2mm} 1} ({\bm x}_0, {\bm x}_1 ; {\bm z}) \, = \notag \\
  & = \, 4 \, \int \frac{d^2 q}{(2\pi)^2} \frac{d^2 q'}{(2\pi)^2} \ 
  e^{ -i {\bm q}\cdot ({\bm z}-{\bm x}_0) +i {\bm q}' \cdot ({\bm
      z}-{\bm x}_1) } \, \frac{{\bm q} \cdot {\bm q}'}{{\bm q}^2 \,
    {\bm q}'^{2}} \, \frac{ \am \, \left( 1 + \am \beta_2 \, \ln
      \frac{Q^2 \, e^{-5/3}}{\mu_{\overline{{\text{MS}}}}^2}\right)}{ \left( 1
      + \am \beta_2 \ln \frac{{\bm q}^2 \,
        e^{-5/3}}{\mu_{\overline{{\text{MS}}}}^2} \right) \, \left( 1 + \am
      \beta_2 \ln \frac{{\bm q}'^2 \, e^{-5/3}}{\mu_{\overline{{\text{MS}}}}^2}
    \right)}
\end{align}
where we have defined a momentum scale $Q$ by
\begin{align}\label{Q}
  \ln \frac{Q^2 \, e^{-5/3}}{\mu_{\overline{{\text{MS}}}}^2} \, \equiv \,
  \frac{{\bm q}^{2} \, \ln \frac{{\bm q}^2 \,
      e^{-5/3}}{\mu_{\overline{{\text{MS}}}}^2} - {\bm q}'^{2} \, \ln
    \frac{{\bm q}'^2 \, e^{-5/3}}{\mu_{\overline{{\text{MS}}}}^2} }{{\bm q}^{2}
    - {\bm q}'^{2}} - \frac{{\bm q}^2 \, {\bm q}'^{2}}{{\bm q} \cdot
    {\bm q}'} \, \frac{\ln ({\bm q}^{2}/{\bm q}'^{2})}{{\bm q}^{2} -
    {\bm q}'^{2}}.
\end{align}
(Indeed $Q^2$ is independent of $\mu_{\overline{{\text{MS}}}}$ as can be easily
seen from \eq{Q}.) As the one-loop running coupling constant is
defined in the $\overline{{\text{MS}}}$ scheme by
\begin{align}
  \as (Q^2) \, = \, \frac{\am}{ 1 + \am \beta_2 \, \ln
    \frac{Q^2}{\mu_{\overline{{\text{MS}}}}^2}}
\end{align}
\eq{Kall1} can be rewritten as
\begin{align}\label{Kall}
  \am \, & {\cal K}_{\text{rc}} ({\bm x}_0, {\bm x}_1 ; {\bm z}) \,
  \equiv \, \am \, {\cal K}^{\text{LO}} ({\bm x}_0, {\bm x}_1 ; {\bm
    z}) + \am^2 \, {\cal K}_{\bigcirc \hspace{-2.2mm} 2} ({\bm x}_0,
  {\bm x}_1 ; {\bm z})+ \am^2 \, {\tilde {\cal K}}_{\bigcirc
    \hspace{-2.2mm} 1} ({\bm x}_0, {\bm x}_1 ; {\bm z}) \, = \notag \\
  & = \, 4 \, \int \frac{d^2 q}{(2\pi)^2} \frac{d^2 q'}{(2\pi)^2} \ 
  e^{ -i {\bm q}\cdot ({\bm z}-{\bm x}_0) +i {\bm q}' \cdot ({\bm
      z}-{\bm x}_1) } \, \frac{{\bm q} \cdot {\bm q}'}{{\bm q}^2 \,
    {\bm q}'^{2}} \, \frac{ \as
    \left({\bm q}^2 \, e^{-5/3} \right) \, \as \left({\bm q}'^2 \,
      e^{-5/3} \right)}{ \as \left(Q^2 \, e^{-5/3} \right)}
\ .
\end{align}
\eq{Kall} is {\sl the first of the two main results of our paper}. It
gives the JIMWLK kernel with the running coupling corrections included
in transverse momentum space. Remarkably, the corrections come in as a
``triumvirate'' of the couplings,\footnote{We note that a similar
  structure containing three coupling constants has been obtained
  independently by Balitsky~\cite{Bal:2006}. Our difference
  from~\cite{Bal:2006} appears to be due to %(i) 
  a different choice of the ``subtraction point''.
%and (ii) a different interpretation of
%  what constitutes a running coupling correction: here the running
%  coupling corrections multiply the LO JIMWLK kernel, while
%  in~\cite{Bal:2006} they are defined as multiplying the LO BK kernel.
} instead of a single coupling constant with some momentum scale!
Despite the surprising form this result is in full agreement with the
expressions found in\cite{Gardi:2006}. To facilitate comparison, we
give a detailed translation in
appendix.~\ref{sec:comparison-with-dispersive}

This provides for an interesting mechanism to reduce the above result to a
simpler underlying structure expected for the purely virtual contributions of
diagrams Fig.~\ref{fig:NLO_all} C and~\ref{fig:NLO_inst} C': There the $z$
integral may be performed and, in the absence of interaction with the target,
this sets ${\bm q}'={\bm q}$.  Then the {\em sum} of the integrands of
transverse and longitudinal contributions reduces to
\begin{align}
  \label{eq:mom-sum-trans-long-virt}
  \am\, \frac{\frac1{\bm q^2}\left\{1 
     +\am \beta_2\, \left[
       1+\ln\left(\frac{{\bm q}^2\,e^{-\frac53}}{\mu^2_{\overline{{\text{MS}}}}} 
\right)\right]\right\} 
 %    \right]
  -\frac1{\bm q^2} \am \beta_2
   }{
    \left(1+\beta_2\am\,
   \ln\left(\frac{{\bm q}^2\,e^{-\frac53}}{\mu^2_{\overline{{\text{MS}}}}}
    \right)\right)^2
   } 
=\frac1{\bm q^2}\ \alpha_s\left({\bm q}^2\,e^{-\frac53}\right)
\end{align}
which clearly corresponds to the exchange of a dressed noninteracting gluon.
This is the counterpart of a cancellation that appears in~\cite{Gardi:2006}
under the same premise. Note that due to the virtual nature of the diagrams in
Figs.~\ref{fig:NLO_all} C and~\ref{fig:NLO_inst} C' the running coupling
corrections can enter only with the sole available transverse momentum scale
${\bm q}^2$ in its argument, as shown in \eq{eq:mom-sum-trans-long-virt}.
Therefore, while different subtraction procedures may yield different
expressions for the scale $Q$, as compared to \eq{Q}, all of these
alternatives must lead to expressions for $Q^2$ that approach ${\bm q}^2$ in
the limit when ${\bm q}'={\bm q}$. This would reduce the ``triumvirate'' of
couplings from \eq{Kall} to the single coupling shown in
\eq{eq:mom-sum-trans-long-virt}, as expected for virtual diagrams here.

 The most intriguing property, however, is that the
``triumvirate'' structure of~(\ref{Kall}) solves the puzzle of how to
successfully perform a BLM scale setting: In~\cite{Gardi:2006} it was observed
that an attempt to perform a BLM scale setting in a perturbative formulation
with a single Borel parameter (corresponding to an approximation in terms of a
single geometric series in our present language) would not lead to a successful
resummation of the dominant contribution. From the present perspective this
would correspond to the attempt to use the joint leading order expansion of
all three couplings in~(\ref{Kall}) to determine a {\em single} scale (instead
of the three separate ones of~(\ref{Kall})) that would give a good
approximation to the triumvirate in terms of a {\em single} geometric series.
Any such attempt would necessarily entail an artificial all orders iteration
of the ``numerator logarithms'' encoded in the ``denominator coupling''
$\alpha_s(Q^2 e^{-\frac53})$ that is clearly absent in the underlying
expression.  This is the source of the spurious and divergent higher inverse
powers of ${\bm q}\cdot{\bm q}'$ or $(\bm x_0-\bm z)\cdot(\bm x_1-\bm z)$
encountered in a naive attempt of deriving a BLM approximation
in~\cite{Gardi:2006}.  The ``triumvirate'' structure will allow for a
successful and transparent BLM approximation of the full coordinate result
given in~\cite{Gardi:2006}.

\eq{Kall} allows one to find the
corresponding BK evolution kernel with the running coupling
corrections included by using
\begin{align}\label{KrcBK}
  K_{\text{rc}} ({\bm x}_0, {\bm x}_1 ; {\bm z}) \, = \, C_F \,
  \sum_{m,n = 0}^1 \, (-1)^{m+n} \, {\cal K}_\text{rc} & ({\bm x}_m,
  {\bm x}_n ; {\bm z}).
\end{align}

To determine the scales of the three physical couplings in \eq{Kall} in
transverse coordinate space we have to evaluate the integrations in the
kernels from Eqs.  (\ref{K1sdr}) and (\ref{K2dr}). In transverse momentum
space each chain of bubbles generates a geometric series, as shown in
\eq{Kall1}.  However, a Fourier-transform of these series is dangerous for
several reasons.  First and foremost the integrals over ${\bm q}$ and ${\bm
  q}'$ in Eqs.  (\ref{K1sdr}) and (\ref{K2dr}) include contributions from the
Landau poles leading to power corrections which are not under perturbative
control. The uncertainties due to power corrections are estimated in
\cite{Gardi:2006} using renormalon techniques
\cite{Beneke:1998ui,Mueller:1992xz}. Our strategy here is to ignore these
contributions concentrating on setting the scale of the running coupling in
transverse coordinate space. Even then our goal is difficult, since, even
though leading powers of $\ln \mu_{\overline{{\text{MS}}}}^2$ terms generate a
geometric series in transverse coordinate space just like in the momentum
space, it is not clear whether the transverse coordinate scale under the
logarithm stays the same in all the terms in the series.  In that sense,
setting the running coupling scale in transverse coordinate space will only be
an approximation of the more exact \eq{Kall} in the sense of a BLM scale
setting.

To study running coupling corrections in transverse coordinate space
we begin by evaluating the Fourier transforms in \eq{K2dr}. Similar to
Appendix \ref{K1FT} we perform the angular integrals first to write
\begin{align}\label{K2dr2}
  \am \, {\cal K}^{\text{LO}} ({\bm x}_0, {\bm x}_1 ; {\bm z}) + \am^2
  \, {\cal K}_{\otwo} ({\bm x}_0, {\bm x}_1 ; {\bm z}) \, = \,
  \frac{\am}{\pi^2} \, \frac{{\bm z}-{\bm x}_0}{|{\bm z}-{\bm x}_0|}
  \cdot \frac{{\bm z}-{\bm x}_1}{|{\bm z}-{\bm x}_1|} \notag \\
  \times \, \int\limits_0^\infty dq \, dq' \, J_1 (q \, |{\bm z}-{\bm
    x}_0|) \, J_1 (q' \, |{\bm z}-{\bm x}_1|) \, \frac{1}{\left( 1 +
      \am \beta_2 \ln \frac{q^2 \, e^{-5/3}}{\mu_{\overline{{\text{MS}}}}^2}
    \right) \, \left( 1 + \am \beta_2 \ln \frac{q'^2 \,
        e^{-5/3}}{\mu_{\overline{{\text{MS}}}}^2} \right)}
\end{align}
with $q = |{\bm q}|$ and $q' = |{\bm q}'|$.  Since our goal is to find
the scale of the strong coupling constant ignoring the power
corrections we can expand the denominators of \eq{K2dr2} into
geometric series obtaining
\begin{align}\label{K2dr3}
  \am \, {\cal K}^{\text{LO}} ({\bm x}_0, {\bm x}_1 ; {\bm z}) + \am^2
  \, {\cal K}_{\otwo} ({\bm x}_0, {\bm x}_1 ; {\bm z}) \, = \,
  \frac{\am}{\pi^2} \, \frac{{\bm z}-{\bm x}_0}{|{\bm z}-{\bm x}_0|}
  \cdot \frac{{\bm z}-{\bm x}_1}{|{\bm z}-{\bm x}_1|} \,
  \sum\limits_{n,m=0}^\infty \, (- \am \beta_2)^{n+m} \notag \\
  \times \, \int\limits_0^\infty dq \, dq' \, J_1 (q \, |{\bm z}-{\bm
    x}_0|) \, J_1 (q' \, |{\bm z}-{\bm x}_1|) \ln^n \frac{q^2 \,
    e^{-5/3}}{\mu_{\overline{{\text{MS}}}}^2} \, \ln^m \frac{q'^2 \,
    e^{-5/3}}{\mu_{\overline{{\text{MS}}}}^2}.
\end{align}
Rewriting the powers of the logarithms in \eq{K2dr3} in terms of
derivatives yields
\begin{align}\label{K2dr4}
  \am \, {\cal K}^{\text{LO}} ({\bm x}_0, {\bm x}_1 ; {\bm z}) + \am^2
  \, {\cal K}_{\otwo} ({\bm x}_0, {\bm x}_1 ; {\bm z}) \, = \,
  \frac{\am}{\pi^2} \, \frac{{\bm z}-{\bm x}_0}{|{\bm z}-{\bm x}_0|}
  \cdot \frac{{\bm z}-{\bm x}_1}{|{\bm z}-{\bm x}_1|} \,
  \sum\limits_{n,m=0}^\infty \, (- \am \beta_2)^{n+m} \notag \\
  \times \, \frac{d^n}{d \lambda^n} \, \frac{d^m}{d \lambda'^m} \,
  \int\limits_0^\infty dq \, dq' \, J_1 (q \, |{\bm z}-{\bm x}_0|) \,
  J_1 (q' \, |{\bm z}-{\bm x}_1|) \left( \frac{q^2 \,
      e^{-5/3}}{\mu_{\overline{{\text{MS}}}}^2} \right)^\lambda \, \left(
    \frac{q'^2 \, e^{-5/3}}{\mu_{\overline{{\text{MS}}}}^2} \right)^{\lambda'}
  \Bigg|_{\lambda, \lambda' =0}.
\end{align}
Performing the $q$- and $q'$-integrals gives
\begin{align}\label{K2dr5}
  \am \, {\cal K}^{\text{LO}} ({\bm x}_0, {\bm x}_1 ; {\bm z}) + \am^2
  \, {\cal K}_{\otwo} ({\bm x}_0, {\bm x}_1 ; {\bm z}) \, = &
  \, \am \, {\cal K}^{\text{LO}} ({\bm x}_0, {\bm x}_1 ; {\bm z}) \,
  \sum\limits_{n,m=0}^\infty \, (- \am \beta_2)^{n+m} \notag \\
  \times \left\{ \frac{d^n}{d \lambda^n} \left[ \left( \frac{4 \,
          e^{-5/3}}{|{\bm z}-{\bm x}_0|^2 \, \mu_{\overline{{\text{MS}}}}^2}
      \right)^\lambda \frac{\Gamma (1+\lambda)}{\Gamma (1 - \lambda)}
    \right] \right\} \Bigg|_{\lambda =0} & \left\{ \frac{d^m}{d
      \lambda^m} \left[ \left( \frac{4 \, e^{-5/3}}{|{\bm z}-{\bm
            x}_1|^2 \, \mu_{\overline{{\text{MS}}}}^2} \right)^{\lambda'}
      \frac{\Gamma (1+\lambda')}{\Gamma (1 - \lambda')} \right]
  \right\} \Bigg|_{\lambda' =0}.
\end{align}
Differentiating with respect to $\lambda$ and $\lambda'$ we write out
the first few terms in the resulting series
\begin{align}\label{K2dr6}
  \am \, {\cal K}^{\text{LO}} ({\bm x}_0, {\bm x}_1 ; {\bm z}) + \am^2
  \, {\cal K}_{\otwo} ({\bm x}_0, {\bm x}_1 ; {\bm z}) \, = &
  \,
  \am \, {\cal K}^{\text{LO}} ({\bm x}_0, {\bm x}_1 ; {\bm z}) \notag \\
  \times \, \Bigg\{ 1 - \am \beta_2 \, \ln \left( \frac{4 \, e^{-5/3 -
        2 \gamma}}{|{\bm z}-{\bm x}_0|^2 \, \mu_{\overline{{\text{MS}}}}^2}
  \right) + & (\am \beta_2 )^2 \, \ln^2 \left( \frac{4 \, e^{-5/3 - 2
        \gamma}}{|{\bm z}-{\bm x}_0|^2 \, \mu_{\overline{{\text{MS}}}}^2} \right) \notag \\
  - (\am \beta_2 )^3 \, & \left[ \ln^3 \left( \frac{4 \, e^{-5/3 - 2
          \gamma}}{|{\bm z}-{\bm x}_0|^2 \, \mu_{\overline{{\text{MS}}}}^2}
    \right) - 4 \, \zeta (3) \right] + \ldots \Bigg\} \, \notag \\
  \times \, \Bigg\{ 1 - \am \beta_2 \, \ln \left( \frac{4 \, e^{-5/3 -
        2 \gamma}}{|{\bm z}-{\bm x}_1|^2 \, \mu_{\overline{{\text{MS}}}}^2}
  \right) + & (\am \beta_2 )^2 \, \ln^2 \left( \frac{4 \, e^{-5/3 - 2
        \gamma}}{|{\bm z}-{\bm x}_1|^2 \, \mu_{\overline{{\text{MS}}}}^2} \right) \notag \\
  - (\am \beta_2 )^3 \, & \left[ \ln^3 \left( \frac{4 \, e^{-5/3 - 2
          \gamma}}{|{\bm z}-{\bm x}_1|^2 \, \mu_{\overline{{\text{MS}}}}^2}
    \right) - 4 \, \zeta (3) \right] + \ldots \Bigg\}.
\end{align}
One can see that the geometric series structure appears to hold up to
the cubic terms in either one of the logarithms. In the sense of a
BLM-type approach \cite{BLM} we approximate the expressions in each of
the curly brackets by a geometric series, obtaining
\begin{align}\label{K2dr7}
  \am \, {\cal K}^{\text{LO}} ({\bm x}_0, {\bm x}_1 ; {\bm z}) + &
  \am^2 \, {\cal K}_{\otwo} ({\bm x}_0, {\bm x}_1 ; {\bm z}) \,
  \approx \,
  {\cal K}^{\text{LO}} ({\bm x}_0, {\bm x}_1 ; {\bm z}) \notag \\
  \times & \, \frac{\am}{ \left[ 1 + \am \beta_2 \, \ln \left( \frac{4 \,
        e^{-5/3 - 2 \gamma}}{|{\bm z}-{\bm x}_0|^2 \,
        \mu_{\overline{{\text{MS}}}}^2} \right) \right] \, \left[ 1 + \am
    \beta_2 \, \ln \left( \frac{4 \, e^{-5/3 - 2 \gamma}}{|{\bm
          z}-{\bm x}_1|^2 \, \mu_{\overline{{\text{MS}}}}^2} \right) \right] }.
\end{align}

Evaluation of the Fourier transforms in \eq{K1sdr} is performed along
similar lines in Appendix \ref{K1dressed}. The result reads
\begin{align}\label{K1sdr1}
  \am^2 \, {\tilde {\cal K}}_{\oone} ({\bm x}_0, {\bm x}_1 ; {\bm z})
  \, \approx \, {\cal K}^{\text{LO}} ({\bm x}_0, {\bm x}_1 ; {\bm z})
  \, \am \beta_2 \, \ln \left( \frac{4 \, e^{-\frac{5}{3} -2
        \gamma}}{R^2 ({\bm x}_0, {\bm
        x}_1 ; {\bm z}) \, \mu_{\overline{{\text{MS}}}}^2} \right) \notag \\
  \times \, \frac{\am}{ \left[ 1 + \am \beta_2 \, \ln \left( \frac{4
          \, e^{-5/3 - 2 \gamma}}{|{\bm z}-{\bm x}_0|^2 \,
          \mu_{\overline{{\text{MS}}}}^2} \right) \right] \, \left[ 1
      + \am \beta_2 \, \ln \left( \frac{4 \, e^{-5/3 - 2
            \gamma}}{|{\bm z}-{\bm x}_1|^2 \,
          \mu_{\overline{{\text{MS}}}}^2} \right) \right] },
\end{align}
with the scale $R$ given by \eq{Rexp}.

Finally, adding Eqs. (\ref{K2dr7}) and (\ref{K1sdr1}) yields
\begin{align}\label{Krc}
  \am \, & {\cal K}_{\text{rc}} ({\bm x}_0, {\bm x}_1 ; {\bm z}) \,
  \approx \, {\cal K}^{\text{LO}} ({\bm x}_0, {\bm x}_1 ; {\bm z}) \,
  \frac{\as \left(\frac{4 \, e^{-5/3 - 2 \gamma}}{|{\bm z}-{\bm
          x}_0|^2}\right) \ \as \left( \frac{4 \, e^{-5/3 - 2
          \gamma}}{|{\bm z}-{\bm x}_1|^2} \right)}{ \as \left( \frac{4
        \, e^{-5/3 - 2 \gamma}}{R^2 ({\bm x}_0, {\bm x}_1 ; {\bm z})}
    \right)}
\end{align}
with the scale $R ({\bm x}_0, {\bm x}_1 ; {\bm z})$ given by \eq{Rexp}
and the couplings calculated in the $\overline{{\text{MS}}}$ scheme.
\eq{Krc} is {\sl the second of the two main results of our paper}. It
gives the JIMWLK kernel with the running coupling constant.  It is
very interesting that the running coupling corrections come in {\sl
  not} through a scale of a single running coupling $\as$ as one would
naively expect, but in the form of a {\sl ``triumvirate''} of the
running couplings shown in \eq{Krc}!  \eq{Krc} can be used to
construct BK kernel with the running coupling constant by employing
\eq{KrcBK}.

%%%%%%%%%%%%%%%%%%%%%%%%%%%%%%%%%%%%%%%%%%%%%%%%%%%%%%%%%%%%%%%%%%%%%%%%%%%%%%%%%

\section{Conclusions}
\label{conc}

To conclude let us state the main results of this work once again. By tracking
the powers of $\as \, N_f$ we have included running coupling corrections into
the JIMWLK and BK evolution equations.  Our all orders result agrees with the
expressions derived with the dispersive method in~\cite{Gardi:2006}, but
renders the result in terms of a ``triumvirate'' of couplings already in the
momentum space expressions~(\ref{Kall}). This allows us to give a concise,
accurate BLM approximation of the perturbative sum in coordinate space in terms
of a corresponding coordinate space ``triumvirate'' shown in \eq{Krc}.
% Interestingly enough the running coupling enters those equations in 
% the form of a ``triumvirate'' shown in \eq{Krc}. 
Our procedure includes one uncertainty, related to choosing the ``subtraction
point'', as was discussed in Sect.~\ref{Subtr}. This ambiguity is akin to
scheme-dependence of the running coupling constant and we believe that the
final result does not depend on the choice of the ``subtraction point'' in a
very crucial way. We picked the subtraction point to be at the transverse
coordinate of the virtual gluon in \fig{fig:NLO1}A.

Our result for the JIMWLK Hamiltonian with the running coupling
constant is 
\begin{align}\label{JIMWLKrc}
  {\cal H}^{\text{rc}} =\frac{1}{2} \, \frac{\as \left(\frac{4 \,
        e^{-5/3 - 2 \gamma}}{|{\bm z}-{\bm x}|^2}\right) \, \as \left(
      \frac{4 \, e^{-5/3 - 2 \gamma}}{|{\bm z}-{\bm y}|^2} \right)}{
    \as \left( \frac{4 \, e^{-5/3 - 2 \gamma}}{R^2 ({\bm x}, {\bm y};
        {\bm z})} \right)} \, {\cal K}^{\text{LO}}_{{\bm x}, {\bm y};
    {\bm z}} \left[ U_{\bm z}^{a b}(i\Bar\nabla^a_{\bm
      x}i\nabla^b_{\bm y} +i\nabla^a_{\bm x} i\Bar\nabla^b_{\bm y}) +
    ( i\nabla^a_{\bm x} i\nabla^a_{\bm y}+i\Bar\nabla^a_{\bm x}
    i\Bar\nabla^a_{\bm y}) \right]
\end{align}
with integrations over $\bm x$, $\bm y$ and $\bm z$ implied.  All the
running couplings should be calculated in the $\overline{{\text{MS}}}$ scheme.

To obtain the BK evolution equation with the running coupling constant
one needs to sum the kernel in \eq{Krc} over all possible connections
of the gluon to the quark and the anti-quark lines, as formally shown
in \eq{KrcBK}. (Alternatively to derive the BK equation one can apply
the JIMWLK Hamiltonian from \eq{JIMWLKrc} to a correlator of two
Wilson lines and take the large-$N_c$ limit.) A straightforward
calculation shows that
\begin{align}
  \lim_{{\bm x}_1 \rightarrow {\bm x}_0} R^2 ({\bm x}_0, {\bm x}_1;
  {\bm z}) \, = \, |{\bm z}-{\bm x}_0|^2 \hspace{1cm} \text{and}
  \hspace{1cm} \lim_{{\bm x}_0 \rightarrow {\bm x}_1} R^2 ({\bm x}_0,
  {\bm x}_1; {\bm z}) \, = \, |{\bm z}-{\bm x}_1|^2,
\end{align}
such that the BK evolution equation with the running coupling
corrections is
\begin{align}\label{eqNrc}
  &\frac{\partial N ({\bm x}_{0}, {\bm x}_1, Y)}{\partial Y} \, = \,
  \frac{C_F}{\pi^2} \, \int d^2 x_2 \notag \\ &\times \, \left[ \as
    \left(\frac{4 \, e^{-5/3 - 2 \gamma}}{x_{20}^2}\right) \,
    \frac{1}{x_{20}^2} - 2 \, \frac{\as \left(\frac{4 \, e^{-5/3 - 2
            \gamma}}{x_{20}^2}\right) \ \as \left( \frac{4 \, e^{-5/3
            - 2 \gamma}}{x_{21}^2} \right)}{ \as \left( \frac{4 \,
          e^{-5/3 - 2 \gamma}}{R^2 ({\bm x}_0, {\bm x}_1; \, {\bm
            x}_2)} \right)} \, \frac{{\bm x}_{20} \cdot {\bm
        x}_{21}}{x_{20}^2 \, x_{21}^2} + \as \left(\frac{4 \, e^{-5/3
          - 2 \gamma}}{x_{21}^2}\right) \, \frac{1}{x_{21}^2}
  \right]  \notag \\
  &\times \, \left[ N ({\bm x}_{0}, {\bm x}_2, Y) + N ({\bm x}_{2},
    {\bm x}_1, Y) - N ({\bm x}_{0}, {\bm x}_1, Y) - N ({\bm x}_{0},
    {\bm x}_2, Y) \, N ({\bm x}_{2}, {\bm x}_1, Y) \right],
\end{align}
where $R^2$ is given by \eq{Rexp}.

It is interesting to explore the limits of \eq{eqNrc}. We will refer
to the original dipole $01$ as the ``parent'' dipole, while the
dipoles $20$ and $21$ generated in one step of the evolution will be
called ``daughter'' dipoles. First of all, when both produced dipoles
are comparable and much larger than the ``parent'' dipole, $x_{20}
\sim x_{21} \gg x_{01}$, the argument of the coupling constant in all
three terms in \eq{eqNrc} would be given by the ``daughter'' dipole
sizes $x_{20} \sim x_{21}$. However, such large dipole sizes should be
cut off by the inverse saturation scale $1/Q_s$, which implies that
the scale for the coupling constant in the IR region of phase space
would be given by $Q_s$ keeping the coupling small and the physics
perturbative. In the other interesting limit when one of the
``daughter'' dipoles is much smaller than the other one, $x_{20} \ll
x_{21} \sim x_{01}$, a simple calculation shows that $R^2 ({\bm x}_0,
{\bm x}_1; \, {\bm x}_2) \approx x_{20}^2$ and the BK kernel in
\eq{eqNrc} becomes
\begin{align}\label{dla}
  \as \left(\frac{4 \, e^{-5/3 - 2 \gamma}}{x_{20}^2}\right) \,
    \frac{1}{x_{20}^2} - 2 \, \as \left( \frac{4 \, e^{-5/3 - 2
          \gamma}}{x_{21}^2} \right) \, \frac{{\bm x}_{20} \cdot {\bm
        x}_{21}}{x_{20}^2 \, x_{21}^2} + \as \left(\frac{4 \, e^{-5/3
          - 2 \gamma}}{x_{21}^2}\right) \, \frac{1}{x_{21}^2}.
\end{align}
Two out of three terms in the kernel have the scale of the coupling
given by the larger dipole size, naively making the evolution ``less
perturbative''. However, in the $x_{20} \ll x_{21}$ limit it is the
first term which dominates \eq{dla}: that term has the running
coupling scale given by the size of the {\sl smaller} dipole, making
the physics perturbative!

%%%%%%%%%%%%%%%%%%%%%%%%%%%%%%%%%%%%%%%%%%%%%%%%%%%%%%%%%%%%%%%%%%%%%%%%%%%%%%%

\section*{Acknowledgments} 

We are greatly indebted to Ian Balitsky for a very useful exchange of
ideas when this work was in progress.  We would like to thank Javier
Albacete, Eric Braaten, Ulrich Heinz, and Robert Perry for many
informative discussions. This work is supported in part by the U.S.
Department of Energy under Grant No.  DE-FG02-05ER41377.

%%%%%%%%%%%%%%%%%%%%%%%%%%%%%%%%%%%%%%%%%%%%%%%%%%%%%%%%%%%%%%%%%%%%%%%%%%%%%%%

\appendix

\renewcommand{\theequation}{A\arabic{equation}}
  \setcounter{equation}{0}
\section{Evaluating the Fourier transforms in \eq{K1int4}}
\label{K1FT}

Here we first calculate the following integral coming from the first
(transverse) term in \eq{K1int4}
\begin{align}\label{A1}
  I_T ({\bm z}, {\bm z}') \, = \, \int \frac{d^2 q}{(2\pi)^2} \frac{d^2
    q'}{(2\pi)^2} \ e^{ -i {\bm q}\cdot {\bm z} +i {\bm q}' \cdot {\bm
      z}'} \, \frac{{\bm q} \cdot {\bm q}'}{{\bm q}^2{\bm q}'^{2}} \,
  \frac{{\bm q}^{2} \, \left( \ln \frac{{\bm
          q}^2}{\mu_{\overline{{\text{MS}}}}^2} - \frac{5}{3} \right) - {\bm
      q}'^{2} \, \left( \ln \frac{{\bm q}'^2}{\mu_{\overline{{\text{MS}}}}^2} -
      \frac{5}{3} \right) }{{\bm q}^{2} - {\bm q}'^{2}},
\end{align}
where ${\bm z}$ and ${\bm z}'$ are some transverse coordinate vectors.
Replacing ${\bm q} \rightarrow i \, {\bm \partial}_z$ and ${\bm q}'
\rightarrow - i \, {\bm \partial}_{z'}$ in the numerator of the first
ratio in the integrand we can integrate over the angles of ${\bm q}$
and ${\bm q}'$ obtaining
\begin{align}\label{A2}
  I_T ({\bm z}, {\bm z}') \, = \, \frac{1}{(2\pi)^2} \, {\bm
    \partial}_z \cdot {\bm \partial}_{z'} \, \int\limits_0^\infty
  \frac{d q}{q} \, \frac{dq'}{q'} \, J_0 (q \, z) \, J_0 (q' \, z') \,
  \frac{{\bm q}^{2} \, \left( \ln \frac{{\bm
          q}^2}{\mu_{\overline{{\text{MS}}}}^2} - \frac{5}{3} \right)
    - {\bm q}'^{2} \, \left( \ln \frac{{\bm
          q}'^2}{\mu_{\overline{{\text{MS}}}}^2} - \frac{5}{3} \right)
  }{{\bm q}^{2} - {\bm q}'^{2}}
\end{align}
with $q = |{\bm q}|$, $q' = |{\bm q}'|$, $z = |{\bm z}|$ and $z' =
|{\bm z}'|$. Bringing the transverse gradients back into the
integrand yields
\begin{align}\label{A3}
  I_T ({\bm z}, {\bm z}') \, = \, \frac{1}{(2\pi)^2} \, \frac{{\bm z}
    \cdot {\bm z}'}{z \, z'} \, \int\limits_0^\infty dq \, dq' \, J_1
  (q \, z) \, J_1 (q' \, z') \, \frac{{\bm q}^{2} \, \ln \frac{{\bm
        q}^2 \, e^{-5/3}}{\mu_{\overline{{\text{MS}}}}^2} - {\bm q}'^{2} \, \ln
    \frac{{\bm q}'^2 \, e^{-5/3}}{\mu_{\overline{{\text{MS}}}}^2} }{{\bm q}^{2}
    - {\bm q}'^{2}}.
\end{align}
To perform $q$ and $q'$ integrations we first rewrite the fraction in
the integrand of (\ref{A3}) as
\begin{align}\label{intgrnd}
  \frac{{\bm q}^{2} \, \ln \frac{{\bm q}^2 \,
      e^{-5/3}}{\mu_{\overline{{\text{MS}}}}^2} - {\bm q}'^{2} \, \ln
    \frac{{\bm q}'^2 \, e^{-5/3}}{\mu_{\overline{{\text{MS}}}}^2} }{{\bm q}^{2}
    - {\bm q}'^{2}} \, = \, \ln \frac{{\bm q}^2 \,
    e^{-5/3}}{\mu_{\overline{{\text{MS}}}}^2} + {\bm q}'^{2} \, \frac{\ln
    \frac{{\bm q}^{2}}{{\bm q}'^{2}}}{{\bm q}^{2} - {\bm q}'^{2}}
\end{align}
and then use the following integral identity
\begin{align}\label{trick}
  \int\limits_0^1 d \beta \, \frac{1}{{\bm q}^{2} (1-\beta) +
    {\bm q}'^{2} \beta} \, = \, \frac{\ln
    \frac{{\bm q}^{2}}{{\bm q}'^{2}}}{{\bm q}^{2} - {\bm q}'^{2}}
\end{align}
to replace the last term in \eq{intgrnd}.\footnote{We thank Ian
  Balitsky for pointing out to us the usefulness of this
  substitution.} \eq{A3} becomes
\begin{align}\label{A4}
  I_T ({\bm z}, {\bm z}') \, = \, \frac{1}{(2\pi)^2} \, \frac{{\bm z}
    \cdot {\bm z}'}{z \, z'} \, \int\limits_0^\infty dq \, dq' \, J_1
  (q \, z) \, J_1 (q' \, z') \, \left\{ \ln \frac{{\bm q}^2 \,
      e^{-5/3}}{\mu_{\overline{{\text{MS}}}}^2} + {\bm q}'^{2} \,
    \int\limits_0^1 d \beta \, \frac{1}{{\bm q}^{2} (1-\beta) + {\bm
        q}'^{2} \beta} \right\}.
\end{align}
Now the $q$ and $q'$ integrations can be performed using standard
formulas for the integrals of Bessel functions yielding
\begin{align}\label{A5}
  I_T ({\bm z}, {\bm z}') \, = \, \frac{1}{(2\pi)^2} \, \frac{{\bm z}
    \cdot {\bm z}'}{{\bm z}^2 \, {\bm z}'^2} \, \left\{ \ln \frac{4 \,
      e^{- \frac{5}{3} -2 \gamma}}{{\bm z}^2 \, \mu_{\overline{{\text{MS}}}}^2}
    + {\bm z}^{2} \, \int\limits_0^1 d \beta \, \frac{1}{{\bm z}^{2}
      (1-\beta) + {\bm z}'^{2} \beta} \right\}.
\end{align}
Finally the $\beta$-integral in \eq{A5} can be done using \eq{trick}
giving
\begin{align}\label{A6}
  I_T ({\bm z}, {\bm z}') \, = \, \frac{1}{(2\pi)^2} \, \frac{{\bm z}
    \cdot {\bm z}'}{{\bm z}^2 \, {\bm z}'^2} \, \frac{{\bm z}^2 \, \ln
    \frac{4 \, e^{- \frac{5}{3} -2 \gamma}}{{\bm z}'^2 \,
      \mu_{\overline{{\text{MS}}}}^2} - {\bm z}'^2 \, \ln \frac{4 \, e^{-
        \frac{5}{3} -2 \gamma}}{{\bm z}^2 \,
      \mu_{\overline{{\text{MS}}}}^2}}{{\bm z}^2 - {\bm z}'^2}.
\end{align}
Now let us use the same technique to perform Fourier transforms
in the second (longitudinal) term in \eq{K1int4}. We want to evaluate
\begin{align}\label{A21}
  I_L ({\bm z}, {\bm z}') \, = \, \int \frac{d^2 q}{(2\pi)^2}
  \frac{d^2 q'}{(2\pi)^2} \ e^{ -i {\bm q}\cdot {\bm z} +i {\bm q}'
    \cdot {\bm z}'} \, \frac{\ln ({\bm q}^{2}/{\bm q}'^{2})}{{\bm
      q}^{2} - {\bm q}'^{2}}.
\end{align}
First we integrate over the angles of $\bm q$ and ${\bm q}'$ 
\begin{align}\label{A22}
  I_L ({\bm z}, {\bm z}') \, = \, \frac{1}{(2\pi)^2} \,
  \int\limits_0^\infty d q \, d q' \, q \, q' \, J_0 (q \, z) \, J_0
  (q' \, z') \, \frac{\ln ({\bm q}^{2}/{\bm q}'^{2})}{{\bm q}^{2} -
    {\bm q}'^{2}}.
\end{align}
Now we can use \eq{trick} to integrate over $q$ and $q'$ obtaining
\begin{align}\label{A23}
  I_L ({\bm z}, {\bm z}') \, = \, \frac{1}{(2\pi)^2} \,
  \int\limits_0^1 \, d \beta \, \frac{1}{z^2 \beta + z'^2 (1-\beta)},
\end{align}
which, using \eq{trick} again we can write as
\begin{align}\label{A24}
  I_L ({\bm z}, {\bm z}') \, = \, \frac{1}{(2\pi)^2} \, \frac{\ln (z^2
    / z'^2)}{z^2 - z'^2}.
\end{align}

%%%%%%%%%%%%%%%%%%%%%%%%%%%%%%%%%%%%%%%%%%%%%%%%%%%%%%%%%%%%%%%%%%%%%%%%%%%%%%%

\renewcommand{\theequation}{B\arabic{equation}}
  \setcounter{equation}{0}
\section{Evaluating the Fourier transforms in~\eq{K1sdr}}
\label{K1dressed}

Here we will try to perform the Fourier transforms in \eq{K1sdr}. We
begin by analyzing the transverse part of the kernel, given by the
first term in the curly brackets in \eq{K1sdr}.  We start by
performing the angular integrations over the angles of $\bm q$ and
${\bm q}'$, which, similar to the way we arrived at \eq{A3}, yield
\begin{align}\label{B1}
  \am^2 \, {\tilde {\cal K}}_{\oone}^T ({\bm x}_0, {\bm x}_1 ;
  {\bm z}) \, = \, \frac{\am^2 \, \beta_2}{\pi^2} \, \frac{{\bm
      z}-{\bm x}_0}{|{\bm z}-{\bm x}_0|} \cdot \frac{{\bm z}-{\bm
      x}_1}{|{\bm z}-{\bm x}_1|} \, \int\limits_0^\infty dq \, dq' \,
  J_1 (q \, |{\bm z}-{\bm
    x}_0|) \, J_1 (q' \, |{\bm z}-{\bm x}_1|) \notag \\
  \times \, \frac{q^{2} \, \ln \frac{q^2 \,
      e^{-5/3}}{\mu_{\overline{{\text{MS}}}}^2} - q'^{2} \, \ln \frac{q'^2 \,
      e^{-5/3}}{\mu_{\overline{{\text{MS}}}}^2} }{q^{2} - q'^{2}} \,
  \frac{1}{\left( 1 + \am \beta_2 \ln \frac{q^2 \,
        e^{-5/3}}{\mu_{\overline{{\text{MS}}}}^2} \right) \, \left( 1 + \am
      \beta_2 \ln \frac{q'^2 \, e^{-5/3}}{\mu_{\overline{{\text{MS}}}}^2}
    \right)}
\end{align}
with ${\tilde {\cal K}}_{\oone}^T$ denoting the transverse part of the
kernel.  Since our intent is to extract the scale of the running
coupling in transverse coordinate space ignoring power corrections, we
expand the denominators in \eq{B1} into geometric series and repeat
the steps which led from \eq{A3} to \eq{A4} writing
\begin{align}\label{B2}
  \am^2 \, & {\tilde {\cal K}}_{\oone}^T ({\bm x}_0, {\bm x}_1 ; {\bm
    z}) \, = \, \frac{\am^2 \, \beta_2}{\pi^2} \, \frac{{\bm z}-{\bm
      x}_0}{|{\bm z}-{\bm x}_0|} \cdot \frac{{\bm z}-{\bm x}_1}{|{\bm
      z}-{\bm x}_1|} \, \sum\limits_{n,m=0}^\infty \, (- \am
  \beta_2)^{n+m} \, \int\limits_0^\infty dq \, dq' \, J_1 (q \, |{\bm
    z}-{\bm
    x}_0|)  \notag \\
  & \times \, J_1 (q' \, |{\bm z}-{\bm x}_1|) \, \left\{ \ln
    \frac{{\bm q}^2 \, e^{-5/3}}{\mu_{\overline{{\text{MS}}}}^2} + {\bm q}'^{2}
    \, \int\limits_0^1 d \beta \, \frac{1}{{\bm q}^{2} (1-\beta) +
      {\bm q}'^{2} \beta} \right\} \, \ln^n \frac{q^2 \,
    e^{-5/3}}{\mu_{\overline{{\text{MS}}}}^2} \, \ln^m \frac{q'^2 \,
    e^{-5/3}}{\mu_{\overline{{\text{MS}}}}^2}.
\end{align}
It is hard to perform $q$ and $q'$ integrations for a general term in
the series characterized by some values of $n$ and $m$. However, we
can try estimating the first correction, i.e., the $n=1, m=0$ term:
\begin{align}\label{B3}
  \frac{\am^2 \, \beta_2}{\pi^2} \, \frac{{\bm z}-{\bm x}_0}{|{\bm
      z}-{\bm x}_0|} \cdot \frac{{\bm z}-{\bm x}_1}{|{\bm z}-{\bm
      x}_1|} \, (- \am \beta_2) \, \int\limits_0^\infty dq \, dq' \,
  J_1 (q \, |{\bm z}-{\bm x}_0|) J_1 (q' \, |{\bm z}-{\bm x}_1|)
  \notag \\ \times \, \left\{ \ln \frac{{\bm q}^2 \,
      e^{-5/3}}{\mu_{\overline{{\text{MS}}}}^2} + {\bm q}'^{2} \,
    \int\limits_0^1 d \beta \, \frac{1}{{\bm q}^{2} (1-\beta) + {\bm
        q}'^{2} \beta} \right\} \, \ln \frac{q^2 \,
    e^{-5/3}}{\mu_{\overline{{\text{MS}}}}^2} .
\end{align}
(The $n=0, m=1$ term will be constructed by replacing ${\bm x}_0
\leftrightarrow {\bm x}_1$ in the result of evaluating (\ref{B3}).)
The $q'$-integral can be easily done in \eq{B3}, along with the
$q$-integral in the first term in the brackets, yielding
\begin{align}\label{B4}
  \frac{\am^2 \, \beta_2}{\pi^2} \, \frac{{\bm z}-{\bm x}_0}{|{\bm
      z}-{\bm x}_0|^2} \cdot \frac{{\bm z}-{\bm x}_1}{|{\bm z}-{\bm
      x}_1|^2} \, (- \am \beta_2) \, \Bigg\{ \ln^2 \left( \frac{4 \,
      e^{-5/3 - 2 \gamma}}{|{\bm z}-{\bm x}_0|^2 \,
      \mu_{\overline{{\text{MS}}}}^2} \right) + |{\bm z}-{\bm x}_0| \, |{\bm
    z}-{\bm x}_1| \notag \\ \times \, \int\limits_0^1 d \beta \,
  \frac{1}{\beta} \, \sqrt{\frac{1-\beta}{\beta}} \,
  \int\limits_0^\infty dq \, q \, J_1 (q \, |{\bm z}-{\bm x}_0|) \,
  K_1 \left( \sqrt{\frac{1-\beta}{\beta}} \, q \, |{\bm z}-{\bm x}_1|
  \right) \, \ln \frac{q^2 \, e^{-5/3}}{\mu_{\overline{{\text{MS}}}}^2}
  \Bigg\}.
\end{align}
To perform the $q$-integral we first re-write the remaining logarithm
as a derivative
\begin{align}\label{B5}
  \frac{\am^2 \, \beta_2}{\pi^2} \, \frac{{\bm z}-{\bm x}_0}{|{\bm
      z}-{\bm x}_0|^2} \cdot \frac{{\bm z}-{\bm x}_1}{|{\bm z}-{\bm
      x}_1|^2} \, (- \am \beta_2) \, \Bigg\{ \ln^2 \left( \frac{4 \,
      e^{-5/3 - 2 \gamma}}{|{\bm z}-{\bm x}_0|^2 \,
      \mu_{\overline{{\text{MS}}}}^2} \right) + |{\bm z}-{\bm x}_0| \, |{\bm
    z}-{\bm x}_1| \notag \\ \times \, \int\limits_0^1 d \beta \,
  \frac{1}{\beta} \, \sqrt{\frac{1-\beta}{\beta}} \, \frac{d}{d
    \lambda} \left[ \int\limits_0^\infty dq \, q \, J_1 (q \, |{\bm
      z}-{\bm x}_0|) \, K_1 \left( \sqrt{\frac{1-\beta}{\beta}} \, q
      \, |{\bm z}-{\bm x}_1| \right) \, \left( \frac{q^2 \,
        e^{-5/3}}{\mu_{\overline{{\text{MS}}}}^2} \right)^\lambda
  \right]\Bigg|_{\lambda =0} \Bigg\}.
\end{align}
Performing the $q$-integration yields
\begin{align}\label{B6}
  \frac{\am^2 \, \beta_2}{\pi^2} \, \frac{{\bm z}-{\bm x}_0}{|{\bm
      z}-{\bm x}_0|^2} \cdot \frac{{\bm z}-{\bm x}_1}{|{\bm z}-{\bm
      x}_1|^2} \, (- \am \beta_2) \, \Bigg\{ \ln^2 \left( \frac{4 \,
      e^{-5/3 - 2 \gamma}}{|{\bm z}-{\bm x}_0|^2 \,
      \mu_{\overline{{\text{MS}}}}^2} \right) + \frac{|{\bm z}-{\bm
      x}_0|^2}{|{\bm z}-{\bm x}_1|^2} \notag \\ \times \,
  \int\limits_0^1 d \beta \, \frac{1}{1-\beta} \, \frac{d}{d
    \lambda} \Bigg[ \left( \frac{4 \, e^{-5/3}}{|{\bm z}-{\bm x}_1|^2
      \, \mu_{\overline{{\text{MS}}}}^2} \, \frac{\beta}{1-\beta}
  \right)^\lambda \, \Gamma (1+\lambda) \, \Gamma (2+\lambda) \notag
  \\ \times \, F \left( 1+\lambda, 2+\lambda; 2; - \frac{\beta \,|{\bm
        z}-{\bm x}_0|^2 }{(1-\beta) \, |{\bm z}-{\bm x}_1|^2} \right)
  \Bigg]\Bigg|_{\lambda =0} \Bigg\}.
\end{align}
Using the definition of hypergeometric functions we write
\begin{align}\label{B7}
  F \left( 1+\lambda, 2+\lambda; 2; z \right) \, = \,
  \frac{1}{1-z} - \lambda \, \frac{1}{1-z} \, \left[ 1 + \ln (1-z) +
    \frac{1}{z} \, \ln (1-z) \right] + o(\lambda^2).
\end{align}
With the help of \eq{B7} the differentiation with respect to $\lambda$
can be easily carried out in \eq{B6}. After integrating over $\beta$
we obtain
\begin{align}\label{B8}
  & - \am \, {\cal K}^{\text{LO}} ({\bm x}_0, {\bm x}_1 ; {\bm z}) \,
  (\am \beta_2)^2 \, \Bigg\{ \ln \left( \frac{4 \, e^{-5/3 - 2
        \gamma}}{|{\bm z}-{\bm x}_0|^2 \, \mu_{\overline{{\text{MS}}}}^2}
  \right) \notag \\ & \times \, \frac{|{\bm z} -{\bm x}_0|^2 \, \ln
    \left( \frac{4 \, e^{-5/3 - 2 \gamma}}{|{\bm z}-{\bm x}_1|^2 \,
        \mu_{\overline{{\text{MS}}}}^2} \right) - |{\bm z}-{\bm x}_1|^2 \, \ln
    \left( \frac{4 \, e^{-5/3 - 2 \gamma}}{|{\bm z}-{\bm x}_0|^2 \,
        \mu_{\overline{{\text{MS}}}}^2}
    \right)}{|{\bm z} -{\bm x}_0|^2 - |{\bm z}-{\bm x}_1|^2} \notag \\
  & + \, \frac{|{\bm z} -{\bm x}_1|^2 \, \left[ \text{Li}_2 \left( 1 -
        \frac{|{\bm z} -{\bm x}_0|^2}{|{\bm z}-{\bm x}_1|^2} \right) -
      \text{Li}_2 (1) \right]- |{\bm z} -{\bm x}_0|^2 \, \left[
      \text{Li}_2 \left( 1 - \frac{|{\bm z} -{\bm x}_1|^2}{|{\bm
            z}-{\bm x}_0|^2} \right) - \text{Li}_2 (1) \right]}{|{\bm
      z} -{\bm x}_0|^2 - |{\bm z}-{\bm x}_1|^2} \Bigg\}.
\end{align}
With the help of ``transverse'' part of \eq{K1int5}, in which we
replace $N_f \rightarrow - 6 \pi \beta_2$, we derive the following
expansion:
\begin{align}\label{B9}
  & \am^2 \, {\tilde {\cal K}}_{\oone}^T ({\bm x}_0, {\bm x}_1 ; {\bm
    z}) \, = \, \am \, {\cal K}^{\text{LO}} ({\bm x}_0, {\bm x}_1 ;
  {\bm z}) \notag \\ & \times \, \Bigg\{ \am \beta_2 \, \frac{|{\bm z}
    -{\bm x}_0|^2 \, \ln \left( \frac{4 \, e^{-5/3 - 2 \gamma}}{|{\bm
          z}-{\bm x}_1|^2 \, \mu_{\overline{{\text{MS}}}}^2} \right) - |{\bm
      z}-{\bm x}_1|^2 \, \ln \left( \frac{4 \, e^{-5/3 - 2
          \gamma}}{|{\bm z}-{\bm x}_0|^2 \, \mu_{\overline{{\text{MS}}}}^2}
    \right)}{|{\bm z} -{\bm x}_0|^2 - |{\bm z}-{\bm x}_1|^2} \notag \\
  & - \, (\am \beta_2)^2 \, \Bigg[ \ln \left( \frac{4 \, e^{-5/3 - 2
        \gamma}}{|{\bm z}-{\bm x}_0|^2 \, \mu_{\overline{{\text{MS}}}}^2}
  \right) \, \frac{|{\bm z} -{\bm x}_0|^2 \, \ln \left( \frac{4 \,
        e^{-5/3 - 2 \gamma}}{|{\bm z}-{\bm x}_1|^2 \,
        \mu_{\overline{{\text{MS}}}}^2} \right) - |{\bm z}-{\bm x}_1|^2 \, \ln
    \left( \frac{4 \, e^{-5/3 - 2 \gamma}}{|{\bm z}-{\bm x}_0|^2 \,
        \mu_{\overline{{\text{MS}}}}^2} \right)}{|{\bm z} -{\bm x}_0|^2 - |{\bm
      z}-{\bm x}_1|^2} \notag \\ & + \frac{|{\bm z} -{\bm x}_1|^2 \,
    \left[ \text{Li}_2 \left( 1 - \frac{|{\bm z} -{\bm x}_0|^2}{|{\bm
            z}-{\bm x}_1|^2} \right) - \text{Li}_2 (1) \right]- |{\bm
      z} -{\bm x}_0|^2 \, \left[ \text{Li}_2 \left( 1 - \frac{|{\bm z}
          -{\bm x}_1|^2}{|{\bm z}-{\bm x}_0|^2} \right) - \text{Li}_2
      (1) \right]}{|{\bm z} -{\bm x}_0|^2 - |{\bm z}-{\bm x}_1|^2}
  \Bigg] + \ldots \Bigg\}.
\end{align}
Here the ellipsis include not only the higher order terms in $\am
\beta_2$, but also the term quadratic in $\am \beta_2$ with the ${\bm
  x}_0 \leftrightarrow {\bm x}_1$ replacement. It can be shown that
the term in the last line of \eq{B9} is numerically small compared to
the other terms in the series. Dropping that term yields
\begin{align}\label{B10}
  & \am^2 \, {\tilde {\cal K}}_{\oone}^T ({\bm x}_0, {\bm x}_1 ; {\bm
    z}) \, \approx \, \am \, {\cal K}^{\text{LO}} ({\bm x}_0, {\bm
    x}_1 ; {\bm z}) \notag \\ & \times \, \am \beta_2 \, \frac{|{\bm z}
    -{\bm x}_0|^2 \, \ln \left( \frac{4 \, e^{-5/3 - 2 \gamma}}{|{\bm
          z}-{\bm x}_1|^2 \, \mu_{\overline{{\text{MS}}}}^2} \right) - |{\bm
      z}-{\bm x}_1|^2 \, \ln \left( \frac{4 \, e^{-5/3 - 2
          \gamma}}{|{\bm z}-{\bm x}_0|^2 \, \mu_{\overline{{\text{MS}}}}^2}
    \right)}{|{\bm z} -{\bm x}_0|^2 - |{\bm z}-{\bm x}_1|^2} \notag \\ 
  & \times \, \Bigg[ 1 - \am \beta_2 \, \ln \left( \frac{4 \, e^{-5/3 -
        2 \gamma}}{|{\bm z}-{\bm x}_0|^2 \, \mu_{\overline{{\text{MS}}}}^2}
  \right) + \ldots \Bigg] \, \Bigg[ 1 - \am \beta_2 \, \ln \left(
    \frac{4 \, e^{-5/3 - 2 \gamma}}{|{\bm z}-{\bm x}_1|^2 \,
      \mu_{\overline{{\text{MS}}}}^2} \right) + \ldots \Bigg].
\end{align}
It appears likely that the higher order corrections would continue the
geometric series in \eq{B10} with the same constants under the
logarithms. Resummation of such series yields the ``transverse'' part
of \eq{K1sdr1}.

Now we have to evaluate the longitudinal (instantaneous) part of
\eq{K1sdr}, given by the last term in the curly brackets in that
equation:
\begin{align}\label{B21}
  \am^2 \, {\tilde {\cal K}}_{\oone}^L ({\bm x}_0, {\bm x}_1 ; {\bm
    z}) \, = \, - \, 4 \, \am^2 \, \beta_2 \, \int \frac{d^2
    q}{(2\pi)^2} \frac{d^2 q'}{(2\pi)^2} \ e^{ -i {\bm q}\cdot ({\bm
      z}-{\bm x}_0) +i {\bm q}' \cdot ({\bm z}-{\bm x}_1) } \notag \\ 
  \times \, \frac{\ln ({\bm q}^{2}/{\bm q}'^{2})}{{\bm q}^{2} - {\bm
      q}'^{2}} \, \frac{1}{\left( 1 + \am \beta_2 \ln \frac{{\bm q}^2
        \, e^{-5/3}}{\mu_{\overline{{\text{MS}}}}^2} \right) \, \left( 1 + \am
      \beta_2 \ln \frac{{\bm q}'^2 \, e^{-5/3}}{\mu_{\overline{{\text{MS}}}}^2}
    \right)}.
\end{align}
Integrating over the angles gives
\begin{align}\label{B22}
  \am^2 \, {\tilde {\cal K}}_{\oone}^L ({\bm x}_0, {\bm x}_1 ; {\bm
    z}) \, = \, - \, \frac{\am^2 \, \beta_2}{\pi^2} \,
  \int\limits_0^\infty dq \, dq' \, q \, q' \, J_0 (q \, |{\bm z}-{\bm
    x}_0|) \, J_0 (q' \, |{\bm z}-{\bm x}_1|) \notag \\
  \times \, \frac{\ln ({q}^{2}/{q}'^{2})}{{q}^{2} - {q}'^{2}} \,
  \frac{1}{\left( 1 + \am \beta_2 \ln \frac{{q}^2 \,
        e^{-5/3}}{\mu_{\overline{{\text{MS}}}}^2} \right) \, \left( 1 + \am
      \beta_2 \ln \frac{{q}'^2 \, e^{-5/3}}{\mu_{\overline{{\text{MS}}}}^2}
    \right)}.
\end{align}
Again the exact integration does not appear possible. Instead we will
expand the running coupling denominators to the linear order in the
logarithms and evaluate the following term
\begin{align}\label{B23}
  \frac{\am^2 \, \beta_2}{\pi^2} \, \int\limits_0^\infty dq \, dq'
  \, q \, q' \, J_0 (q \, |{\bm z}-{\bm x}_0|) \, J_0 (q' \, |{\bm
    z}-{\bm x}_1|) \, \frac{\ln ({q}^{2}/{q}'^{2})}{{q}^{2} -
    {q}'^{2}} \, \am \beta_2 \ln \frac{{q}^2 \,
    e^{-5/3}}{\mu_{\overline{{\text{MS}}}}^2},
\end{align}
which we rewrite using \eq{trick} as
\begin{align}\label{B24}
  \frac{\am^2 \, \beta_2}{\pi^2} \, \am \beta_2 \, \int\limits_0^1
  d \beta \, \int\limits_0^\infty dq \, dq' \, q \, q' \, J_0 (q \,
  |{\bm z}-{\bm x}_0|) \, J_0 (q' \, |{\bm z}-{\bm x}_1|) \,
  \frac{1}{{q}^{2} (1-\beta) + {q}'^{2} \beta} \, \ln \frac{{q}^2 \,
    e^{-5/3}}{\mu_{\overline{{\text{MS}}}}^2}.
\end{align}
Performing the $q'$-integral first yields
\begin{align}\label{B25}
  \frac{\am^3 \, \beta_2^2}{\pi^2} \, \int\limits_0^1 \frac{d
    \beta}{\beta} \, \frac{d}{d \lambda} \, \int\limits_0^\infty dq \,
  q \, J_0 (q \, |{\bm z}-{\bm x}_0|) \, K_0 \left(
    \sqrt{\frac{1-\beta}{\beta}} \, q \, |{\bm z}-{\bm x}_1| \right)
  \, \left( \frac{{q}^2 \, e^{-5/3}}{\mu_{\overline{{\text{MS}}}}^2}
  \right)^\lambda \, \Bigg|_{\lambda =0},
\end{align}
where we have again replaced the logarithm by a derivative of a power.
Integrating over $q$ we obtain
\begin{align}\label{B26}
  \frac{\am^3 \, \beta_2^2}{\pi^2} \, \int\limits_0^1 \frac{d
    \beta}{1 - \beta} \, \frac{1}{|{\bm z}-{\bm x}_1|^2} \, \frac{d}{d
    \lambda} \, \left[ \left( \frac{4 \, e^{-5/3}}{|{\bm z}-{\bm
          x}_1|^2 \, \mu_{\overline{{\text{MS}}}}^2} \, \frac{\beta}{1 - \beta}
    \right)^\lambda \, \Gamma^2 (1+\lambda) \right. \notag \\ \times
  \, \left. F \left( 1+\lambda, 1+\lambda; 1; - \frac{\beta \,|{\bm
          z}-{\bm x}_0|^2 }{(1-\beta) \, |{\bm z}-{\bm x}_1|^2}
    \right) \right] \Bigg|_{\lambda =0}.
\end{align}\label{B27}
Using the expansion of the hypergeometric function
\begin{align}
  F \left( 1+\lambda, 1+\lambda; 1; z \right) \, = \, \frac{1}{1-z} -
  \lambda \, \frac{2}{1-z} \, \ln \left( 1-z \right) + o(\lambda^2)
\end{align}
we can perform the differentiation with respect to $\lambda$ in
\eq{B26} and integrate over $\beta$ to get
\begin{align}\label{B28}
  \frac{\am^3 \, \beta_2^2}{\pi^2} \, \frac{1}{|{\bm z}-{\bm x}_0|^2 -
    |{\bm z}-{\bm x}_1|^2} \, \Bigg\{ \ln \frac{|{\bm z}-{\bm
      x}_0|^2}{|{\bm z}-{\bm x}_1|^2} \, \ln \left( \frac{4 \, e^{-5/3
        - 2 \gamma}}{|{\bm z}-{\bm x}_0|^2 \, \mu_{\overline{{\text{MS}}}}^2}
  \right) \notag \\ + \, \text{Li}_2 \left( 1 - \frac{|{\bm z} -{\bm
        x}_0|^2}{|{\bm z}-{\bm x}_1|^2} \right) - \text{Li}_2 \left( 1
    - \frac{|{\bm z} -{\bm x}_1|^2}{|{\bm z}-{\bm x}_0|^2} \right)
  \Bigg\}.
\end{align}
Similar to the above one can show that the dilogarithms in \eq{B28}
are numerically small and can be neglected compared to the rest of the
expression. The Fourier transforms in the leading term in the
expansion of running coupling corrections in \eq{B21} was performed in
obtaining \eq{K1int5} (see also the derivation of \eq{A24}). That
result, combined with \eq{B28}, allows us to write
\begin{align}\label{B29}
  & \am^2 \, {\tilde {\cal K}}_{\oone}^L ({\bm x}_0, {\bm x}_1 ; {\bm
    z}) \, = \, - \, \frac{\am^2 \, \beta_2}{\pi^2} \, \frac{\ln
    \frac{|{\bm z}-{\bm x}_0|^2}{|{\bm z}-{\bm x}_1|^2}}{|{\bm z}-{\bm
      x}_0|^2 - |{\bm z}-{\bm x}_1|^2} \notag \\ & \times \, \Bigg[ 1
  - \am \beta_2 \, \ln \left( \frac{4 \, e^{-5/3 - 2 \gamma}}{|{\bm
        z}-{\bm x}_0|^2 \, \mu_{\overline{{\text{MS}}}}^2} \right) + \ldots
  \Bigg] \, \Bigg[ 1 - \am \beta_2 \, \ln \left( \frac{4 \, e^{-5/3 -
        2 \gamma}}{|{\bm z}-{\bm x}_1|^2 \, \mu_{\overline{{\text{MS}}}}^2}
  \right) + \ldots \Bigg].
\end{align}
This expansion again demonstrates the emerging geometric series when
higher order fermion loops are included in ${\tilde {\cal
    K}}_1^{\text{NLO}}$. Resumming those series to all orders we
obtain the ``longitudinal'' part of \eq{K1sdr1}.

Finally, adding \eq{B10} and \eq{B29} together yields
\begin{align}\label{B30}
  \am^2 \, {\tilde {\cal K}}_{\oone} ({\bm x}_0, {\bm x}_1 ; {\bm z})
  \, =& \, \am^2 \, {\tilde {\cal K}}_{\oone}^T ({\bm x}_0, {\bm x}_1
  ; {\bm z}) + \am^2 \, {\tilde {\cal K}}_{\oone}^L ({\bm x}_0, {\bm
    x}_1 ; {\bm z}) \notag \\ \approx & \, \am \, {\cal K}^{\text{LO}}
  ({\bm x}_0, {\bm x}_1 ; {\bm z}) \, \am \beta_2 \, \ln \left(
    \frac{4 \, e^{-\frac{5}{3} -2 \gamma}}{R^2 ({\bm x}_0, {\bm x}_1 ;
      {\bm z}) \, \mu_{\overline{{\text{MS}}}}^2} \right) \notag \\ 
  \times \, \Bigg[ 1 - \am \beta_2 \, \ln \left( \frac{4 \, e^{-5/3 -
        2 \gamma}}{|{\bm z}-{\bm x}_0|^2 \,
      \mu_{\overline{{\text{MS}}}}^2} \right) & + \ldots \Bigg] \,
  \Bigg[ 1 - \am \beta_2 \, \ln \left( \frac{4 \, e^{-5/3 - 2
        \gamma}}{|{\bm z}-{\bm x}_1|^2 \,
      \mu_{\overline{{\text{MS}}}}^2} \right) + \ldots \Bigg],
\end{align}
which, after resumming the geometric series gives \eq{K1sdr1}, as
desired.

%%%%%%%%%%%%%%%%%%%%%%%%%%%%%%%%%%%%%%%%%%%%%%%%%%%%%%%%%%%%%%%%%%%%%%%%%
\renewcommand{\theequation}{C\arabic{equation}}
  \setcounter{equation}{0}
\section{Comparison with dispersive calculation of~\cite{Gardi:2006}}
\label{sec:comparison-with-dispersive}

Here we demonstrate explicitly that our all orders expression in
Sect.~\ref{all_orders}, Eqs.~(\ref{Kall1}) to~(\ref{Kall}), agree with the
results found in~\cite{Gardi:2006}. In~\cite{Gardi:2006} the running coupling
effects were presented in terms of transverse and longitudinal Borel functions
$B^T$ and $B^L$ with Borel parameter $u$. With our convention for the kernel
and the shorthand notation ${\bm r}_1=\bm x_0-\bm z$, ${\bm r}_2=\bm x_1-\bm z$ and the
replacement ${\bm p}\to{\bm q}$, $\bm q\to\bm q'$ to match notations in this
paper, we quote the expressions of~\cite{Gardi:2006} as
\begin{subequations}
    \label{eq:mom-space-Borels}
\begin{align}
\label{eq:mom-space-trans-res}
  {\cal K}_{\bm x,\bm y;\bm z}\, B^T(u,{\bm r}_1\mu_{\overline{{\text{MS}}}},{\bm r}_2\mu_{\overline{{\text{MS}}}}) = &\, 
e^{\frac53 u} \int \frac{d^2p\,d^2q}{(2\pi)^2}\
e^{i{\bm q}\cdot{\bm r}_1}\  e^{-i{{\bm q}'}\cdot{\bm r}_2} \
\frac{{\bm q}\cdot{{\bm q}'}}{{\bm q}^2\, {{\bm q}'}^2}\
\frac{{{\bm q}'}^2\left(\frac{{\bm q}^2}{\mu_{\overline{{\text{MS}}}}^2}\right)^{-u}
-{\bm q}^2\left(\frac{{{\bm q}'}^2}{\mu_{\overline{{\text{MS}}}}^2}\right)^{-u}}{{{\bm q}'}^2-{\bm q}^2}
 \\
  {\cal K}_{\bm x,\bm y;\bm z}\, B^L(u,{\bm r}_1\mu_{\overline{{\text{MS}}}},{\bm r}_2\mu_{\overline{{\text{MS}}}}) = &\, 
e^{\frac53 u} \int \frac{d^2p\,d^2q}{(2\pi)^2}\
e^{i{\bm q}\cdot{\bm r}_1} \ e^{-i{{\bm q}'}\cdot{\bm r}_2} \
\frac{  \left(\frac{{\bm q}^2}{\mu_{\overline{{\text{MS}}}}^2}\right)^{-u} 
           - \left(\frac{{{\bm q}'}^2}{\mu_{\overline{{\text{MS}}}}^2}\right)^{-u} }{
   {\bm q}^2-{{\bm q}'}^2}
\ .
\end{align}
\end{subequations}
Everything else follows from the definitions for the Borel representation of
what is called the coupling function $R({\bm r}_1 \Lambda,{\bm r}_2 \Lambda)$
in~\cite{Gardi:2006} (expressed in terms of the QCD scale $\Lambda$) which
takes the form of a sum of transverse and longitudinal contributions
\begin{align}
  \label{eq:Rdef}
    R({\bm r}_1 \Lambda,{\bm r}_2 \Lambda) = 
  \frac{1}{\beta_0}
  \int_0^{\infty}du & \ T(u) 
\left(\frac{\mu_{\overline{{\text{MS}}}}^2}{\Lambda^2}\right)^{-u}  
\Big(B^T(u,{\bm r}_1 \mu_{\overline{{\text{MS}}}},{\bm r}_2 \mu_{\overline{{\text{MS}}}})
+B^L(u,{\bm r}_1\mu_{\overline{{\text{MS}}}},{\bm r}_2\mu_{\overline{{\text{MS}}}})\Big)
\end{align}
where for one-loop running as employed in this paper $T(u)$ is to be set to
one. The notation for the $\beta$-function coefficients is such that
$\beta_2=\beta_0/\pi$. All that is left to do to
compare~\eqref{eq:mom-space-Borels} to our present results is to perform the
Borel integral.  Up to an overall factor of
$\am/\pi$, this amounts to replacing the
Borel powers $\left(\frac{{\bm
      a}^2}{\mu_{\overline{{\text{MS}}}}^2}\right)^{-u}$ by the corresponding
geometric series
$\frac1{1+\frac{\beta_0\am}{\pi}\ln\left(\frac{{\bm
        a}^2}{\mu_{\overline{{\text{MS}}}}^2\,e^{\frac53}}\right)}$ for both
${\bm a}={\bm q}$ and ${\bm a}={\bm q}'$. All that
is left to do is to factor out a common denominator $\frac1{
  \left(1+\frac{\beta_0\am}{\pi} \ln\left(\frac{{\bm q}^2\,e^{-\frac53}}{\mu_{\overline{{\text{MS}}}}^2}\right)\right)
  \left(1+\frac{\beta_0\am}{\pi} \ln\left(\frac{{{\bm
            q}'}^2\,e^{-\frac53}}{\mu_{\overline{{\text{MS}}}}^2}\right)\right)
}$. One finds
\begin{subequations}
  \label{eq:mom-diagrammatic}
 \begin{align}
    \label{eq:mom-diagrammatic-transv}
    {\cal K}_{\bm x,\bm y;\bm z}\, 
    R^T(u,{\bm r}_1\mu_{\overline{{\text{MS}}}},{\bm r}_2\mu_{\overline{{\text{MS}}}}) = &\, 
\frac{\am}{\pi}\int \frac{d^2q\,d^2q'}{(2\pi)^2}
\ e^{i{\bm q}\cdot{\bm r}_1}\ e^{-i{{\bm q}'}\cdot{\bm r}_2}
\notag \\ & \times
\frac{{\bm q}\cdot{{\bm q}'}}{{\bm q}^2\, {{\bm q}'}^2}\
\frac{
\frac{{{\bm
      q}'}^2\left(1+\frac{\beta_0\am}{\pi}
    \ln\left(\frac{{{\bm q}'}^2}{\mu_{\overline{{\text{MS}}}}^2\,e^{\frac53}}\right)
\right)
-{\bm q}^2\left(1+\frac{\beta_0\am}{\pi}\ln\left(\frac{{\bm q}^2}{\mu_{\overline{{\text{MS}}}}^2\,e^{\frac53}}\right)
\right)}{{{\bm q}'}^2-{\bm q}^2}
}{
   \left(1+\frac{\beta_0\am}{\pi}
     \ln\left(\frac{{\bm q}^2}{\mu_{\overline{{\text{MS}}}}^2\,e^{\frac53}}\right)\right)
   \left(1+\frac{\beta_0\am}{\pi}
     \ln\left(\frac{{{\bm q}'}^2}{\mu_{\overline{{\text{MS}}}}^2\,e^{\frac53}}\right)\right)
   }
\displaybreak[0]
\notag \\ = \, 
\frac{\am}{\pi}\int \frac{d^2q\,d^2q'}{(2\pi)^2}
\ e^{i{\bm q}\cdot{\bm r}_1}\ e^{-i{{\bm q}'}\cdot{\bm r}_2} & \,
%\notag \\ & \times
\frac{{\bm q}\cdot{{\bm q}'}}{{\bm q}^2\, {{\bm q}'}^2}\
\frac{
1+
\frac{\beta_0\am}{\pi}
\frac{{{\bm q}'}^2
  \ln\left(\frac{{{\bm q}'}^2}{\mu_{\overline{{\text{MS}}}}^2\,e^{\frac53}}\right)
-{\bm q}^2
\ln\left(\frac{{\bm q}^2}{\mu_{\overline{{\text{MS}}}}^2\,e^{\frac53}}\right)}{{{\bm q}'}^2-{\bm q}^2}
}{
   \left(1+\frac{\beta_0\am}{\pi}
     \ln\left(\frac{{\bm q}^2}{\mu_{\overline{{\text{MS}}}}^2\,e^{\frac53}}\right)\right)
   \left(1+\frac{\beta_0\am}{\pi}
     \ln\left(\frac{{{\bm q}'}^2}{\mu_{\overline{{\text{MS}}}}^2\,e^{\frac53}}\right)\right)
   }
\intertext{ and}
{\cal K}_{\bm x,\bm y;\bm z}\, 
R^L(u,{\bm r}_1\mu_{\overline{{\text{MS}}}},{\bm r}_2\mu_{\overline{{\text{MS}}}}) = &
\frac{\am}{\pi}\int \frac{d^2q\,d^2q'}{(2\pi)^2}
\ e^{i{\bm q}\cdot{\bm r}_1}\ e^{-i{{\bm q}'}\cdot{\bm r}_2}
\notag \\ & \times
\frac{
\frac{
  \left(1+\frac{\beta_0\am}{\pi}
    \ln\left(\frac{{{\bm q}'}^2}{\mu_{\overline{{\text{MS}}}}^2\,e^{\frac53}}\right)
\right)
-\left(1+\frac{\beta_0\am}{\pi}
  \ln\left(\frac{{\bm q}^2}{\mu_{\overline{{\text{MS}}}}^2\,e^{\frac53}}\right)
\right)}{{\bm q}^2-{{\bm q}'}^2}
}{
   \left(1+\frac{\beta_0\am}{\pi}
     \ln\left(\frac{{\bm q}^2}{\mu_{\overline{{\text{MS}}}}^2\,e^{\frac53}}\right)\right)
   \left(1+\frac{\beta_0\am}{\pi}
     \ln\left(\frac{{{\bm q}'}^2}{\mu_{\overline{{\text{MS}}}}^2\,e^{\frac53}}\right)\right)
   }
\notag \\ = \,
\frac{\am}{\pi}\int \frac{d^2q\,d^2q'}{(2\pi)^2}
\ e^{i{\bm q}\cdot{\bm r}_1}\ e^{-i{{\bm q}'}\cdot{\bm r}_2} & \,
 \frac{\frac{\beta_0\am}{\pi}
   \frac{
\ln\left(\frac{{{\bm q}'}^2}{\mu_{\overline{{\text{MS}}}}^2\,e^{\frac53}}\right)
   -\ln\left(\frac{{\bm
         q}^2}{\mu_{\overline{{\text{MS}}}}^2\,e^{\frac53}}\right)
}{{\bm q}^2-{{\bm q}'}^2}}{
   \left(1+\frac{\beta_0\am}{\pi}
     \ln\left(\frac{{\bm q}^2}{\mu_{\overline{{\text{MS}}}}^2\,e^{\frac53}}\right)\right)
   \left(1+\frac{\beta_0\am}{\pi}
     \ln\left(\frac{{{\bm q}'}^2}{\mu_{\overline{{\text{MS}}}}^2\,e^{\frac53}}\right)\right)
   }
\ .
 \end{align}
\end{subequations}
The sum of these contributions is in full agreement with~(\ref{Kall1}) as
advertised.

%%%%%%%%%%%%%%%%%%%%%%%%%%%%%%%%%%%%%%%%%%%%%%%%%%%%%%%%%%%%%%%%%%%%%%%%%%%%%%%

%\bibliography{newletter,rc}                   %<-----------BIBLIOGRAPHIE
%\bibliographystyle{JHEP}  

\providecommand{\href}[2]{#2}\begingroup\raggedright\endgroup

\end{document}